\documentclass[11pt,a4paper,aps,nofootinbib]{article}
\usepackage{jcappub}
\usepackage[utf8]{inputenc}
\usepackage{color}
\usepackage{amsmath, bm, physics}
\usepackage{amssymb, mathtools}
\usepackage{subcaption}
\usepackage{hyperref}

\title{Prospects for detecting anisotropies and polarization of the stochastic gravitational wave background with ground-based detectors}

\author[a,1]{G. Metasti,\note{Corresponding author.}}
\author[a]{C.~R. Contaldi,}
\author[b,c]{and M. Peloso}

\affiliation[a]{Blackett Laboratory, Imperial College London, SW7 2AZ, United Kingdom}
\affiliation[b]{Dipartimento di Fisica e Astronomia ``G. Galilei", Universit\`a degli Studi di Padova, via Marzolo 8, I-35131 Padova, Italy}
\affiliation[c]{INFN, Sezione di Padova, via Marzolo 8, I-35131 Padova, Italy}

\emailAdd{g.mentasti21@imperial.ac.uk}

\abstract{
We build an analytical framework to study the observability of anisotropies and a net chiral polarization of the Stochastic Gravitational Wave Background (SGWB) with a generic network of ground-based detectors. We apply this formalism to perform a Fisher forecast of the performance of a network consisting of the current interferometers (LIGO, Virgo and KAGRA) and planned third-generation ones, such as the Einstein Telescope and Cosmic Explorer. Our results yield limits on the observability of anisotropic modes, spanning across noise- and signal-dominated regimes. We find that if the isotropic component of the SGWB has an amplitude close to the current limit, third-generation interferometers with an observation time of $10$ years can measure multipoles (in a spherical harmonic expansion) up to $\ell = 8$ with ${\cal O }\left( 10^{-3} - 10^{-2} \right)$ accuracy relative to the isotropic component, and an ${\cal O }\left( 10^{-3} \right)$ amount of net polarization. For weaker signals, the accuracy worsens as roughly the inverse of the SGWB amplitude.
}

\begin{document}

\maketitle

\tableofcontents

\section{Introduction}

Following the first observation of the signal originating from the merging of a massive, compact system \cite{PhysRevLett.116.061102}, Gravitational Wave (GW) 
science has become one of the most rapidly growing fields in fundamental physics. Since then, many more detections have been made thanks to the constantly improving sensitivity of the existing detector network that now includes LIGO, Virgo, and KAGRA \cite{PhysRevD.104.022004}. The sensitivity of detectors is expected to keep improving in the coming years. In the next ten to twenty years, a new generation of ground-based interferometers, such as Einstein Telescope (ET) \cite{Maggiore:2019uih} and Cosmic Explorer (CE) \cite{CosmicExplorer} will also become operational. These detectors will work alongside improved versions of the current ones. The expectation is that these instruments will be several orders of magnitude more sensitive to gravitational wave strain \cite{Maggiore:2019uih,Amalberti:2021kzh} than existing ones. This sensitivity increase will significantly enrich our catalogue of merger events of many different origins.

We still await the detection of a Stochastic Gravitational Wave Background (SGWB) of any origin. Today, only upper limits exist as obtained from analyzing LIGO--Virgo--KAGRA data. The energy density per logarithmic frequency interval of a scale-invariant and an astrophysical-sourced SGWB are constrained to be, respectively, $\Omega_{\rm GW} \lesssim 5\times 10^{-9}$ \cite{PhysRevD.104.022004} and $\Omega_{\rm GW} \lesssim 8.6\times 10^{-9}$ \cite{LIGOisotropic2022,LIGOanisotropic2022,astrostochRegimbau,astrostochsiskalek}.

There are several well-motivated theoretical reasons to expect that astrophysical and cosmological phenomena can generate an SGWB. On the one hand, at frequencies probed by ground-based detectors (1Hz-10kHz), an astrophysical component due to the superposition of many unresolved sources, primarily black holes and neutron star binary coalescences, is expected. On the other hand, several mechanisms can produce a cosmological SGWB. Among them are several non-minimal models of inflation, phase transitions, and topological defects (see, e.g. \cite{firstorder1,LISAcosmogroup,Bartolo_2022,ValbusaDallArmi:2022htu}).

Detection of an SGWB will generate significant interest in determining the originating mechanism. Characterizing the SGWB will involve several efforts and may be complicated by the presence of multiple underlying source mechanisms. Firstly, to understand the nature of an SGWB and distinguish separate components, the frequency dependence of the isotropic part of $\Omega_{\rm GW}$ will be analyzed \cite{Bartolo_2022}. Secondly, a measurement of any anisotropies in the SGWB will also be of interest for characterization and component separation. In particular, any directional dependence will assist in distinguishing between galactic and extra-galactic components and searching for any correlation with known tracers of structure \cite{SGWB-astro-sources,SGWB-astro-correl,SGWB-astro-correl-1,Ricciardone:2021kel}. Furthermore, anisotropies in the astrophysical background are expected to be correlated with the large-scale structure due to how the GW originate and how they propagate through a perturbed universe~\cite{Contaldi:2016koz,Cusin:2017fwz,Bartolo:2019oiq,ValbusaDallArmi:2020ifo}. Thirdly, crucial information on the nature of the SGWB might be obtainable from studying its polarization. Some cosmological mechanisms, for instance, a coupling between a pseudo-scalar inflaton and a gauge field~\cite{Sorbo:2016rzu}, can result in a fully circularly polarized SGWB~\cite{Crowder:2012ik,Martinovic:2021hzy}. A degree of polarization might also be present in the astrophysical SGWB due to Poisson fluctuations in the (finite) number of unresolved sources~\cite{ValbusaDallArmi:2023ydl}. 

It is worth noting that an SGWB could be, in general, non-stationary in that its statistical properties could vary in time due to the non-stationarity of the source. In this case, the analysis of such a background is still possible in principle \cite{popcorn}. For our purpose, we assume that the non-stationary signal of the SGWB can be analyzed and removed from the data stream, considering the fully stationary component of the background.

Previous work \cite{Allen:1996gp} studied the instrumental response to an anisotropic SGWB by two L-shaped ground-based detectors. The formalism was extended in \cite{Mentasti:2020yyd} and \cite{anisotropy-DomckePeloso} to include the case of a polarized SGWB. This {\sl paper} extends the aforementioned formalism to a more general one. We consider anisotropies and an asymmetry in polarizations of the SGWB within a unified framework.
We generalize the results of \cite{Allen:1996gp} to a generic network of L-shaped and triangular-shaped detectors. We give explicit analytic expressions for the response of the network to the monopole, the dipole, and the quadrupole in a  spherical harmonics expansion of the SGWB, extending some results of \cite{anisotropy-DomckePeloso} (these expressions significantly speed up the numerical forecasts). We provide a pedagogical derivation of the variance of the measurement, also paying attention to the contributions for the variance of the signal, to assert the level of accuracy  that can ever be ideally achieved by a given network configuration of arbitrary low noise \cite{mentasti2023}. Finally, we apply this formalism to forecast the measurement of the anisotropies and the polarization of the SGWB with a network of existing and third-generation ground-based interferometers.

The paper is structured so readers only interested in the results can find a self-contained and complete presentation and discussion in  section~\ref{sec:shortsummary}. The remainder of the paper serves to introduce and develop the mathematical tools necessary for this forecast. In section~\ref{sec:theSGWB}, we review the formalism for describing SGWBs. Sections~\ref{sec:DataStream} and~\ref{sec:measure} introduce the formalism and notation for the observables involved in ground-based networked detectors. The formalism is given for a generic network made of an arbitrary number of triangular-shaped detectors and L-shaped detectors. Section~\ref{sec:Anisotropies} applies this formalism to the detection of anisotropic multipoles of the SGWB, whilst section~\ref{sec:Polarization} includes a similar analysis for the detection of a net polarization. In both cases, we individually highlight the signal- and noise-dominated regimes before obtaining general Fisher forecasts for the observational uncertainty. Many of the technical derivations underlying the main results are elaborated in appendices~\ref{app:sigvar}-\ref{app:Gamma-MDQ}.  

\section{Main results}
\label{sec:shortsummary}

\begin{figure}[ht!]
	\centerline{
		\includegraphics[width=0.8\textwidth,angle=0]{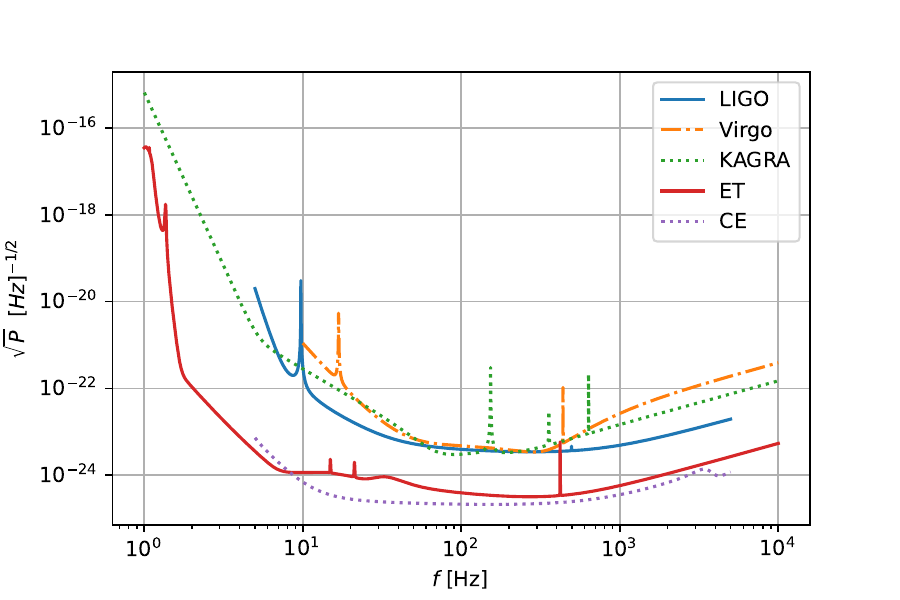}
	}
	\caption{ Predicted (square root of the) Power Spectral Density (PSD) of Advanced LIGO, Advanced Virgo, KAGRA, Einstein Telescope, and Cosmic Explorer as a function of the frequency. Data were taken from ref. \cite{LVKcurves}}
	\label{fig:allpsd}
\end{figure}

This section presents our main results on the forecast of measurements of the anisotropies and the net polarization of the SGWB with a network of ground-based interferometers. Our formalism assumes a generic set of L-shaped and triangularly-shaped GW detectors, with noise specified by eqs.~\eqref{variance-ijLX}, \eqref{variance-ijLL},  and \eqref{variance-ij}.
In our evaluations and explicit studies, we consider a noise model according to the advanced versions of the existing LIGO, Virgo, and Kagra (denoted collectively as LVK in our figures) and the design sensitivity of Einstein Telescope (ET) and Cosmic Explorer (CE).
We show the noise Power Spectral Densities for these instruments in figure~\ref{fig:allpsd}, where the numerical values are taken from ref. \cite{LVKcurves}. Our numerical results are obtained by integrating in frequency in the range $f\in [1,1000]$ Hz.

The section is divided into three parts. Subsections~\ref{subsec:result_anisotropic} and~\ref{subsec:result_polarized} present our results for, respectively, anisotropies and a net polarization of the SGWB. These results are discussed in subsection~\ref{subsec:result_discussion}, which also serves as a concluding section of our paper, as the remainder of this work is devoted to deriving the methodology used to obtain the results presented here.

\subsection{Results: forecast observations of anisotropic modes}
\label{subsec:result_anisotropic}

We consider a Gaussian SGWB characterized by
\begin{equation}
\left\langle h^*(f,\hat n)h(f,\hat n) \right\rangle=\frac{3H_0^2}{32\pi^3f^3}\, \Omega_{0}\left(\frac{f}{f_0}\right)^\alpha\sum_{\ell m}\delta^{GW}_{\ell m}Y_{\ell m}(\hat n)\,,
\label{intro:PSD_def}
\end{equation}
which has a factorized dependence on frequency $f$ and on the GWs arrival direction ${\hat n}$.
The parameter $\Omega_0$ is the standard fractional energy density in GWs at the pivot frequency $f_0$ (in our analysis, we fix $f_0 = 100$ Hz). In our explicit evaluations, we fix the spectral tilt $\alpha$ to two typical values considered in the literature, namely to $\alpha=0$, as for a (nearly) scale-invariant cosmological inflationary signal, and to $\alpha = 2/3$, as expected for the stochastic background due to the black hole-black hole and black hole-neutron star binary system inspirals \cite{popcorn}. 

The last factor in eq.~\eqref{intro:PSD_def} encodes the non-trivial angular structure in terms of a sum over spherical harmonics $Y_{\ell m}$ (where the $z-$axis is taken to coincide with the rotation axis of the Earth), which generalizes the more common isotropic studies in which only the monopole term is assumed. Due to the normalization of the spherical harmonics, we fix $\delta_{00}=\sqrt{4\pi}$ so that the coefficients $\delta^{GW}_{\ell m}$ encode the relative strength of the various multipoles relative to the monopole. In our explicit analysis, we assume that only one given multipole is present, with a magnitude much smaller than one, corresponding to a nearly isotropic signal. 

For each interferometer $i$ in the network (and for each channel in the case of triangular-shaped detectors, which are made of multiple instruments), we take the data stream $m_i(t)$ that results from the superposition of the GW signal and instrumental noise, 
\begin{align}
m_i(t)=s_i(t)+n_i(t) \;. 
\end{align}
We (window-)Fourier transform these functions and take every possible cross-correlation of the data streams at different sites. We then integrate the cross-correlations in time, using the assumption of a stationary SGWB, so that the statistics underlying the response of the network of detectors is periodic with period $T_e=1$ day. We apply optimal filters, building a set of unbiased observables 
\begin{equation}\label{CmExp}
{\cal C}_{ij,m}\propto \int_{-\infty}^{\infty}df m_i(f)m_j(f)\,Q_m(f) \;\;,\;\; i \neq j \;, 
\end{equation}
whose expectation value depends on the parameters in eq.~\eqref{intro:PSD_def}.  We consider typical values $\alpha=0$ (from cosmology) and $\alpha=2/3$ (from astrophysics) for the tilt of the signal. We construct a $\chi^2$ as a function of the amplitude of the signal $\Omega_0$ and of the anisotropy parameter $\delta_{\ell m}^{GW}$ under consideration (namely, for a specific choice of $\ell$ and $m$). From the $\chi^2$, we forecast the uncertainties under which the network will measure these two quantities under consideration.

\begin{figure*}[ht!]
        \centering
        \begin{subfigure}[b]{0.45\textwidth}
            \includegraphics[width=\textwidth,trim=18 25 18 22, clip]{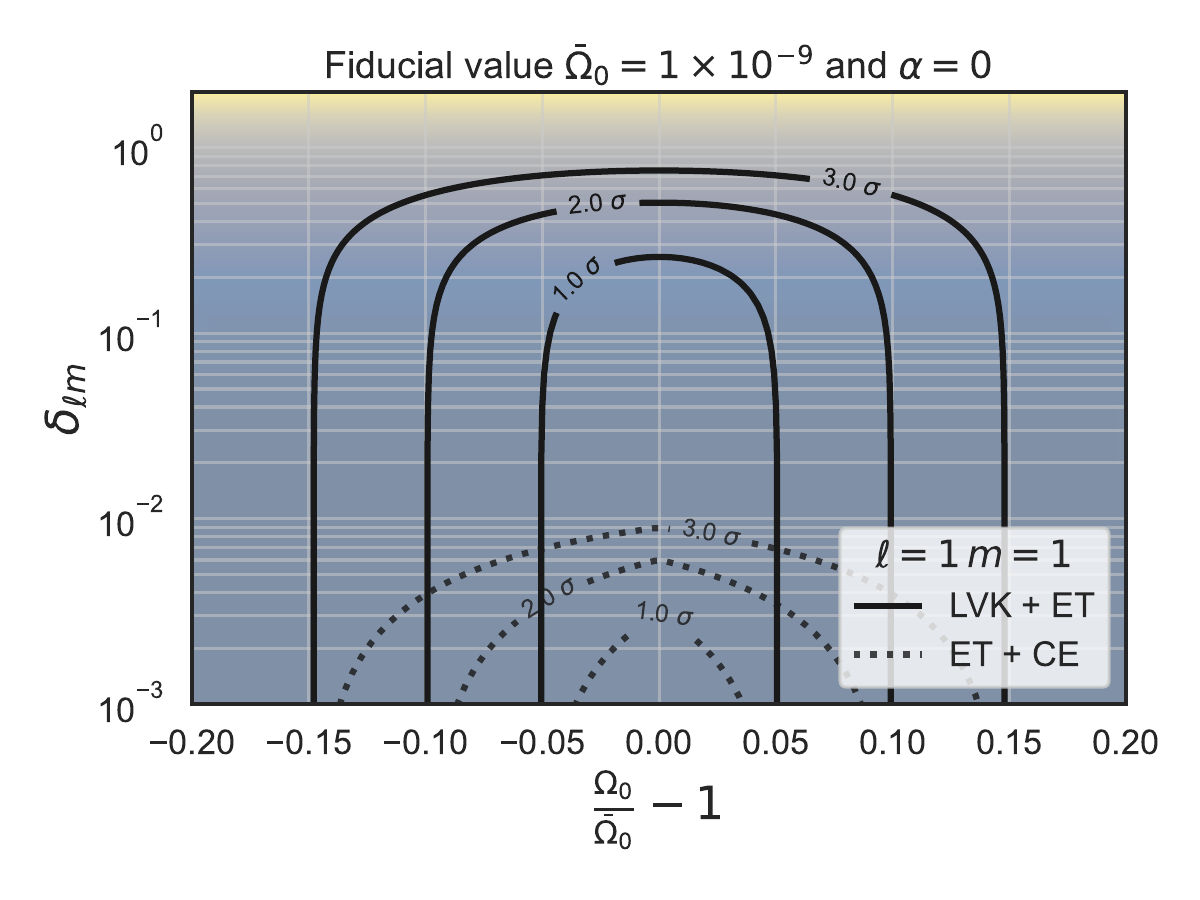}
            %\caption[Network2]%
            %{{\small Network 1}}    
        \end{subfigure}
        \hfill
        \begin{subfigure}[b]{0.45\textwidth}  
            \centering 
            \includegraphics[width=\textwidth,trim=18 25 18 22, clip]{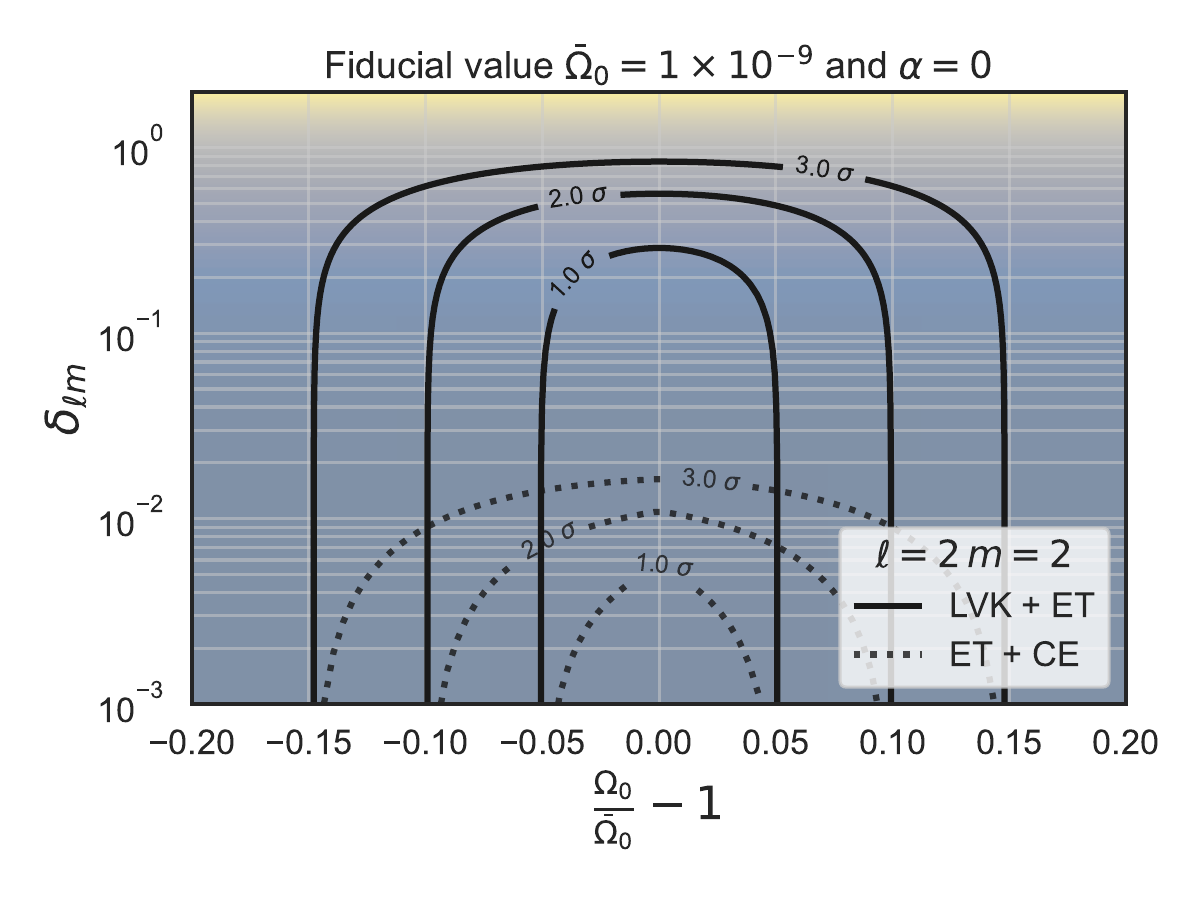}
        \end{subfigure}
        \vskip\baselineskip
        \begin{subfigure}[b]{0.45\textwidth}   
            \centering 
            \includegraphics[width=\textwidth,trim=18 25 18 22, clip]{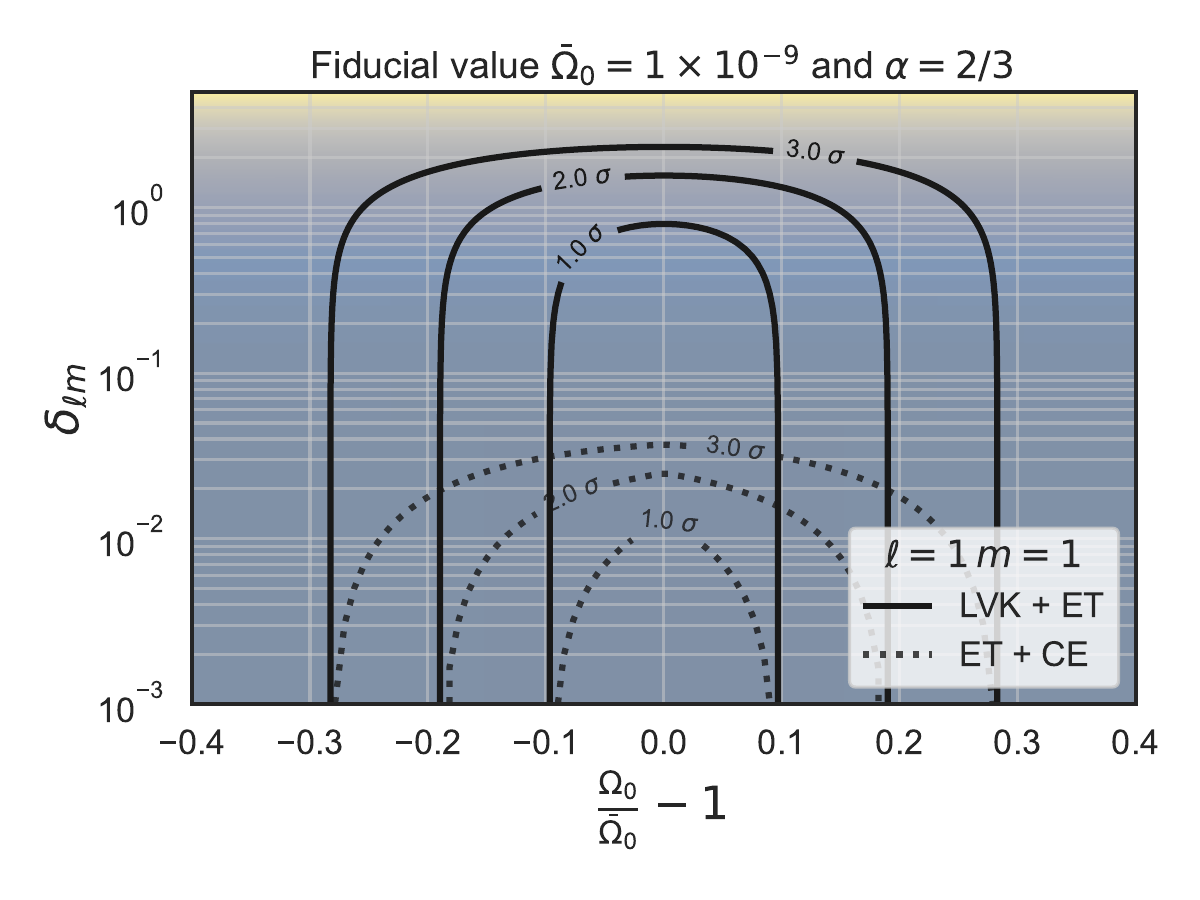}
        \end{subfigure}
        \hfill
        \begin{subfigure}[b]{0.45\textwidth}   
            \centering 
            \includegraphics[width=\textwidth,trim=18 25 18 22, clip]{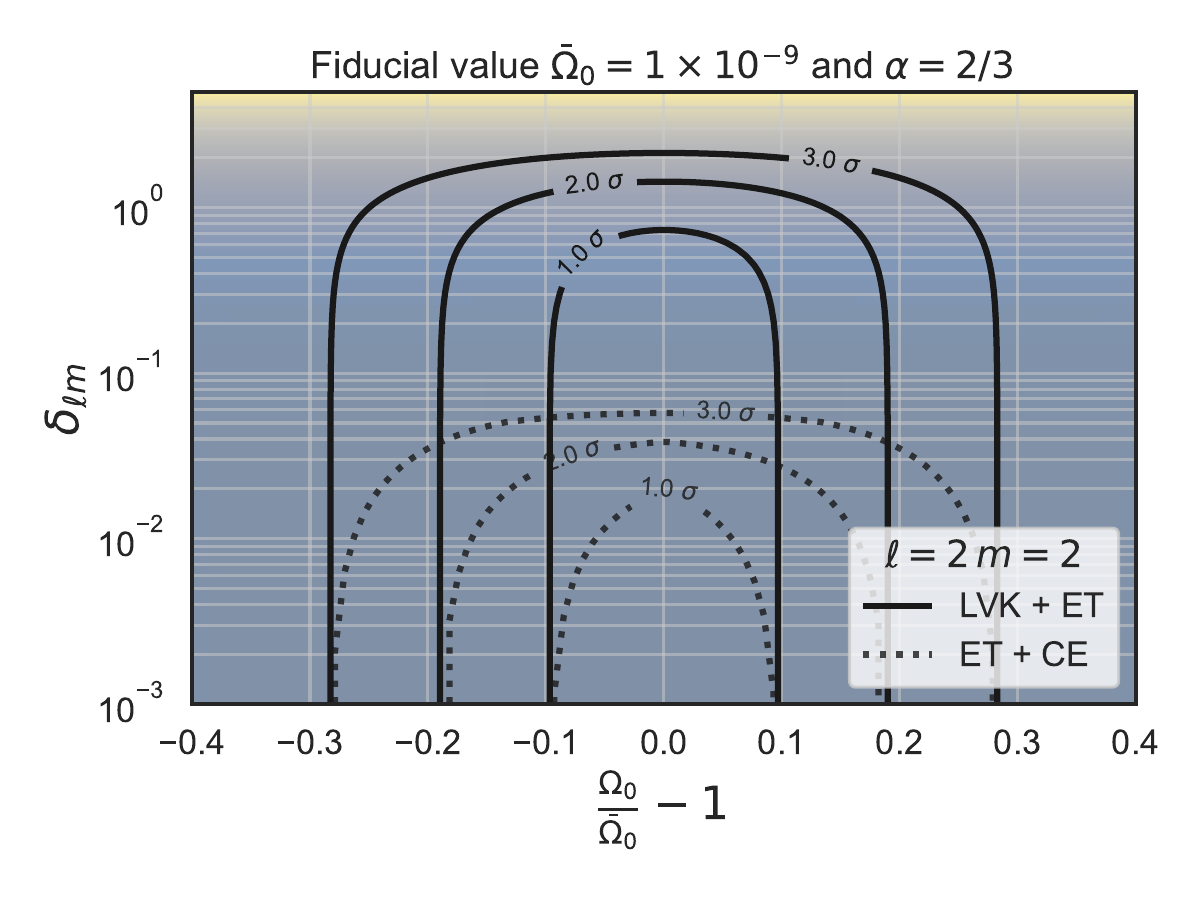}
        \end{subfigure}
        \caption{ Forecast $1\sigma$, $2\sigma$ and $3\sigma$ contours as a function of the amplitude $\Omega_0$ and of one anisotropy parameter $\delta^{GW}_{\ell m}$ for a fiducial isotropic model with fiducial amplitude ${\bar \Omega}_0 = 10^{-9}$, for an observation time of $T=10$ yrs in the LVK+ET (solid lines) and in the ET-CE network (dotted lines) network. The left (respectively, right) panels study the dipole coefficient $\ell = m = 1$ (resp., the quadrupole coefficient $\ell = m = 2$). The upper (resp., lower) panels assume a scale-invariant signal (resp., a blue signal with $\alpha = 2/3$).}
        \label{fig:contour_anisotropies}
    \end{figure*}

In our explicit evaluations, we consider a fiducial isotropic signal, $\delta^{GW}_{\ell m} = 0$ (apart from the monopole term), with amplitude ${\bar \Omega}_0$. We then compute the $\chi^2$ as discussed above, as a function of
$\Omega_0$ and $\delta^{GW}_{\ell m}$. In figure~\ref{fig:contour_anisotropies} we show the $1-,  2-$, and $3-\sigma$ contours obtained from, an increase of, respectively, 
$\delta \chi^2 = 1 ,\, 4 ,\, 9$, with respect to the value obtained for the fiducial signal.~\footnote{The coefficients  $\delta^{GW}_{\ell m}$ of the spherical harmonics decomposition of eq.~\eqref{intro:PSD_def} are in general complex. The limits obtained in our analyses refer to their absolute values.} A measurement outside these contours would reject the hypothesis of isotropy of the signal at the corresponding confidence level. The different panels of the figure correspond to a different multipole analysed and to a different choice of spectral tilt. In all cases the fiducial amplitude is ${\bar \Omega}_0 = 10^{-9}$, which is close to the current upper bound \cite{LIGOisotropic2022}. In all panels, we consider measurements involving third-generation detectors (either correlated with the existing second-generation ones or among themselves), as measurements from the current detectors alone, do not constrain the anisotropy parameters below one.

\begin{figure*}[ht!]

        \centering
        \begin{subfigure}[b]{0.47\textwidth}
            \includegraphics[width=\textwidth,trim=18 25 18 22, clip]{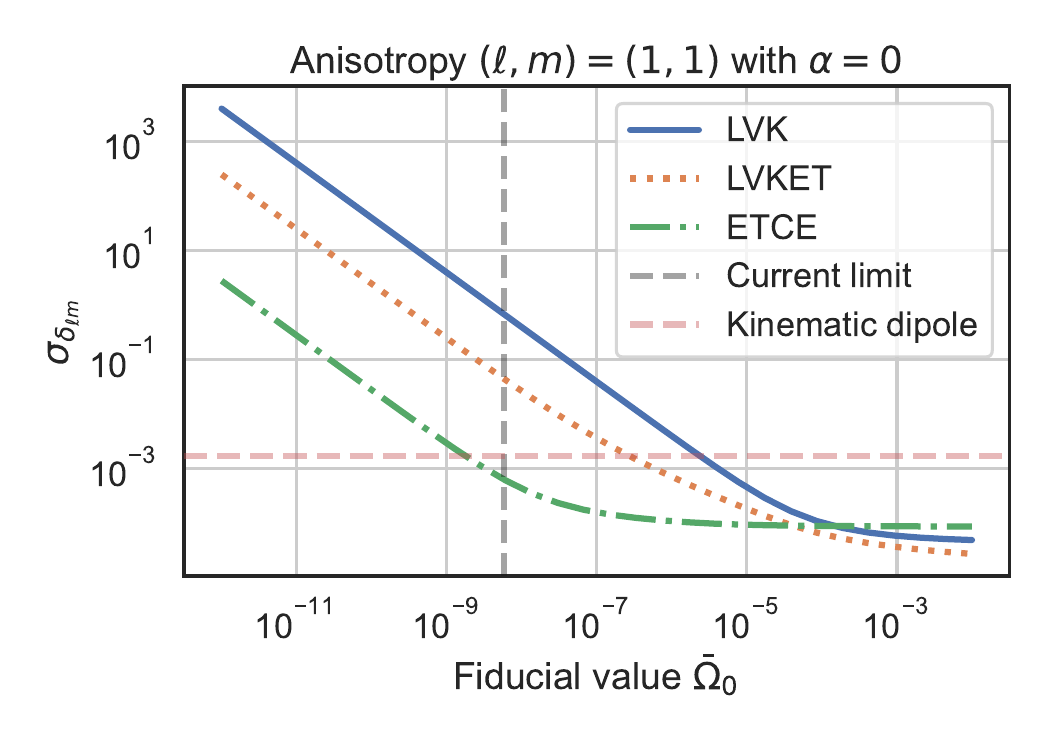}  
        \end{subfigure}
        \hfill
        \begin{subfigure}[b]{0.47\textwidth}  
            \centering 
            \includegraphics[width=\textwidth,trim=18 25 18 22, clip]{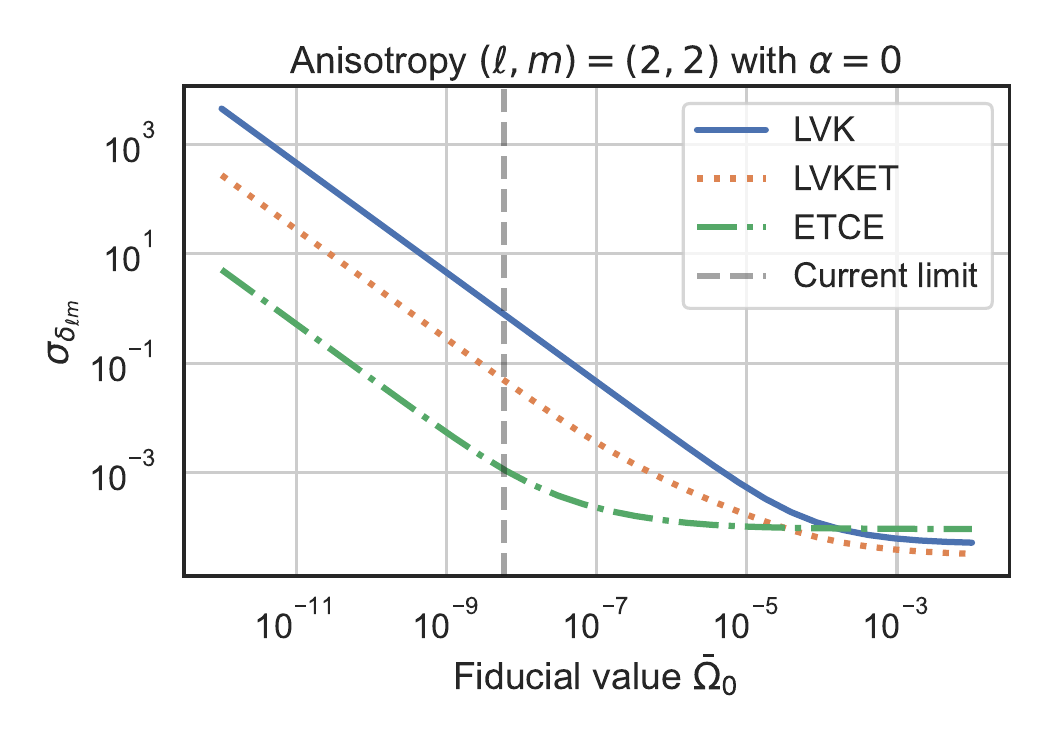}
        \end{subfigure}
        \vskip\baselineskip
        \begin{subfigure}[b]{0.47\textwidth}   
            \centering 
            \includegraphics[width=\textwidth,trim=18 25 18 22, clip]{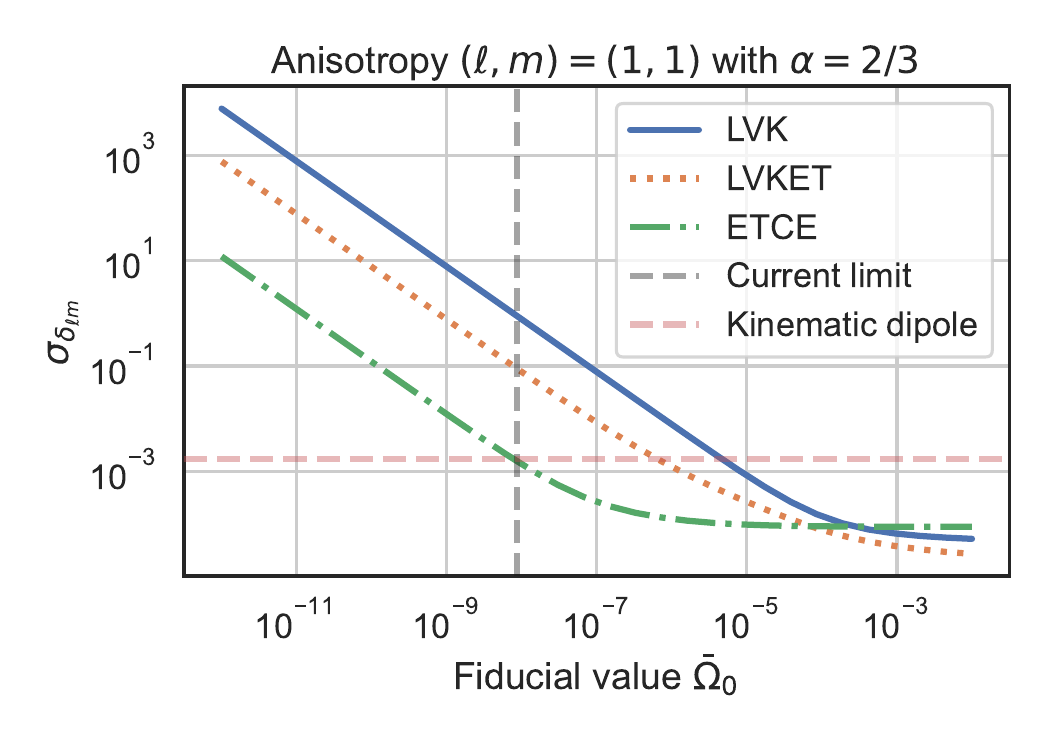}
        \end{subfigure}
        \hfill
        \begin{subfigure}[b]{0.47\textwidth}   
            \centering 
            \includegraphics[width=\textwidth,trim=18 25 18 22, clip]{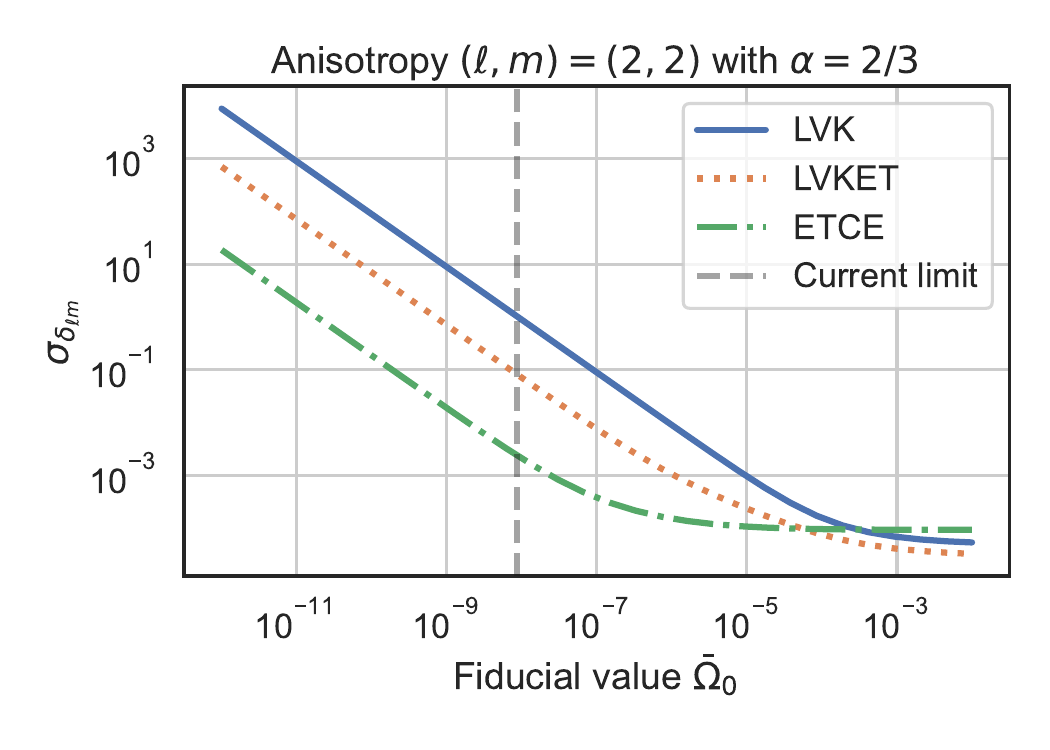}
        \end{subfigure}
        \caption{\small Forecast error in the measurement of the anisotropic coefficient $\delta_{\ell m}^{GW}$ as a function of the fiducial amplitude $\bar\Omega_0$ of the signal for a total observation time of $T=10$ yr and for  different networks. The left (resp. right) panels study the dipole coefficient $\ell = m = 1$ (resp., the quadrupole coefficient $\ell = m = 2$). The upper (resp., lower) panels assume a scale-invariant signal (resp., a blue signal with $\alpha = 2/3$). The horizontal and vertical lines indicate, respectively, the expected value of the kinematic dipole and the current upper bound on the amplitude of $\Omega_0$ \cite{LIGOisotropic2022}.}
        \label{fig:dipole_and_quadrupole}
    \end{figure*}

Marginalizing the results of such analyses over $\Omega_0$, we can then obtain the uncertainty on the measurement of the anisotropy parameter alone. In figure~\ref{fig:dipole_and_quadrupole}, we plot the forecast error in the measurements of a dipole ($\ell = m = 1$) and a quadrupole ($\ell = m = 2$) coefficient, as a function of the fiducial amplitude of $\Omega_0$ and for different networks. 

The vertical line in each panel shows the current upper limit on $\Omega_0$ 
\cite{LIGOisotropic2022}. We also show results in the excluded range of $\Omega_0$ because we want to observe the change of slope in the forecast  uncertainty of the anisotropy. The left (resp., right) portion of the plot is in a regime in which the instrumental noise (resp., the signal) dominates in the measurement (and, at the technical level, in the variance that is present at the denominator of the $\chi^2$). 

One specific targeted anisotropy that we study in figure~\ref{fig:dipole_and_quadrupole} is the one arising from our particular motion. If we assume that the SGWB is isotropic in the CMB rest-frame, then our peculiar motion with velocity $v$ in this frame produces an anisotropic signal analogous to that in the CMB, characterized by a kinematic dipole (suppressed by $v$ wrt the monopole), kinematic quadrupole (suppressed by $v^2$) and so on \cite{LISACosmologyWorkingGroup:2022kbp}~\footnote{In principle, the kinematic effect also changes the frequency dependence of the signal in the various multipoles, invalidating the factorization of eq.~\eqref{intro:PSD_def}). There are two possible contributions to this, one arising from the spectral dependence of $\Omega_{GW} \left( f \right)$ and one from an aberration that mixes different multipoles. The first effect is not present for a power-law $\Omega_{GW}\left( f \right)$ \cite{LISACosmologyWorkingGroup:2022kbp}, as we consider in eq.~\eqref{PowerLawOmega}. The second effect  is suppressed by the smallness of the anisotropies (as it is proportional to spatial gradients of the SGWB energy density \cite{Cusin:2022cbb}). Therefore our analysis of the kinematic effect based on eq. (\ref{intro:PSD_def}) remains accurate. For more general frequency shapes and for a large anisotropic astrophysical SGWB, see \cite{Cusin:2022cbb,ValbusaDallArmi:2022htu}.} As $v/c \simeq 1.2 \times 10^{-3}$ \cite{10.1093/ptep/ptaa104}, the dipole is the largest coefficient induced by this kinematic effect. In the frame of reference chosen by our decomposition in multipoles, the kinematic dipole aligns closely with the $\ell=1$, $m=-1$ multipole, and it has a predicted value of $\delta^{GW}_{1,-1} \simeq 1.7 \times 10^{-3}$
\cite{Allen:1996gp}, that we indicate with a horizontal line in figure~\ref{fig:dipole_and_quadrupole}. 

\begin{figure}[!ht]

\includegraphics[scale=0.4,trim= 20 0 0 0 ,clip]{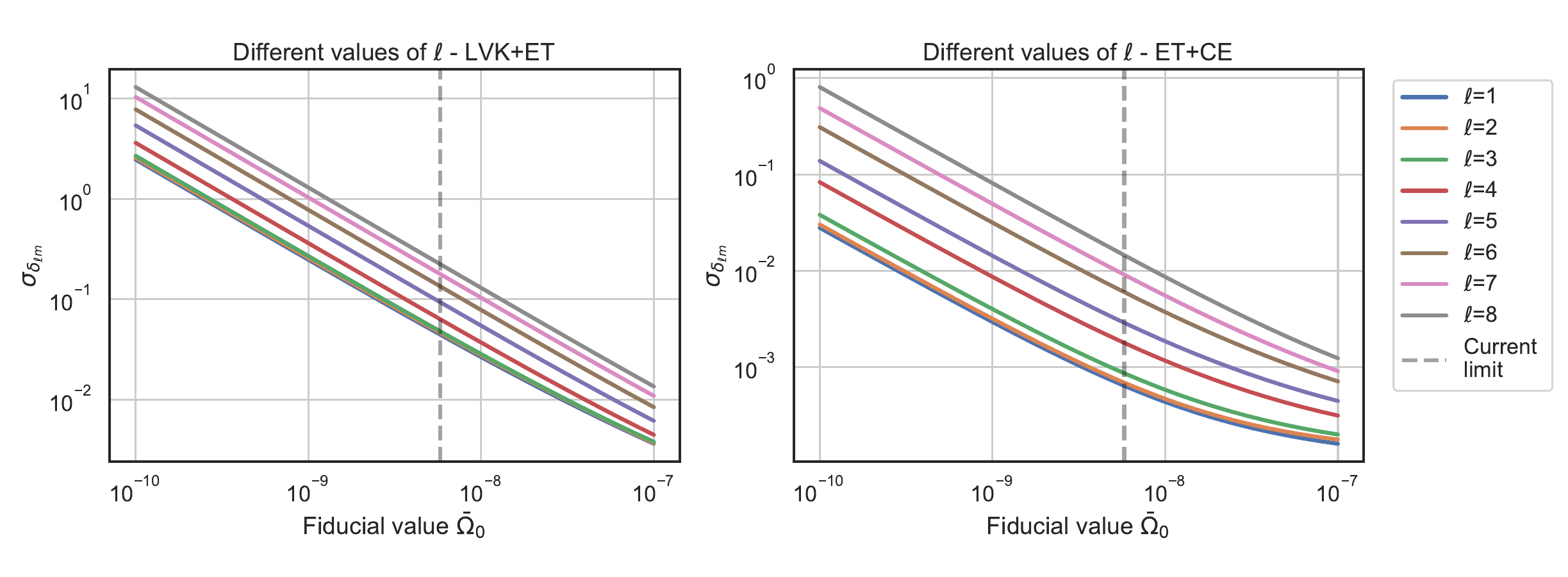}\\

\caption{ Forecast error in the measurement of the coefficients $\delta_{\ell m}$ as a function of the fiducial amplitude $\bar\Omega_0$. Each line corresponds to a different $\ell$ and the value of $m$ that results in the smallest $\sigma_{\delta_{\ell m}}$ for that $\ell$). An observation time $T=10$ yr with the LVK + ET (left panel) and the ET + CE (right panel) network is assumed. The vertical line indicates the current upper bound on the amplitude of $\Omega_0$ \cite{LIGOisotropic2022}.}
	\label{fig:anisotropies_different_l}
\end{figure}

Higher multiples are studied in figure~\ref{fig:anisotropies_different_l}, where we show the forecast errors on various $\delta_{\ell m}^{GW}$, after marginalizing over $\Omega_0$. Each line in the figure is obtained from a separate analysis. For each $\ell$ varying from $1$ to $8$, we make separate analyses for each value of $m$, and we then show in the figure the result for the value of $m$ that is best reconstructed for that $\ell$. 

\subsection{Results: forecast observations of net polarization}
\label{subsec:result_polarized}

We now consider a Gaussian SGWB characterized by 
\begin{equation}
\left\langle h_\lambda^*(f,\hat n)h_\lambda(f,\hat n) \right\rangle=\frac{3H_0^2}{32\pi^3f^3}\, \Omega_{0}\left(\frac{f}{f_0}\right)^\alpha\left(1+p\lambda\right)\,,
\label{intro:PSD_def_pol}
\end{equation}
where the parameter $p$ encodes the amount of net circular polarization of the SGWB, this parameter ranges from $-1$, corresponding to a left-handed background, to $+1$, corresponding to a right-handed background. Within this range, the choice $p=0$ corresponds to an unpolarized background. 

\begin{figure}[!ht]
\centering
\includegraphics[scale=0.4]{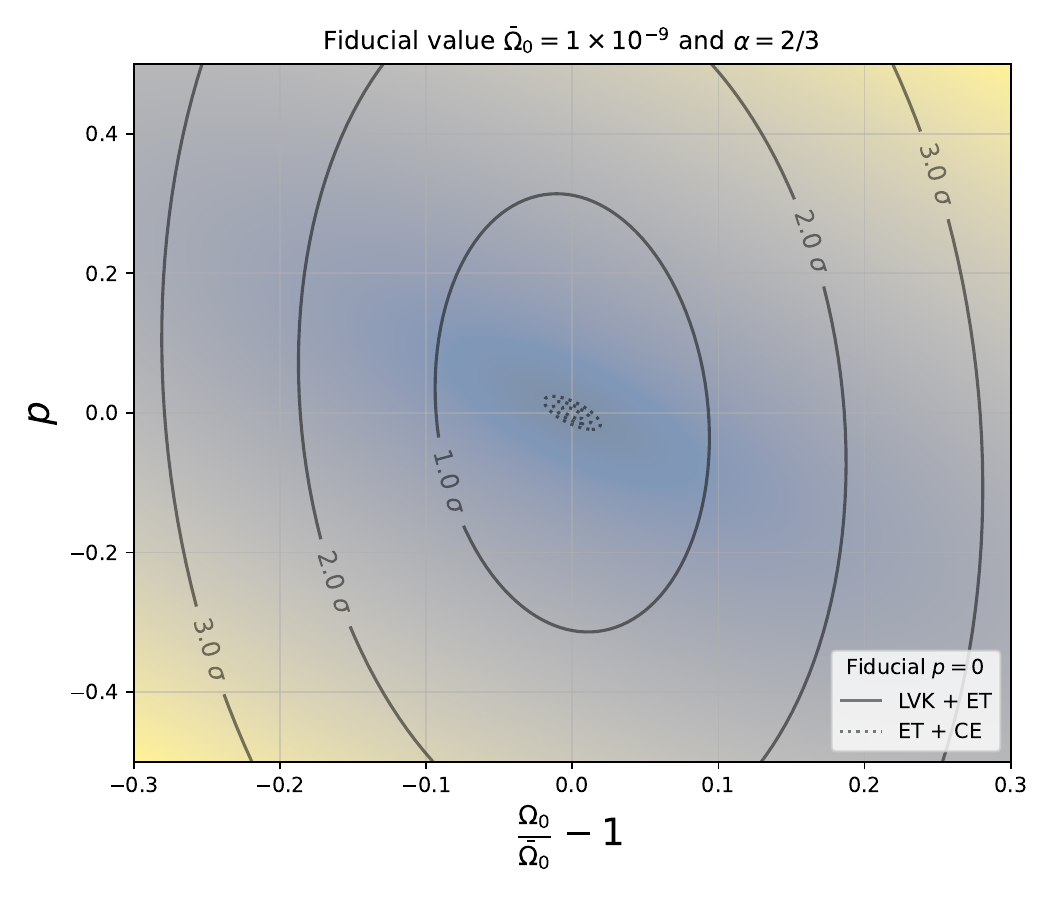}
\includegraphics[scale=0.4]{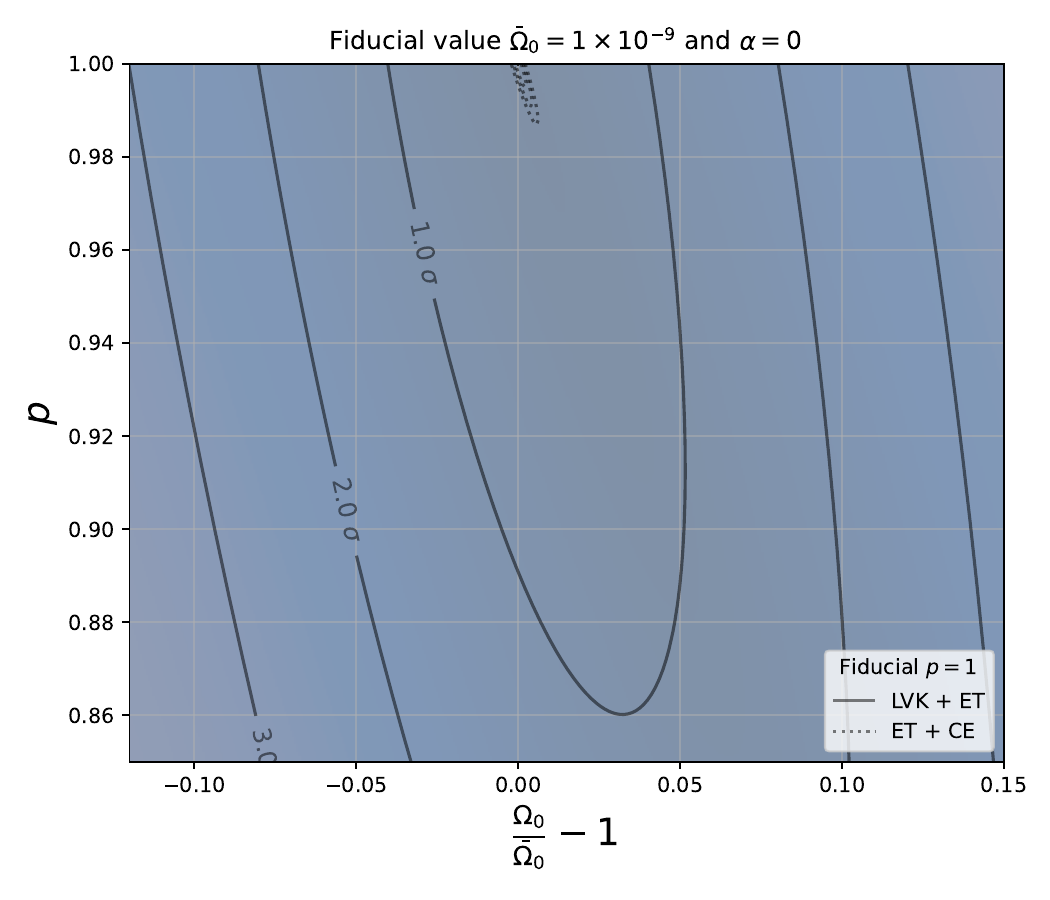}\\

\caption{Forecast $1\sigma$, $2\sigma$ and $3\sigma$ contours as a function of the amplitude $\Omega_0$ and of the polarization parameter $p$ for a fiducial amplitude ${\bar \Omega}_0 = 10^{-9}$, for an observation time of $T=10$ yrs in the LVK+ET (solid lines) and in the ET-CE network (dotted lines, barely visible in the two panels). The left (resp., right) panel assumes a fiducial unpolarized signal with $\alpha = 2/3$, (resp., a fully circularly polarized scale-invariant signal).}
	\label{fig:contours_polarization}
\end{figure}

The study of the net polarization proceeds in the same way as for the  anisotropic search.  We form cross-correlations of the data streams at different sites and build an unbiased observable as a function of the two varying parameters $\Omega_0$ and $p$. figure~\ref{fig:contours_polarization} shows the $1-$, $2-$, and $3-\sigma$ contours in the $\left\{ \Omega_0 ,\, p \right\}$ plane for measurements with the LVK+ET and ET+CE networks. As for the anisotropy, we do not show contour plots from the LVK network alone as they do not yield significant constraints on $p$.

\begin{figure}[!ht]
\centering
\includegraphics[scale=0.4,trim=18 25 18 22, clip]{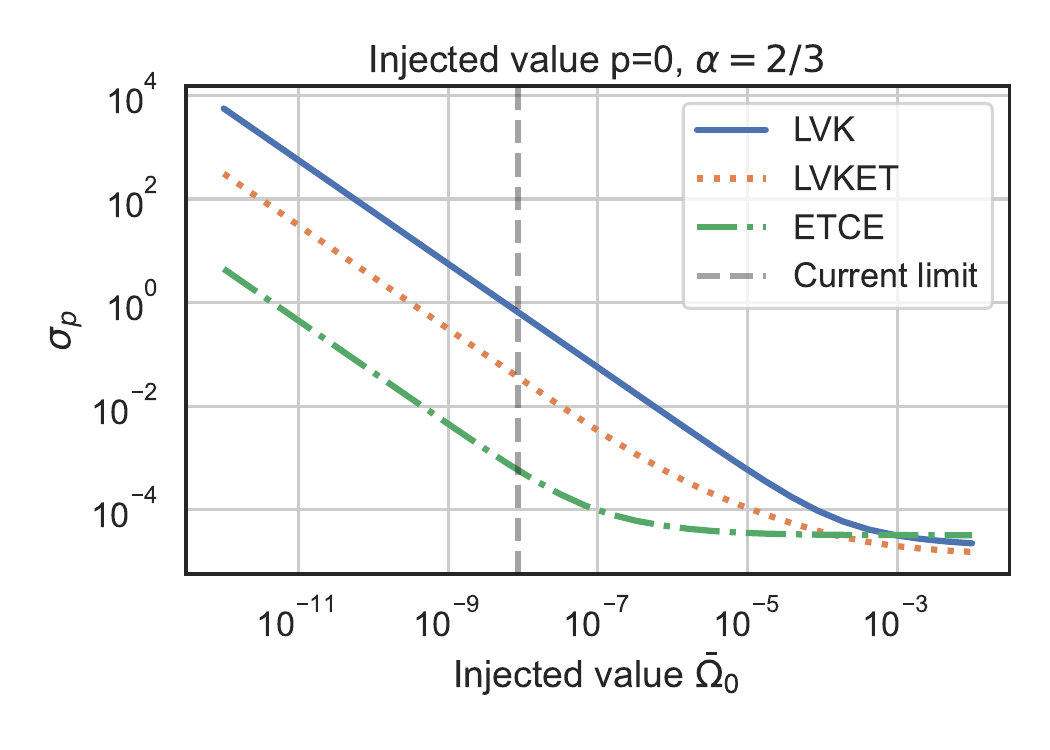}
\includegraphics[scale=0.4,trim=18 25 18 22, clip]{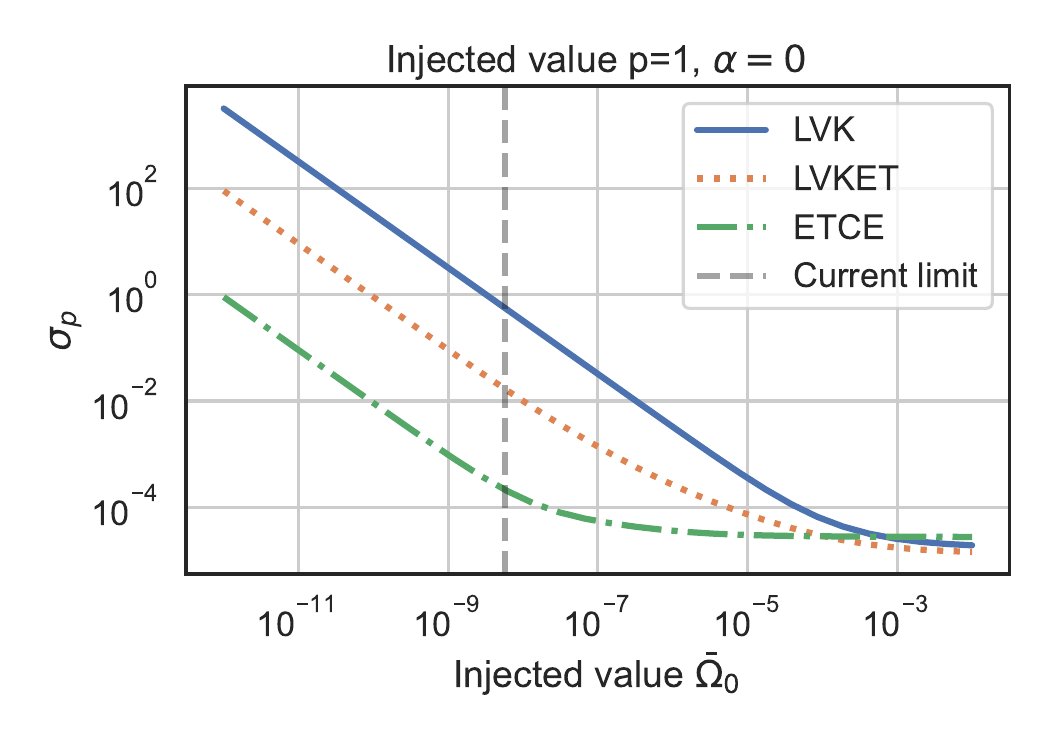}\\

\caption{Forecast error in the measurement of the polarization factor $p$ as a function of the fiducial amplitude $\bar\Omega_0$ for a total observation time of $T=10$ yrs. and different networks. The left (resp., right) panel assumes an unpolarized fiducial signal with $\alpha = 2/3$, (resp., a fully circularly polarized scale-invariant signal). The vertical line indicates the current upper bound on the amplitude of $\Omega_0$ \cite{LIGOisotropic2022}. }
	\label{fig:polarization}
\end{figure}

In figure~\ref{fig:polarization}, we then show forecasts on the net polarization factor $p$ for different values of the fiducial amplitude $\Omega_0$, after marginalizing over $\Omega_0$ in the analysis. The left panel (resp., right panel) assumes an astrophysical unpolarized background (resp., a cosmological fully polarized background), and the different lines shown correspond to different network configurations. 

\subsection{Discussion} 
\label{subsec:result_discussion}

In this work, we outline in full detail a mathematical framework to forecast the sensitivity of a network of L-shaped and triangular-shaped ground-based detectors to an SGWB with anisotropies or with net circular polarization.  Above, we have shown some results obtained when this formalism is applied to different networks of present and next-generation ground-based detectors. 

Our study of anisotropies in subsection~\ref{subsec:result_anisotropic} assumes that the monopole dominates the SGWB signal, which is expected to be true  for the vast majority of astrophysical and cosmological models \cite{LISACosmologyWorkingGroup:2022kbp}. Moreover, we assumed that, besides the monopole, only one additional multipole is present in the expansion of eq.~\eqref{intro:PSD_def}, where, parametrically, the coefficient $\delta^{GW}_{\ell m}$ controls the ratio between the (directionally-dependent) contribution to the fractional SGWB energy density (per logarithmic interval in frequency) of that multipole relative to the monopole. The  amplitude of the monopole term is controlled by the standard parameter $\Omega_0$, which gives the fractional GW energy density (per logarithmic interval in frequency) $\Omega_0$ at the pivot scale $f_0=100 \, {\rm Hz}$. Our analysis provides the forecast uncertainties on determining the two parameters $\Omega_0$ and $\delta^{GW}_{\ell m}$. We also presented the uncertainty on $\delta^{GW}_{\ell m}$ by marginalizing the results of this analysis over $\Omega_0$. 

Not surprisingly, as visible in figures~\ref{fig:dipole_and_quadrupole} and~\ref{fig:anisotropies_different_l}, the sensitivity to the relative size of the anisotropic component $\delta^{GW}_{\ell m}$ worsens as the overall magnitude of the signal decreases, namely as we decrease the fiducial value of $\Omega_0$. More precisely, as the variance of the measurement is noise-dominated in this range, the sensitivity scales as the inverse of this quantity. As a reference, we show in the figures the current limit on $\Omega_0$ \cite{LIGOisotropic2022}. The SGWB obtained in astrophysical models may be close to this upper limit \cite{PhysRevD.103.043002,Ezquiaga_2021}, while predictions for the cosmological SGWB are much more uncertain and model dependent \cite{Caprini2018}. Our results indicate that if the SGWB saturates the current bound, the existing LIGO-Virgo-KAGRA network at the designed sensitivity will barely be able to constrain even anisotropies of order one significantly. On the other hand, the next generation detectors, such as Einstein Telescope and Cosmic Explorer, might be able to probe a ${\cal O }\left( 10^{-3} \right)$ anisotropy at the largest angular scales (lowest $\ell$; figure~\ref{fig:anisotropies_different_l} shows how the sensitivity worsens for higher multipoles).

As a comparison, anisotropies in the astrophysical SGWB can be estimated to be of the order of $10^{-2}$ \cite{Cusin:2019jpv,Bellomo:2021mer} The anisotropies due to the propagation of primordial GWs are expected to be below this level \cite{Contaldi:2016koz,Cusin:2017fwz,Bartolo:2019oiq,Bartolo_2022}, with a quadrupole amplitude $\delta^{GW}_{2m} \simeq \sqrt{C_2} \simeq 3 \times 10^{-5}$ \cite{LISACosmologyWorkingGroup:2022kbp}. \footnote{Cosmological perturbations might be greater in certain cases, see e.g. \cite{Kumar:2021ffi,Malhotra:2022ply,Dimastrogiovanni:2021mfs}.} Finally, the kinematic dipole, induced by the peculiar motion of the Earth, is of ${\cal O }\left( 10^{-3} \right)$, as we already discussed above in relation to figure~\ref{fig:dipole_and_quadrupole}. 

To summarize, the anisotropies induced by astrophysical sources and the kinematic effect might be probed by the Einstein Telescope - Cosmic Explorer combination, while the measurement of the anisotropy from the cosmological propagation appears to be more challenging. Significantly improving over this will require more than just decreasing the instrumental noise. Even in the ideal limit of vanishing instrumental noise, the sensitivity to the subdominant anisotropic component is limited by the variance of the dominant monopole contribution \cite{mentasti2023}. The curves shown in figures~\ref{fig:dipole_and_quadrupole} and~\ref{fig:polarization} show a marked change in their slope from the left part at small ${\bar \Omega_0}$ and the right part at large ${\bar \Omega_0}$.
This change in the slope is due to the transition between a regime in which the variance of the measurement is dominated by the instrumental noise to a regime in which it is dominated by the signal.\footnote{While both contributions to the variance are fully included in the results presented in this section, in subsection~\ref{subsec:noise} we show that the variance can be written in terms of relatively simple analytical expressions in the limits in which it is fully dominated by either contribution.} For the network configurations we have discussed, the measurement is in the noise-dominated regime once the signal is constrained to be below the current upper limit.
Assuming that futuristic experiments will reach an ideal level of negligible instrumental noise, the variance of their measurements will be due solely to that of the signal. The sensitivity to the anisotropy will then reach the level that can be seen from the high-${\bar \Omega}_0$ portion of our figures. Our results assume an observation period of $10$ years. Improving over these limits will then require increasing the observation period $T$, or the number of detectors $N$ \cite{mentasti2023}.  We recall that the expected error decreases as the square root of the measurements. Therefore it scales as $T^{-1/2}$ and as $N^{-1}$ since the number of cross-correlations between different detectors scales as $N^2$.~\footnote{This statement is strictly true in the limit of vanishing contribution to the variance of measurements taken at different pairs. As we discuss in subsection~\ref{subsec:noise}, we verified that the off-diagonal contributions of the variance between different pairs are indeed negligible with respect to the diagonal ones between the same pair (see the structure in eq.~\eqref{SigmaNoiseOrSignal}. We expect that this would no longer be true in the limit of large $N$.} 

In subsection~\ref{subsec:result_polarized}, we presented the results of a study of the forecast sensitivity to a net circular polarization, which, as we discussed in the Introduction, is predicted to be of order one in some inflationary mechanisms, and which might also be present, although to a significantly smaller amount, in the astrophysical signal. figure~\ref{fig:polarization} shows that if the amplitude of the SGWB is close to the current limit, an order one net polarization can be marginally probed by the existing detectors at design sensitivity. The Einstein Telescope - Cosmic Explorer network can improve over this by about three orders of magnitude. As shown in the figure, the forecast sensitivities worsen as the amplitude $\Omega_0$ decreases below the current limit. 

Our results are in agreement with \cite{Martinovic_Mandic_2021}, where a frequency-dependent power spectrum and a generic power law tilt have been investigated, and with \cite{ValbusaDallArmi:2023ydl}, where the predicted values of $\Omega_0$ for an astrophysical background and the noise curves (as in figure 1 of that paper) produce an expected SNR for a network made of ET and CE which is in agreement with our forecast error in the measurement of a net amount of polarization.
A more precise comparison with these results, particularly for what concerns \cite{Martinovic_Mandic_2021}, may be done by allowing the polarization parameter $p$ and the power spectrum tilt $\alpha$ to depend on frequency, which is feasible in principle in our case. However, the comparison between our two panels in figure~\ref{fig:polarization} indicates that the result does not vary very much in the realistic region of the parameter space, thus supporting the agreement between our results and those of \cite{Martinovic_Mandic_2021} on the polarization.

The analyses carried out in this paper can be extended in several directions. Our formalism can be immediately employed for any planned network and to compare the performance of different configurations (as done, for instance, in the recent ET study \cite{Branchesi:2023mws}). Another natural step will be to go beyond the assumption of factorized angular frequency dependence of the SGWB, along with considering simultaneously more than a single multipole $(\ell,m)$ in the analysis of the anisotropies. It may also be interesting to produce the same kind of forecasts for LISA \cite{amaroseoane2017laser}, the space-based triangular-shaped GW interferometer planned to be launched on $\sim$2037, improving over the estimates of \cite{Bartolo_2022} that assume  statistical isotropy.

\section{The stochastic gravitational wave background}
\label{sec:theSGWB}
We summarise here the basic formalism for describing an anisotropic SGWB following the notation and conventions of earlier work \cite{Allen:1996gp, Mentasti:2020yyd}. We consider gravitational waves at $x^\alpha$, the location of the detectors, as small perturbations of the metric tensor on a Minkowski background in such a way that
\begin{equation}\label{eq:hmunu}
	g_{\mu\nu}(x^\alpha)=\eta_{\mu\nu}+h_{\mu\nu}(x^\alpha)\,,
\end{equation}
where $\eta_{\mu\nu}$ is the (mostly positive signature) metric tensor for Minkowski spacetime and $h_{\mu\nu}$ is the gravitational wave, which is a small perturbation of the background metric, $h\ll\eta$.

It is convenient to completely fix the gauge freedom through the standard choice $\partial_\mu h^\mu_\nu - \frac{1}2 \partial_\nu \, h^\mu_\mu= h_{00} = h_{0a} = 0$ (greek indices run from 0 to 3, while latin indices run from 1 to 3). With this choice, the GW term has non-vanishing entries only in the spatial components, where it is transverse ($\partial_a h_{ab} = 0$) and traceless ($h_{aa}=0$), and where we decompose in plane waves
\begin{align}
	h_{ab}(t,\vec x)&=\int_{-\infty}^{\infty}df\int d^2 {\hat n} \, e^{2\pi i f \left( t - {\hat n} \cdot \vec{x} \right)} 
	\sum_{s=+,\times} h_s (f,\hat n) \, e^s_{ab} \left( {\hat n} \right) \equiv\sum_s h_{ab,s}(t,\vec x)\,,
	\label{eqGWTTpc}
\end{align}
where $\hat n$ is the unit vector in $\mathrm{R}^3$, and $e^+_{ab}(\hat n)$, $e^\times_{ab}(\hat n)$ are the two polarization basis tensors, defined by
\begin{align}\label{matrPolarizz}
	e_{ab}^+(\hat n)=p_ap_b-q_aq_b \;\;,\;\; 
	e_{ab}^\times(\hat n)=p_aq_b+q_ap_b\,.
\end{align}
For our choice of spherical coordinates, the direction vector is defined as
\begin{align}
	\hat n=\cos\phi\sin\theta\,\hat x +\sin\phi\sin\theta\,\hat y +\cos\theta\,\hat z\,.
\end{align}
The direction vector is aligned with the direction of propagation of the plane wave ($\theta, \phi$ are polar angular coordinates, while $\hat x,\hat y, \hat z$ are unit vectors along the three Cartesian axes), while $\hat p$ and $\hat q$ are two directions perpendicular to $\hat n$ and to each other, that are conventionally chosen as in \cite{Allen:1996gp}
\begin{align}
	\hat p =\sin\phi\hat x -\cos\phi\hat y \;\;,\;\; 
	\hat q=\cos\phi\cos\theta\hat x +\sin\phi\cos\theta\hat y -\sin\theta\hat z\,.
\end{align}
One can also consider right-handed ($\lambda = +1$) and left-handed ($\lambda = -1$) circular polarizations rewriting    eq.~\eqref{eqGWTTpc} as 
\begin{align}
	h_{ab}(t,\vec x)=\int_{-\infty}^{\infty}df\int d^2 {\hat n} \, e^{2\pi i f \left( t - {\hat n} \cdot \vec{x} \right)} 
	\sum_{\lambda=\pm 1} h_\lambda (f,\hat n) \, e^\lambda_{ab} \left( {\hat n} \right) \;\;,\;\; 
	e^\lambda_{ab} \equiv \frac{e^+_{ab} + i \, \lambda \, e^\times_{ab}}{\sqrt{2}} 
	\,. 
	\label{eqGWTT}
\end{align}
 
The plane wave solution of eq.~\eqref{eqGWTT} is defined in a reference system fixed with respect to the expanding cosmological background and in which the spatial pattern of the perturbations of the stochastic background is assumed to be statistically time-independent.
The $\hat{z}$ vector of this system is chosen to point along the direction of the Earth's rotation axis.
To describe a stochastic source, we treat the complex amplitude $h_\lambda(f,\hat n)$ as a random Gaussian variable with zero mean. This means that its statistics are completely specified by its variance. The dependence of the stochastic background on frequency and direction may be stated solely in terms of the expectation value of the two-point correlator for the random variable $h_\lambda(f,\hat n)$ as
\begin{equation}
	\left\langle h^*_\lambda \left( f, {\hat n} \right) h_{\lambda'} \left( f' {\hat n}' \right) \right\rangle = \delta_{\lambda\lambda'} \delta_D^{(2)} \left( {\hat n} - {\hat n}' \right) \delta_D \left( f-f' \right) \mathcal{H}_\lambda \left( \left\vert f \right\vert, {\hat n}\right)\,,	
	\label{eq2points}
\end{equation}
where $\delta_D^{(2)} \left( {\hat n} - {\hat n}' \right)$ is a covariant two-dimensional Dirac delta-function on the unit two-sphere and $\delta_D \left( f-f' \right)$ the Dirac delta function on the frequency space.

In principle, such a source has spectral properties which arbitrarily depend upon amplitude and frequency. For simplicity, here we consider a factorized dependence 
\begin{equation}\label{PowerSpectrumFactorized}
	\mathcal{H}_\lambda \left( \left\vert f \right\vert, {\hat n}\right)=H_\lambda \left( \left\vert f \right\vert \right) P_\lambda \left( {\hat n} \right)\,,
\end{equation}
which amounts to assuming that sources of the SGWB situated along a different line of sights emit with the same "average" spectrum in frequency. It is conventional to define the fractional energy density in GW per logarithmic frequency interval and per polarisation as~\footnote{In the decompositions of eqs.~\eqref{eqGWTTpc} and \eqref{eqGWTT} the frequency can have both positive and negative values. The power $\mathcal{H}_\lambda$ in eq.~\eqref{eq2points} is a function only of the magnitude of the frequency so that one can conveniently define the energy density $\rho_{GW,\lambda}$ and the fractional energy density $\Omega_{GW,\lambda}$ in terms of positive frequencies only.} 
\begin{equation}
	\Omega_{GW,\lambda} \left( f \right)\equiv \frac{f}{\rho_{\rm crit} \left( f \right)}\frac{d\rho_{GW,\lambda}}{df} = \sum_\lambda \frac{16\pi^3}{3H_0^2}f^3H_\lambda \left(|f| \right)  \,,
\label{OmegaGW2}
\end{equation}
where $\rho_{\rm crit}$ is the current energy density of the universe, which we assume to be flat, while $H_0$ is the Hubble constant. In the second step, we have used the fact that only the monopole contributes to the total GW energy density (after integrating over the arrival direction). 

Next, we decompose the angular power spectrum in terms of spherical harmonics, writing it as
\begin{equation}\label{eqPOmega}
	P_\lambda(\hat n)=\sum_{\ell=0}^{+\infty}\sum_{m=-\ell}^{\ell} \, \delta^{GW}_{\ell m,\lambda} \, Y_{lm}(\hat n) \;\;\;,\;\;\; \delta^{GW}_{00,\lambda} = \sqrt{4 \, \pi} \,, 
\end{equation}
where the value of $\delta^{GW}_{00,\lambda}$ is set considering the convention for the monopole $Y_{00} = \frac{1}{\sqrt{4 \pi}}$.

In this expression, we have allowed the coefficients of the decomposition $\delta^{GW}_{\ell m,\lambda}$ to depend on the polarization.  However, due to the lack of strong theoretical motivation for the generation of large-scale polarization-dependent anisotropies, for simplicity, we assume that 
\begin{equation}
	\delta^{GW}_{\ell m, 1}=\delta^{GW}_{\ell m,-1}\equiv \delta^{GW}_{\ell m} \;\; \Rightarrow\;\; P_{+1} \left( {\hat n} \right) = P_{-1} \left( {\hat n} \right) \equiv P \left( {\hat n} \right) \;. 
	\label{plm-unpol}
\end{equation}
On the other hand, we allow the monopole term to depend on the polarization in a frequency-independent amount: 
\begin{align}
	&H_\lambda \left( \vert f \vert \right)\equiv \left( 1+ \lambda \, p \right) \, H \left( \vert f \vert \right)  \;\;\;,\;\;\; -1 \leq p \leq 1 \;, 
	\label{def-pH_LR}
\end{align}
where $p =+1$ (respectively, $p=-1$) corresponds to a fully right-handed (respectively, left-handed) circularly polarized SGWB, while $p=0$ corresponds to an unpolarized background. With this parametrization, the quantity $H \left( \vert f \vert \right)$ can be immediately related to the GW fractional energy density with the standard expression 
\begin{equation}
\Omega_{GW}(f)\equiv\sum_\lambda \Omega_{GW,\lambda} (f) =\frac{32\pi^3}{3H_0^2}f^3H(|f|) \;, 
\end{equation}
irrespective of the degree of polarization of the SGWB (namely, for any value of $p$). We also note that one can write 
\begin{align}
	&\Omega_{GW,\lambda}\left( f \right) = \frac{1 + \lambda \, p}{2} \, \Omega_{GW} \left( f \right) \;. 
\end{align}
Finally, we assume that, in the range which affects the measurements, the GW energy density has a power-law frequency dependence 
\begin{equation}
	\Omega_{\rm GW} \left( f \right) = {\Omega}_{0} \left( \frac{f}{f_0} \right)^\alpha \;, 
	\label{PowerLawOmega}
\end{equation}
where ${\Omega}_{0}$ is the fractional energy density at the pivot scale of $f_0=100 \, {\rm Hz}$, and where, as discussed in subsection~\ref{subsec:result_anisotropic} we fix $\alpha = 0 ,\, 2/3$.  From the above relations, it is clear that the same power-law dependence applies to $H_\lambda \left( \vert f \vert \right)$.

\section{Data stream of ground-based interferometers}
\label{sec:DataStream}

As mentioned above, we consider a network of $N$ L-shaped ground-based detectors (for instance, LIGO, Virgo, and KAGRA) and $M$ ground-based detectors with triangular configuration (for instance Einstein Telescope). For each L-shaped interferometer of the network, we measure the difference $\Delta T$ for light-travel times across separate arm paths. The measurement is affected by the instrument noise and any GW signal present,
\begin{equation}
	m_{i} \left( t \right) = \frac{\Delta T_{i}}{T_0} = n_{i} \left( t \right) + s_{i} \left( t \right) \;\;\;,\;\;\; i =1,\dots,N \;,
	\label{mnsLshaped}
\end{equation}
where $T_0$ is the time the laser would need to travel back and forth without any signal and noise.

We define the set of measurements made by each triangular-shaped interferometer of the network similarly. In this case, each triangular detector has an equilateral triangular configuration with three Michelson interferometers at its vertices $\alpha = X,\, Y ,\, Z$, therefore
\begin{equation}
	m_{i\alpha} \left( t \right) = \frac{\Delta T_{i\alpha}}{T_0} = n_{i\alpha} \left( t \right) + s_{i\alpha} \left( t \right) \;\;\;,\;\;\; i =1,\dots,M \;\;\;,\;\;\; \alpha =X,\, Y ,\, Z \;.
	\label{mns}
\end{equation}
We then assume that the noise of each network instrument is not correlated with the one affecting the others. We also assume that the noise of each detector is Gaussian distributed with zero mean. These assumptions imply a vanishing correlation between L-shaped and triangular-shaped detectors 
\begin{equation}
	\left\langle {\tilde n}_{i\alpha}^* \left( f \right)  {\tilde n}_{j} \left( f' \right) \right\rangle \equiv 0 \;, 
	\label{variance-ijLX}
\end{equation}
where $i =1,\dots,M$ and $j=1,\dots,N$. For the L-shape detectors, we can then write
\begin{equation}
	\left\langle {\tilde n}_{i}^* \left( f \right)  {\tilde n}_{j} \left( f' \right) \right\rangle \equiv \frac{\delta_{i j}}{2} \delta_D \left( f - f' \right) P_{i} \left( \left\vert f \right\vert \right) \;,
	\label{variance-ijLL}
\end{equation}
where $i =1,\dots,N$ and $j=1,\dots,N$.

For the variance of the noise of the triangular-shaped detectors, accounting for the fact that every interferometer shares one arm with each of the other two interferometers (of the same detector), we write
\begin{equation}
	\left\langle {\tilde n}_{i\alpha}^* \left( f \right)  {\tilde n}_{j\beta} \left( f' \right) \right\rangle \equiv \frac{\delta_{i j}}{2} \delta_D \left( f - f' \right) N_{\alpha \beta, i} \left( \left\vert f \right\vert \right) \;,
	\label{variance-ij}
\end{equation}
where $i =1,\dots,M$, $j=1,\dots,M$. If the $i$-th detector has an exact equilateral configuration, with three identical instruments at the vertices, the noise matrix at each site is a $3 \times 3$ matrix having the same element $N_{d, i}$ in all diagonal entries and the same element $N_{o, i}$ in all off-diagonal entries.  This matrix can be diagonalized in the three channels
\begin{align}
m_{iO} \equiv c_{O\alpha} \, m_{i\alpha} \;\;,\;\;\;\;O = A ,\, E ,\, T \;\;,\;\; \alpha = X ,\, Y ,\, Z \,, 
\label{AET-def}
\end{align}
(we note that the same combinations $c_{O \alpha}$ are taken in each detector $i$)  with
\begin{equation}
	c \equiv  \left( \begin{array}{ccc} 
		\frac{4}{3 \sqrt{3}}  & - \frac{2}{3 \sqrt{3}} & - \frac{2}{3 \sqrt{3}} \\ 
		0 & - \frac{2}{3} & \frac{2}{3} \\ 
		\left( \frac{2}{3} \right)^{3/2} &\left( \frac{2}{3} \right)^{3/2} & \left( \frac{2}{3} \right)^{3/2} 
	\end{array} \right) \;.   
\label{def-cmat}
\end{equation} 
These linear combinations, already introduced in \cite{Adams:2010vc} for the LISA experiment, which also has an equilateral configuration (the explicit frequency dependence of $N_{d, i}$ and $N_{o, i}$ for LISA can be found for example in ref.~\cite{Smith:2019wny}) were used for ET in \cite{Mentasti:2020yyd}.

We use these channels as they diagonalize the noise matrix, which simplifies the computation of the SNR that we perform below. In fact, combining  eqs.~\eqref{variance-ij} and \eqref{AET-def}, one obtains 
\begin{eqnarray} 
	&& \left\langle {\tilde n}_{iO}^* \left( f \right)  {\tilde n}_{jO'} \left( f' \right) \right\rangle 
	= \frac{1}{2} \delta_D \left( f - f' \right) \, \delta_{ij} \, \delta_{OO'} \, N_{O, i} \left( \left\vert f \right\vert \right) \;, 
	\label{variance-AET}
\end{eqnarray} 
with 
\begin{equation}
	N_{A, i} \left( f \right) = N_{E, i} \left( f \right) = \frac{8}{9} \left[ N_{d, i} \left( f \right) -  N_{o, i} \left( f \right) \right] \;\;,\;\;
	N_{T, i} \left( f \right) =  \frac{8}{9} \left[ N_{d, i} \left( f \right) + 2  N_{o, i} \left( f \right) \right] \;. 
	\label{noise-AET}
\end{equation}

As in \cite{Mentasti:2020yyd}, we consider this choice of channels in the present work since they diagonalize the noise matrix. Furthermore, as explicitly verified in \cite{Mentasti:2020yyd}, the $A-$ and $E-$channels behave as $90^\circ$ interferometers for small frequencies (a limit that holds in the relevant frequency range for Einstein Telescope and instruments of similar dimensions~\footnote{The condition for the low frequency / short arm limit is $2 \pi f L \ll 1$, where $L$ is the arm length. In the diagonalization of eq.~\eqref{def-cmat} we changed the normalization of each channel with respect to \cite{Adams:2010vc}, so that, in the low frequency / short arm limit, the $A-$ and $E-$channels behave as $90^\circ$ degrees interferometers, with unit-norm arm factors $d_{iA,E}^{ab}$ given in eq.~\ref{d-def}.}. In this regime, the contribution of the $T-$channel to the signal vanishes, so we disregard it in our computations \cite{LISACosmologyWorkingGroup:2022kbp}.

\section{Measurement by a network of detectors}
\label{sec:measure}

Similarly to what has been done in \cite{Allen:1996gp} and \cite{Mentasti:2020yyd}, we start by introducing the time-dependent Fourier transform of eq.~\eqref{mnsLshaped} for the measurement in the $i$-th L-shaped interferometer 
\begin{equation}
	{\tilde m}_{i} \left( f ,\, t \right) \equiv \int_{t-\tau/2}^{t+\tau/2} d t' {\rm e}^{-2 \pi i f t'} \, m_{i} \left( t' \right)  \;.
	\label{tmOLshaped}
\end{equation}
We label the measurements in triangular-shaped interferometers with an additional index $O$ accounting for the channel considered (namely, the A- or the E-channel for that interferometer), and we also transform the measurements according to 
\begin{equation}
	{\tilde m}_{iO} \left( f ,\, t \right) \equiv \int_{t-\tau/2}^{t+\tau/2} d t' {\rm e}^{-2 \pi i f t'} \, m_{iO} \left( t' \right) \;. 
	\label{tmO}
\end{equation}
In these relations, the integration in time is performed over a timescale $\tau$ much greater than the inverse of the smallest frequency of the GWs that we want to study but sufficiently small that we can neglect the effect due to the rotation of the Earth during this time. We introduce the Fourier transforms of the signals ${\tilde s}_i$ and ${\tilde s}_{jO}$, and of the noise  ${\tilde n}_i$,  and ${\tilde n}_{jO}$ contributions in an analogous manner.

We can then define the set of estimators
\begin{align}\label{C-def}
{\cal C}_{ij} \left(t \right)&\equiv\int_{-\infty}^{\infty}df{\tilde m}_{i}^* \left( f ,\, t \right) {\tilde m}_{j} \left( f ,\, t \right) \, 
	{\tilde Q}_{ij} \left( f \right) \;,\nonumber\\
{\cal C}_{ij,O} \left(t \right)&\equiv\int_{-\infty}^{\infty}df{\tilde m}_{iO}^* \left( f ,\, t \right) {\tilde m}_{j} \left( f ,\, t \right) \, 
	{\tilde Q}_{ij,O} \left( f \right) \;,\nonumber\\
	{\cal C}_{ij,OO'} \left(t \right)&\equiv\int_{-\infty}^{\infty}df{\tilde m}_{iO}^* \left( f ,\, t \right) {\tilde m}_{jO'} \left( f ,\, t \right) \, 
	{\tilde Q}_{ij,OO'} \left( f \right) \;,
\end{align}
where the functions ${\tilde Q}_{ij,OO'} \left( f \right) $, ${\tilde Q}_{ij,O} \left( f \right) $ and ${\tilde Q}_{ij} \left( f \right)$ are weights (in the sum over channels, interferometers of the network, and frequencies) that will be chosen later to maximize the effect of the signal over the noise. Note that in the definition of the estimator~\eqref{C-def} we do not consider the correlations between instruments located at the same site. This is done because it is difficult to characterize all the noise contributions for collocated instruments properly. 

From now on, to shorten the notation, we use the expression $(,OO')$ to indicate all possible combinations of L-shaped and triangular-shaped detectors collectively. For instance, the three expressions~\eqref{C-def} can be collectively written as 
\begin{equation}
{\cal C}_{ij(,OO')} \left(t \right)  \equiv \int_{-\infty}^{\infty}df{\tilde m}_{i(,O)}^* \left( f ,\, t \right) {\tilde m}_{j(,O')} \left( f ,\, t \right) \, {\tilde Q}_{ij,(,OO')} \left( f \right) \;\;\;,\;\;\; i \neq j \;. 
\label{C-def-compact}
\end{equation} 
We assume that the statistical properties of the signal and the noise do not change with time. Then for an anisotropic SGWB, the statistics of the signal are periodic, with periodicity given by the rotation period $T_e = \frac{2 \pi}{\omega_e}$ of the Earth. It is then convenient to perform the Fourier decomposition 
\begin{equation}
	{\cal C}_{ij(,OO')}  \left( t \right)   = \sum_{m=-\infty}^\infty 
	{\cal C}_{ij(,OO'),m} \, {\rm e}^{i m \omega_e t} \;\;\;\;,\;\;\;\; 
	{\cal C}_{ij(,OO'),m}  \equiv  \frac{1}{T} \int_0^{T} dt \, {\rm e}^{-i m \omega_e t} \, 
	{\cal C}_{ij(,OO')}  \left( t \right) \;,
	\label{C-m}
\end{equation} 
where we take the observation time $T$ to be an integer multiple of one day $T_e$.

Under the assumption of Gaussianity, we can define the Likelihood function of the estimated parameters of the theory $\theta$ and their fiducial value $\bar\theta$ for the set of measurements ${\cal C}_m$ provided by the network:
	\begin{align}\label{likel_def}
	&\mathcal{L}( \theta,\bar\theta)\equiv\frac{1}{N(\bar\theta)}e^{-\frac{1}{2}\chi^2(\theta,\bar\theta)} \;,\nonumber\\
	&\chi^2(\theta,\bar\theta)\equiv\sum_{m',m''}\sum_{\substack{i\neq j\\(,OO')}}\sum_{\substack{k\neq l\\(,O''O''')}} \left\langle r^*_{ij(,OO'),m'}(\theta,\bar\theta) \right\rangle\Sigma_{ij(,OO'),kl(,O''O'''),m'm''}^{-2}(\bar\theta) \left\langle r_{kl(,O''O'''),m''}(\theta,\bar\theta) \right\rangle \;,
\end{align}
where we introduced the quantities
\begin{align} \label{rijm}
	&\Sigma_{ij(,OO'),kl(,O''O'''),m'm''}^2(\hat \theta)\equiv\left\langle r^*_{ij(,OO'),m'}(\bar\theta,\bar\theta)\,r_{kl(,O''O'''),m''}(\bar\theta,\bar\theta) \right\rangle \;,\nonumber\\
&r_{ij(,OO'),m}(\theta,\bar\theta)\equiv	\mathcal{C}_{ij(,OO'),m}(\theta)-\langle\mathcal{C}_{ij(,OO'),m}(\bar\theta)\rangle \;.
\end{align}
We stress that the sums over $ij(,OO')$, $kl(,O''O''')$ are meant over all the possible combinations of the observables built-in eq.~\eqref{C-def}.
The $\chi^2$ in eq.~\eqref{likel_def} is a well-defined $\chi^2$ since it has a global minimum in $\theta=\bar\theta$. 

The set of parameters for our analysis is  $\theta \equiv (\Omega_{0}, \delta^{GW}_{\ell m},p,\alpha)$, where we recall that $\Omega_{0}$ is the amplitude of the monopolar power spectrum as in eq.~\eqref{PowerLawOmega}, $\delta^{GW}_{\ell m}$ is the ratio between the ($\ell, m$) multipole and the monopole, as in eq.~\eqref{eqPOmega}, $p$ is the polarization factor as defined in eq.~\eqref{def-pH_LR}, and $\alpha$ is the tilt of the power spectrum in eq.~\eqref{PowerLawOmega}.

\subsection{Expectation value of the signal} 
\label{subsec:signal}

In appendix~\ref{app:sigvar} we show that the expectation value of eq.~\eqref{C-m} is
\begin{align} 
	\left\langle C_{ij(,OO'),m} \right\rangle =\sum_\lambda \frac{8 \pi \, \tau}{5}  
	\int_{-\infty}^{+\infty} d f \, H_\lambda \left( \vert f \vert \right)   \sum_{\ell = \vert m \vert}^\infty \delta^{GW}_{\ell m,\lambda} \gamma_{\ell m,ij(,OO'),\lambda}\left( f \right) Q_{ij(,OO'),m} \left( f \right) \;,
	\label{C1-res} 
\end{align}
where, following \cite{Allen:1996gp}, the overlap functions are defined as ~\footnote{The factor $\frac{5}{8 \pi}$ is conventional, and it has the purpose of eliminating an overall factor in the monopole term; see \cite{Mentasti:2020yyd} for further details.}
\begin{align}
	&\gamma_{\ell m,ij(,OO'),\lambda}(f) \equiv \frac{5}{8 \pi} \int d^2 {\hat n}{\rm e}^{2 \pi i f {\hat n} \cdot \Delta \vec x_{ij}} \, Y_{\ell m} \left( {\hat n} \right) 
	e_{ab}^{*\lambda} \left(  {\hat n} \right) e_{cd}^\lambda \left(  {\hat n} \right) \, d_{i(,O)}^{ab}  \, d_{j(,O')}^{cd} \;. 
	\label{gamma}
\end{align}
It is useful to rewrite the $\gamma$ coefficients of eq.~\eqref{gamma} as
\begin{align}
	&\gamma_{\ell m,ij(,OO'),\lambda}(f) \equiv \gamma_{\ell m,\lambda,abcd}(\kappa_{ij},\hat s_{ij}) \times d_{i(O)}^{ab}  \, d_{j(O')}^{cd} \;, 
	\label{gamma-dec}
\end{align} 
where we define the tensors 
\begin{align}
	d_{i(\alpha)}^{ab} \equiv  \frac{\hat u_{i(\alpha)}^a\hat u_{i(\alpha)}^b-\hat v_{i(\alpha)}^a\hat v_{i(\alpha)}^b}{2} \;\;,\;\;  
	d^{ab}_{iO} \equiv c_{O\alpha} \, d_{i\alpha}^{ab} \;, 
\label{d-def}
\end{align}
where $\hat u_{i}$, $\hat v_{i}$ are the arm directions of the L-shaped interferometers, while $\hat u_{i\alpha}$, $\hat v_{i\alpha}$ denote the arm directions of the triangular-shaped interferometers (we recall that $\alpha = X ,\, Y ,\, Z$, while $O = A ,\, E ,\, T$, and  the matrix $c$ is given in eq.~\eqref{def-cmat}. 

In eq.~\eqref{gamma-dec} we also defined
\begin{equation}
\kappa_{ij} \equiv 2\pi f \left| \Delta \vec x_{ij}\right| 
\;\;\;,\;\;\;
\hat s_{ij} \equiv \frac{\Delta \vec x_{ij}}{\left| \Delta \vec x_{ij}\right|} \;\;,\;\; {\rm with} \;\; 
\Delta \vec x_{ij}=\vec x_i-\vec x_j
\label{eq:deltaxij} \;. 
\end{equation}
The decomposition of eq.~\eqref{gamma-dec} introduces the coefficients 
\begin{align}
	\gamma_{\ell m,\lambda,abcd}(\kappa,\hat s) \equiv \frac{5}{8 \pi} \int d^2 {\hat n}{\rm e}^{i\kappa {\hat n} \cdot \hat s} \, Y_{\ell m} \left( {\hat n} \right) 
	e_{ab}^{*\lambda} \left(  {\hat n} \right) e_{cd}^\lambda \left(  {\hat n} \right) \;, 
	\label{ga-lm-ab-cd-allf}
\end{align}
that can be evaluated just once for all for any multipole of interest, independently of the pair of detectors under consideration. These coefficients can be evaluated numerically or analytically. In appendix~\ref{app:Gamma-MDQ}, we provide the explicit analytic expressions for the monopole, dipole, and quadrupole contributions\footnote{The  expressions for the monopole and dipole can be found in \cite{anisotropy-DomckePeloso}, and we report them here for completeness, while the expression for the quadrupole is an original result of this paper. The explicit analytic expressions for the non-vanishing coefficients in the zero frequency limit can be found in \cite{Mentasti:2020yyd}.}, and we explain how this computation can be extended to higher multipoles.

\subsection{Covariance matrix and optimal \texorpdfstring{$\chi^2$}{}}
\label{subsec:noise}

In appendix~\ref{appVarSignal}, we evaluate the covariance matrix from the first of eqs.~\eqref{rijm} for an arbitrary level of signal and noise. We show that the covariance assumes the structure (indices $OO'$ accounting for the channels of each triangular-shaped detector are omitted for brevity)
\begin{align}
\label{SigmaNoiseOrSignal}
\Sigma_{ij,kl,m'm''}^2\simeq\delta_{m'm''}(\delta_{ik}\delta_{jl}+\delta_{il}\delta_{jk})\int_{-\infty}^{+\infty} df V_{ij}(f) \left\vert Q_{ij,m'}(f)\right\vert^2 \;, 
\end{align}
where 
\begin{align}
V_{ij}(f)&=\left(\frac{8\pi}{5}\right)^2\frac{\tau^2\delta_{00,GW}^2}{T}\Bigg[\sum_\lambda H_\lambda(f)\gamma_{ii,\lambda}(f)\sum_{\lambda'} H_{\lambda'}(f)\gamma_{jj,\lambda'}(f)\nonumber\\
&+\sum_\lambda H_\lambda(f)\gamma_{ij,\lambda}(f)\sum_{\lambda'} H_{\lambda'}(f)\gamma_{ji,\lambda'}(f)\nonumber\\
&+\frac{1}{\sqrt{4\pi}}\frac{5}{16\pi}\left(\sum_\lambda H_\lambda(f)\gamma_{ii,\lambda}(f)P_j(f)+\sum_\lambda H_\lambda(f)\gamma_{jj,\lambda}(f)P_i(f)\right)\nonumber\\
&+\frac{1}{4\pi}\left(\frac{5}{16\pi}\right)^2P_i(f)P_j(f)\Bigg] \;, 
\label{Vij-full}
\end{align}
where we defined $\gamma_{ij,\lambda}(f)\equiv\gamma_{ij,00,\lambda}(f)$ taken from eq.~\eqref{gamma}. The terms proportional to $\gamma_{jj,\lambda}$ in this last expression are due to the contribution of the dominant isotropic component of the signal, while the terms proportional to $P_i$ (which are the Power Spectral Density plotted in figure~\ref{fig:allpsd}) are due to the noise. 

The tensorial structure in eq.~\eqref{SigmaNoiseOrSignal} is exact in the noise-dominated regime since the noise is diagonal across different instruments and channels. On the contrary, the variance of the signal has contributions that are diagonal in each pair of detectors, and off-diagonal contributions that originate from products of correlators taken at different pairs of detectors. Only the former contributions respect the tensorial structure of eq.~\eqref{SigmaNoiseOrSignal}. We verified, by a direct inspection of about $100$ correlators in the networks that we considered in section~\ref{sec:shortsummary}, that the latter contributions are negligible. We refer the reader to the discussion after eq.~\eqref{Sigma-ab} for more details on this. 

Hereafter, we consider an anisotropic power spectrum as in eq.~\eqref{eqPOmega} such that
\begin{equation}\label{onlyoneplm}
	P(\hat n )=1+\delta^{GW}_{\ell m}Y_{lm}(\hat n) \;, 
\end{equation}
with $\ell$ and $m\neq 0$. In this case the only measurements $C_{ij,m'}$ in eq.~\eqref{C1-res} with a non-vanishing expectation value are $C_{ij,0}$ and $C_{ij,m}$. Plugging this expression, along with eqs.~\eqref{C1-res} and \eqref{SigmaNoiseOrSignal}, into the second equation of eqs.~\eqref{likel_def} one obtains the following expression for the $\chi^2$: 
\begin{align}
\chi^2&=\sum_{ij}
\sum_{\{m',m''\}=\{0,m\}} 
\int df\langle r_{ij,m'}^*(f)\rangle\, Q_{ij,m'}^*(f) 
\Sigma_{ij,m',m''}^{-2}
\int df\langle r_{ij,m''}(f)\rangle\, Q_{ij,m''}(f)= \nonumber\\
&=\sum_{ij}\left[\frac{\left|\int df\langle r_{ij,0}(f)\rangle\, Q_{ij,0}(f)\right|^2}{\int df V_{ij}(f)\, \left\vert Q_{ij,0}(f)\right\vert^2}+\frac{\left|\int df\langle r_{ij,m}(f)\rangle\, Q_{ij,m}(f)\right|^2}{\int df V_{ij}(f)\, \left\vert Q_{ij,m}(f)\right\vert^2}\right] \;,
\label{chisquared_nonopt_0m}
\end{align}
where we used the fact that the covariance matrix is diagonal in the indices $(m',m'')$, as we show in appendix~\ref{app_variance:subsec_generalnetwork} and we have defined the functions $r_{ij,m}(f)$ through
\begin{align}
r_{ij,m}=\int_{-\infty}^{\infty} df r_{ij,m}(f) Q_{ij,m}(f)\,,
\end{align}
with the coefficients $r_{ij,m}$ introduced in eq.~\eqref{rijm}.

The optimization of $Q_{ij,0}(f)$ and $Q_{ij,m}(f)$ (namely, the choice of the filter functions that maximize the signal-to-noise ratio) is performed by equating to zero the first functional derivative of the $\chi^2$ over $Q_{ij,0}(f)$ and $Q_{ij,m}(f)$. Doing so, the optimal filters are found to be 
\begin{align}
Q^*_{ij,0}(f)=\frac{\langle r_{ij,0}(f)\rangle}{V_{ij}(f)} \;\;\;,\;\;\; Q^*_{ij,m}(f)=\frac{\langle r_{ij,m}(f)\rangle}{V_{ij}(f)} \;,  
\label{filters}
\end{align}
leading to the optimal $\chi^2$:
\begin{align}\label{chisquared_opt_Cm}
	\chi^2_{\text{opt}}=\sum_{ij}\left[\int df \frac{|\langle r_{ij,0}(f)\rangle|^2}{V_{ij}(f)}+\int df \frac{|\langle r_{ij,m}(f)\rangle|^2}{V_{ij}(f)}\right] \;. 
\end{align}
This relation, with the numerator from eqs.~\eqref{rijm} and \eqref{C1-res}, and the denominator from eq.~\eqref{Vij-full} is our main analytic result. It can be employed to study an SGWB which is both anisotropic (with the anisotropy dominated by one multipole) and polarized. In the following two sections, we study how this expression simplifies in cases in which only one of these two properties is present.

\section{Detection of anisotropies}
\label{sec:Anisotropies}

As the first application of the formalism developed in the previous section, we quantify the ability to detect anisotropies in the SGWB with a network of ground-based detectors. We do so through a Fisher forecast. To achieve this, in subsection~\ref{subsec_anisotropies_signal_exp} we write explicitly how the expectation value of the measurements depends on the relevant parameters for the study of the anisotropies.
On the other hand, in subsection~\ref{subsec_anisotropies_fisher} we use the $\chi^2$ to obtain a well-posed posterior function for the parameters controlling the anisotropy in the Gaussian limit.

\subsection{Expectation value of the signal}\label{subsec_anisotropies_signal_exp}
We firstly consider an unpolarized power spectrum by setting $p=0$ in eq.~\eqref{def-pH_LR} and assuming that the tilt of the power spectrum $\alpha$ in eq.~\eqref{PowerLawOmega} has a known value. Under this assumption, from eq.~\eqref{rijm}, \eqref{C1-res}, \eqref{PowerLawOmega}, and \eqref{onlyoneplm} we obtain 
\begin{align}
	\langle r_{ij,0}(f)\rangle&=\frac{3 H_0^2 \tau}{20 \pi^2}\delta^{GW}_{00}(\Omega_0-\bar\Omega_0) \left(\frac{f}{f_0}\right)^\alpha f^{-3}  \gamma_{ij}(f) \;, \nonumber\\
	\langle r_{ij,m}(f)\rangle&=\frac{3 H_0^2 \tau}{20 \pi^2}(\Omega_0 \delta^{GW}_{\ell m}-\hat	\Omega_0\hat \delta^{GW}_{\ell m}) \left(\frac{f}{f_0}\right)^\alpha f^{-3}  \gamma_{ij,lm}(f) \;, 
\end{align}
where we introduced the shorthand notation for
\begin{align}
\gamma_{ij,\ell m}(f)\equiv\sum_{\lambda}\gamma_{\ell m,ij,\lambda}(f) \;\;,\;\; 
\gamma_{ij}(f)\equiv\gamma_{ij,00}(f) \;. 
\end{align}

\subsection{Covariance matrix and optimal \texorpdfstring{$\chi^2$}{}}\label{subsec_anisotropies_variance_all}

With an explicit evaluation of eq.~\eqref{SigmaNoiseOrSignal} under the assumption of an unpolarized background $(p=0)$, the variance matrix, evaluated for a generic fiducial value $\bar\Omega_0$ (i.e. without assuming either the signal- or the noise-dominated regime), reads
\begin{align}\label{variance_anisotropies_COMPLETE}
\Sigma&_{ij,kl,m'm''}^2\simeq\delta_{m'm''}(\delta_{ik}\delta_{jl}+\delta_{il}\delta_{jk})\frac{\tau^2}{T}\left(\frac{8\pi}{5}\right)^2\delta^{2}_{00,GW}\int_{-\infty}^{+\infty} df H^2(f)\left\vert Q_{ij,m'}(f)\right\vert^2 \nonumber\\
&\;\;\;\;\;\;\;\; \times\Bigg[\gamma_{ii}(f)\gamma_{jj}(f)+\gamma_{ij}(f)\gamma_{ji}(f)+\frac{1}{\sqrt{4\pi}}\frac{5}{16\pi}\frac{\gamma_{ii}(f)P_j(f)+\gamma_{jj}(f)P_i(f)}{H(f)}\nonumber\\
&\;\;\;\;\;\;\;\; +\frac{1}{4\pi}\left(\frac{5}{16\pi}\right)^2\frac{P_i(f)P_j(f)}{H^2(f)}\Bigg]\;. 
\end{align}
Setting in this expression the filters to maximize the $\chi^2$ as in eq.~\eqref{filters} (see appendix~\ref{appVarSignal} for details), we obtain
\begin{align}\label{chisq_opt_explicit_COMPLETE}
\chi^2_{\rm opt}&= \frac{T}{1 \text{year}}	\left[\left(\frac{\Omega_0}{\bar\Omega_0} -1\right)^2I_{00}+\frac{1}{4\pi}\left|\frac{\Omega_0}{\bar\Omega_0} \delta^{GW}_{\ell m}-\hat \delta^{GW}_{\ell m}\right|^2I_{\ell m}\right] \;, 
\end{align}
where we defined (a power law frequency dependence as in eq.~\eqref{PowerLawOmega} is assumed)
\begin{align}\label{Ilm_COMPLETE}
I_{00/\ell m}&\equiv \sum_{i\neq j}\int_{0}^{\infty}\frac{df}{\text{Hz}}6.31\times 10^7 \frac{| \gamma_{ij,00 /\ell m}(f)|^2}{D_{ij}(f)}\;,\nonumber\\
D_{ij}(f)&\equiv \gamma_{ii}(f)\gamma_{jj}(f)+\gamma_{ij}(f)\gamma_{ji}(f)+\frac{\gamma_{ii}(f)P_j(f)+\gamma_{jj}(f)P_i(f)}{\bar\Omega_0\frac{16\pi\sqrt{4\pi}}{5}\left(\frac{f}{f_0}\right)^\alpha f^{-3}\frac{3H_0^2}{32\pi^3}}+\nonumber\\
&\;\;\;\;+\frac{P_i(f)P_j(f)}{\bar\Omega_0^2\left(\frac{16\pi\sqrt{4\pi}}{5}\right)^2\left(\frac{f}{f_0}\right)^{2\alpha} f^{-6}\left(\frac{3H_0^2}{32\pi^3}\right)^2}\;.
\end{align}
We note that the result depends linearly on the total observation time $T$. In contrast, the ``auxiliary time'' $\tau$ (the interval duration over which the time stream is Fourier transformed) cancels out from the combined expression. 
Furthermore, in this regime, the coefficients $I_{00}$ and $I_{\ell m}$ do not depend on the specific frequency shape of $\Omega_{GW}(f)$ but solely on the geometry of the network (i.e. the baselines between the detectors and the orientation of their arms). This is because the anisotropy coefficients do not depend on frequency, cf. eq.~\eqref{PowerSpectrumFactorized}, so the frequency shape cancels out in the ratio  between the expected signal and the variance. 

We also note that the final $\chi^2$ can be rewritten just in terms of the relative amplitude of $\Omega_{0}$ with respect to its fiducial value $\bar\Omega_{0}$, namely on the ratio 
\begin{align}
\omega\equiv\frac{\Omega_0}{\bar\Omega_0} \;, 
\end{align}
whose fiducial value is, by definition, $\bar\omega=1$.
As a last remark, a direct inspection of $\gamma_{ij,\ell m}(f)$  in  eq.~\eqref{chisq_fisher_omplm} shows that this $\chi^2$ is symmetric under the redefinition $m\to-m$.

We conclude this subsection with two expressions showing how eq.~\eqref{Ilm_COMPLETE} simplifies in the noise-dominated and the signal-dominated regime. These expressions are shown for clarity and completeness. However, they are not used in our explicit evaluations, where the general expression for the covariance is used to make a Fisher forecast for detecting the anisotropies. In the noise-dominated regime, eq.~\eqref{Ilm_COMPLETE} reduces to
\begin{align}\label{Ilm_NOISE}
&I^N_{00/\ell m} \equiv \bar\Omega_0^2\sum_{i\neq j}\int_{0}^{\infty} \frac{df}{\text{Hz}}1.28 \times 10^{-66-4 \alpha} \,\left( \frac{f}{1 \, {\rm Hz} } \right)^{2\alpha-6} \frac{| \gamma_{ij,00/\ell m}(f)|^2}{P_i(f)P_j(f)\text{Hz}^2} \;. 
\end{align}
while in the signal-dominated case, we have
\begin{align}\label{Ilm_SIGNAL}
	&I_{00/\ell m}^S=\sum_{i\neq j}\int_{0}^{\infty}\frac{df}{\text{Hz}}6.31\times 10^7 \frac{| \gamma_{ij,00 /\ell m}(f)|^2}{\left[\gamma_{ii}(f)\gamma_{jj}(f)+\gamma_{ij}(f)\gamma_{ji}(f)\right]} \;. 
\end{align}

\subsection{Fisher forecast}\label{subsec_anisotropies_fisher}

In eq.~\eqref{chisq_opt_explicit_COMPLETE} we introduced the $\chi^2$ as a function of the parameters ($\omega$ and $\delta^{GW}_{\ell m}$) and of their fiducial values ($\bar\omega=1, \hat \delta^{GW}_{\ell m}$). We consider the null case of $\hat \delta^{GW}_{\ell m}=0$ to obtain the range of measurements that is compatible with this hypothesis at a given confidence level (C.L.) so that a measurement outside this range can be considered as an evidence of an anisotropic signal at that C.L.

The Fisher analysis consists of Taylor expanding the $\chi^2$ about an extremum (namely, about the fiducial values) to quadratic order in the parameters. One only needs the Fisher matrix
\begin{align}
\mathcal{F}_{\alpha\beta} \equiv \left. \frac 1 2\frac{\partial^2\chi^2_{\rm opt}}{\partial\alpha\partial\beta}\right|_{\substack{\alpha=\hat\alpha\\\beta=\hat \beta}} \;, 
\end{align}
which, for the set of parameters  $(\omega,\delta^{GW}_{\ell m})$ of our interest, evaluates to 
\begin{align}
	\mathcal{F}=\begin{pmatrix} \mathcal{F}_{\omega\omega} & \mathcal{F}_{\omega \,\delta_{\ell m}^{GW}} \\ \mathcal{F}_{\delta_{\ell m}^{GW}\omega} & \mathcal{F}_{\delta_{\ell m}^{GW}\delta_{\ell m}^{GW}}
	\end{pmatrix}=\frac{T}{1\text{year}}\begin{pmatrix} I_{00} & 0 \\ 0 & \frac{1}{4\pi}I_{\ell m} 
	\end{pmatrix}  \;. 
\end{align}
Correspondingly, the expanded $\chi^2$ is
\begin{align}\label{chisq_fisher_omplm}
	\chi^2_{F} (\omega,\delta^{GW}_{\ell m}) &=\begin{pmatrix} \omega-1\\ \delta^{GW}_{\ell m}- \hat \delta^{GW}_{\ell m}
	\end{pmatrix}^T
	\cdot
	\begin{pmatrix} \mathcal{F}_{\omega\omega} & \mathcal{F}_{\omega \,\delta_{\ell m}^{GW}} \\ \mathcal{F}_{\delta_{\ell m}^{GW}\omega} & \mathcal{F}_{\delta_{\ell m}^{GW}\delta_{\ell m}^{GW}}
	\end{pmatrix}\cdot
	\begin{pmatrix} \omega-1\\ \delta^{GW}_{\ell m}- \hat \delta^{GW}_{\ell m}
	\end{pmatrix}=\nonumber\\
 &=\frac{T}{1 \text{year}}\left[\left(\omega-1\right)^2 I_{00}+\frac{1}{4\pi}|\delta^{GW}_{\ell m}|^2 I_{\ell m}\right] \;, 
\end{align}
where the $I_{00}$ and $I_{\ell m}$ elements are given in eq.~\eqref{Ilm_COMPLETE} for the general case but can be approximated to eq.~\eqref{Ilm_NOISE} in the noise-dominated case and to eq.~\eqref{Ilm_SIGNAL} in the signal-dominated case, depending on the specific network of instruments used. From the $\chi^2$ one obtains the posterior distribution on the parameters $(\omega,\delta^{GW}_{\ell m})$
\begin{align}\label{posterior_anisotropies}
P(\omega,\delta^{GW}_{\ell m})\propto \Pi(\omega,\delta^{GW}_{\ell m})e^{-\frac 1 2 \chi^2_{fish}(\omega,\delta^{GW}_{\ell m})} \;, \end{align}
where $\Pi(\omega,\delta^{GW}_{\ell m})$ is the prior function on the parameters, and the proportionality is given up to a constant normalization factor. 

If one then wishes to marginalize over all the possible values of $\omega$, the forecasted posterior function (assuming a flat prior) on the parameter $\delta^{GW}_{\ell m}$ is 
\begin{align}
P\left(\delta^{GW}_{\ell m}\right)=\sqrt{\frac{\mathcal{F}_{\delta\delta}}{2\pi}} \exp(-\frac{\delta_{\ell m,GW}^2}{2}\mathcal{F}_{\delta\delta}) \;.
\end{align}
Under these assumptions, we define the error of the measurement of the anisotropic coefficient $(\ell,m)$ as $\sigma_{\delta_{\ell m}}=\mathcal{F}_{\delta\delta}^{-1}$. 

These expressions are used to obtain the numerical results of subsection~\ref{subsec:result_anisotropic}. 

\section{Detection of a net polarization}
\label{sec:Polarization}

In this section, we apply the above formalism to probe a net polarization of the SGWB. Analogously to the previous section, this analysis aims to obtain a Fisher forecast of the measurement of a net polarization with a network of interferometers. To do so, in subsection~\ref{subsec_polarization_signal_exp}, we derive the expectation value of the measurement as a function of the theoretical parameters (the amplitude of the signal and its polarization). We evaluate the covariance and Fisher matrices in subsection~\ref{subsec_polarization_all}. In subsection~\ref{subsec_polarization_fisher}, we finally derive the explicit form of a posterior function on the parameters used to describe the polarized SGWB in the Gaussian limit.

\subsection{Expectation value of the signal}\label{subsec_polarization_signal_exp} 

We evaluate eqs.~\eqref{C1-res}, \eqref{SigmaNoiseOrSignal}, and \eqref{chisquared_opt_Cm} in the case of a power spectrum described by eqs.~\eqref{def-pH_LR} and eq.~\eqref{PowerLawOmega}, for a generic value of net polarization $p$ (ranging from $-1$ to $+1$) and assuming a completely isotropic background, i.e. $\delta^{GW}_{\ell m}=0$ for every $(\ell,m)\neq(0,0)$. In this case, the only coefficient $\mathcal{C}_{ij,m}(f)$ which has a non-zero expectation value is $\mathcal{C}_{ij,0}(f)$, therefore we are only interested in computing
\begin{align}
	\langle r_{ij,0}(f)\rangle=\frac{8\pi \tau}{5}\delta^{GW}_{00}\left(\frac{f}{f_0}\right)^\alpha f^{-3}\frac{3H_0^2}{32\pi^3}\left(\Omega_0\sum_\lambda \gamma_{ij,\lambda}(f)(1+p\lambda)-\bar\Omega_0\sum_\lambda \gamma_{ij,\lambda}(f)(1+\hat p\lambda)\right)\;,
\end{align}
where $\gamma_{ij,\lambda}(f)=\gamma_{ij,00,\lambda}(f)$. Then, separating the response functions in their parity invariant and parity-violating (introduced in appendix~\ref{subapp:parity_gammas}), 
\begin{align}
\gamma_{ij,\lambda}(f)=\gamma_{ij,I}(f)+\lambda\gamma_{ij,V}(f) \;, 
\end{align}
the expectation value of the signal can then be rewritten as
\begin{align}\label{r0_polarization_expvalue}
\langle r_{ij,0}(f)\rangle=\frac{16\pi \tau}{5}\delta^{GW}_{00} \left(\frac{f}{f_0}\right)^\alpha f^{-3}\frac{3H_0^2}{32\pi^3}\left((\Omega_0-\bar\Omega_0)\gamma_{ij,I}(f)+(p\,\Omega_0-\bar p\,\bar\Omega_0)\gamma_{ij,V}(f)\right)\;.
\end{align}

\subsection{Covariance matrix and optimal \texorpdfstring{$\chi^2$}{}}\label{subsec_polarization_all}

The   covariance evaluates to
\begin{align}
\Sigma_{ij,kl}^2 &=(\delta_{ik}\delta_{jl}+\delta_{il}\delta_{jk})\frac{\tau^2}{T}\left(\frac{8\pi}{5}\right)^2\delta^{2}_{00,GW}\int_{-\infty}^{+\infty}df\,H^2(f)\left|Q_{ij}(f)\right|^2\times\nonumber\\
&\times\Bigg[\gamma_{ii,I}(f)\gamma_{jj,I}(f)+\left(\gamma_{ij,I}(f)+\hat p\gamma_{ij,V}(f)\right)^2+\nonumber\\
&+\frac{1}{\sqrt{4\pi}}\frac{5}{32\pi}\frac{\gamma_{ii,I}(f)P_j(f)+\gamma_{jj,I}(f)P_i(f)}{H(f)}+\frac{1}{4\pi}\left(\frac{5}{32\pi}\right)^2\frac{P_i(f)P_j(f)}{H^2(f)}\Bigg]\;,
\label{covariance-full-polarization}
\end{align}
since $\gamma_{ii,V}(f)=0$, and therefore
\begin{align}\label{chisq_opt_explicit_ALL_pol}
\chi^2&=\frac{T}{1 \text{year}}\sum_{i\neq j}\int_{0}^{\infty}\frac{df}{\text{Hz}}6.31\times 10^7 \frac{\left[(\omega-1)\gamma_{ij,I}(f)+(\omega p-\bar p)\gamma_{ij,V}(f) \right]^2}{D_{ij}(f)}\;,
\end{align}
with
\begin{align}\label{Dij_general}
D_{ij}(f)&=\Bigg[\gamma_{ii,I}(f)\gamma_{jj,I}(f)+\gamma_{ij,I}^2(f)+\bar p^2\gamma_{ij,V}^2(f)+2\bar p\gamma_{ij,I}(f)\gamma_{ij,V}(f)+\nonumber\\
&+\frac{\gamma_{ii,I}(f)P_j(f)+\gamma_{jj,I}(f)P_i(f)}{\bar\Omega_0\frac{32\pi\sqrt{4\pi}}{5}\left(\frac{f}{f_0}\right)^{\alpha} f^{-3}\frac{3H_0^2}{32\pi^3}}+\frac{P_i(f)P_j(f)}{\bar\Omega_0^2\left(\frac{32\pi\sqrt{4\pi}}{5}\right)^2\left(\frac{f}{f_0}\right)^{2\alpha} f^{-6}\left(\frac{3H_0^2}{32\pi^3}\right)^2}\Bigg]\;. 
\end{align}
The vector of parameters will be $\theta=\left(\begin{smallmatrix}\omega \\ p\end{smallmatrix}\right)$, and we can then perform the standard Fisher approximation 
\begin{align}
	\chi^2_F(\omega,p)&\simeq\begin{pmatrix}
	    \omega-1\\
     p-\bar p
	\end{pmatrix}^T \mathcal{F}(\hat p)\begin{pmatrix}
	    \omega-1\\
     p-\bar p
	\end{pmatrix} \;\;\;,\;\;\; 
	\mathcal{F}_{ab} \equiv \left. \frac 1 2\frac{\partial^2\chi^2(\omega,p)}{\partial a \partial b}\right|_{\substack{a=\bar a\\b=\bar b}}\;.
\end{align}
The Fisher matrix evaluates to (recall that $\bar \omega=1$ by definition)
\begin{align}\label{F_omp_general}
\mathcal{F}_{\omega\omega}&=\frac{T}{1 \text{year}}\sum_{i\neq j}\int_{0}^{\infty}\frac{df}{\text{Hz}}6.31\times 10^7 \frac{(\gamma_{ij,I}(f)+\bar p\gamma_{ij,V}(f))^2}{D_{ij}(f)}\;,\nonumber\\
\mathcal{F}_{\omega p}&= \frac{T}{1 \text{year}}\sum_{i\neq j}\int_{0}^{\infty}\frac{df}{\text{Hz}}6.31\times 10^7 \frac{(\gamma_{ij,I}(f)+\bar p\gamma_{ij,V}(f))\gamma_{12,V}(f)}{D_{ij}(f)}\;,\nonumber\\
\mathcal{F}_{pp}&=\frac{T}{1 \text{year}}\sum_{i\neq j}\int_{0}^{\infty}\frac{df}{\text{Hz}}6.31\times 10^7 \frac{\gamma_{ij,V}^2(f)}{D_{ij}(f)}\;.
\end{align}

Analogously to what was done in the previous section, we conclude this subsection with two parts, where we report the explicit evaluation of the Fisher matrix of eq.~\eqref{F_omp_general} in the noise-dominated and the signal-dominated regimes. These are presented for completeness even if the full formula in eq.~\eqref{F_omp_general} is used to perform the numerical results in our analysis. In particular, in the noise-dominated regime eq.~\eqref{F_omp_general} reduces to
\begin{align}
\mathcal{F}_{\omega\omega}^N&=\frac{T}{1 \text{year}}\sum_{i\neq j}\int_{0}^{\infty}\frac{df}{\text{Hz}}1.28 \times 10^{-66-4 \alpha} \,\left( \frac{f}{1 \, {\rm Hz} } \right)^{2\alpha-6} \frac{(\gamma_{ij,I}(f)+\bar p\gamma_{ij,V}(f))^2}{P_i(f)P_j(f)\text{Hz}^2}\;,\nonumber\\
\mathcal{F}_{\omega p}^N&= \frac{T}{1 \text{year}}\sum_{i\neq j}\int_{0}^{\infty}\frac{df}{\text{Hz}}1.28 \times 10^{-66-4 \alpha} \,\left( \frac{f}{1 \, {\rm Hz} } \right)^{2\alpha-6} \frac{(\gamma_{ij,I}(f)+\bar p\gamma_{ij,V}(f))\gamma_{12,V}(f)}{P_i(f)P_j(f)\text{Hz}^2}\;,\nonumber\\
\mathcal{F}_{pp}^N&=\frac{T}{1 \text{year}}\sum_{i\neq j}\int_{0}^{\infty}\frac{df}{\text{Hz}}1.28 \times 10^{-66-4 \alpha} \,\left( \frac{f}{1 \, {\rm Hz} } \right)^{2\alpha-6} \frac{\gamma_{ij,V}^2(f)}{P_i(f)P_j(f)\text{Hz}^2}\,,
\end{align}
while in the signal-dominated case, we have
\begin{align}
\mathcal{F}_{\omega\omega}^S&=\frac{T}{1 \text{year}}\sum_{i\neq j}\int_{0}^{\infty}\frac{df}{\text{Hz}}6.31\times 10^7 \frac{(\gamma_{ij,I}(f)+\bar p\gamma_{ij,V}(f))^2}{\gamma_{ii,I}(f)\gamma_{jj,I}(f)+\left(\gamma_{ij,I}(f)+\bar p\gamma_{ij,V}(f)\right)^2}\;,\nonumber\\
\mathcal{F}_{\omega p}^S&= \frac{T}{1 \text{year}}\sum_{i\neq j}\int_{0}^{\infty}\frac{df}{\text{Hz}}6.31\times 10^7 \frac{(\gamma_{ij,I}(f)+\bar p\gamma_{ij,V}(f))\gamma_{12,V}(f)}{\gamma_{ii,I}(f)\gamma_{jj,I}(f)+\left(\gamma_{ij,I}(f)+\bar p\gamma_{ij,V}(f)\right)^2}\;,\nonumber\\
\mathcal{F}_{pp}^S&=\frac{T}{1 \text{year}}\sum_{i\neq j}\int_{0}^{\infty}\frac{df}{\text{Hz}}6.31\times 10^7 \frac{\gamma_{ij,V}^2(f)}{\gamma_{ii,I}(f)\gamma_{jj,I}(f)+\left(\gamma_{ij,I}(f)+\bar p\gamma_{ij,V}(f)\right)^2}\;.
\end{align}

\subsection{Fisher forecast}
\label{subsec_polarization_fisher}

Finally, we can compute the posterior function of the parameters $(\omega,p)$, obtaining an expression analogous to eq.~\eqref{posterior_anisotropies}
\begin{align}
P(\omega,p)\propto \Pi(\omega,p)e^{-\frac{1}{2}\chi^2_{F}(\omega,p)}\;.
\end{align}
Under the assumption of a uniform prior on the parameter $p$, after marginalizing over all the possible values of $\omega$, the posterior distribution on the single parameter p will be
\begin{align}
P(p)\propto \exp(-\frac{p^2}{2}\frac{\mathcal{F}_{\omega\omega}\mathcal{F}_{pp}-\mathcal{F}_{\omega p}^2}{\mathcal{F}_{\omega\omega}})\;.
\end{align}
Note that since the Fisher matrix depends on $\bar p$, the posterior functions do as well. As a last remark, we note that thanks to the Schwarz inequality, we have
\begin{align}
\det(F)=\mathcal{F}_{\omega\omega}\mathcal{F}_{pp}-\mathcal{F}_{p\omega}^2\ge 0\;,
\end{align}
which means that the Fisher matrix is always definite positive, and therefore, the quadratic approximation for the $\chi^2$ is always well defined.
In this case, the error in the measurement of the polarization factor $p$, marginalized over the possible values of the amplitude of the logarithmic energy density $\Omega_0$, is given by
\begin{align}
\sigma_p=\left(\frac{\mathcal{F}_{\omega\omega}\mathcal{F}_{pp}-\mathcal{F}_{\omega p}^2}{\mathcal{F}_{\omega\omega}}\right)^{-\frac 1 2}\;.
\end{align}
These expressions are used to obtain the numerical results of subsection~\ref{subsec:result_polarized}. 

\acknowledgments 

We thank Vuk Mandic and Angelo Ricciardone for useful discussions. G.M. acknowledges support from the Imperial College London Schr\"odinger Scholarship scheme. C.R.C. acknowledges support under a UKRI Consolidated Grant ST/T000791/1. M.P. is supported by Istituto Nazionale di Fisica Nucleare (INFN) through the Theoretical Astroparticle Physics (TAsP) and the Inflation, Dark Matter and the Large-Scale Structure of the Universe (InDark) project.

\appendix
\section{Signal expectation value}
\label{app:sigvar}

As explained in section~\ref{sec:measure}, we work under the assumption that the measurements at different sites are noise-uncorrelated. For this reason, the expectation value of the estimator ${\cal C}_{ij,m}$ in eq.~\eqref{C-m} (the ones for ${\cal C}_{ij,O,m}$ and ${\cal C}_{ij,OO',m}$ are defined analogously) contains the contribution coming from the correlator of the signal component in the data stream
\begin{align}
	\left\langle {\cal C}_{ij,m}  \right\rangle =  \frac{1}{T} \int_0^{T} dt \, {\rm e}^{-i m \omega_e t} \,   \int_{-\infty}^{+\infty} df \left\langle  {\tilde s}_{i}^* \left( f ,\, t \right) {\tilde s}_{j} \left( f ,\, t \right) \right\rangle 
	{\tilde Q}_{ij}(f)\;, 
	\label{<C>}
\end{align}
between two different sites, $i \neq j$. 

To show explicitly how this component depends on the amplitude of the SGWB, we recall that, to first order in the GW perturbation, a light beam starting from $\vec{x}$ at the unperturbed time $t - 2 L$, arriving at $\vec{x} + L \, {\hat l}$, and returning back to $\vec{x}$ at the unperturbed time $t$ covers the distance in a time 
\begin{equation}
	T_{\rm return} = 2 L + \frac{{\hat l}^a \, {\hat l}^b}{2} \int_0^L d s \, h_{ab} \left( t - 2 L + s ,\, \vec{x} + s \, {\hat l}  \right) +   \frac{{\hat l}^a \, {\hat l}^b}{2} \int_0^L d s \, h_{ab} \left( t -  L + s ,\, \vec{x} + L \, {\hat l}  - s \, {\hat l}  \right)\;, 
	\label{Tret} 
\end{equation}
where $T_0 = 2 L$ is the return travel time in unperturbed Minkowski spacetime. Since our ground-based interferometers (existing and planned) are in the short arm regime, namely $2 \pi \, f \, L \ll 1$, we can approximate the GW appearing in eq.~\eqref{Tret} as $h_{ab} \left( t ,\, \vec{x} \right)$, which is constant along the integral, and therefore 
\begin{equation} 
	T_{\rm ret} = 2 L + \frac{{\hat l}^a \, {\hat l}^b}{2} \times 2 \, L \, h_{ab} \left( t ,\, \vec x \right)\;.
\end{equation}
Here we denote by $\vec{x}_{i\alpha} \left( t \right)$ the location of the vertex $\alpha$ of the $i-$th triangular-shaped detector and by $\vec{x}_{j} \left( t \right)$ the location of the $j-$th L-shaped detector of the network. In a similar manner, ${\hat u}_{i\alpha} \left( t \right)$ and  ${\hat v}_{i\alpha} \left( t \right)$ are the unit vectors in the directions of the two arms that start from this vertex of each triangular-shaped detector, as defined in \cite{Mentasti:2020yyd}. These quantities are time-dependent due to the rotation of the Earth about its axis. Under these conventions, the signal at the time $t$ at the vertex $\alpha$ of the $i$-th triangular-shaped detector, located at $\vec{x}_{i\alpha}\left( t \right)$, is 
\begin{equation}
	s_{i\alpha} (t) = d_{i\alpha}^{ab} (t) h_{ab}( t,\vec x_i(t)) \;\;,\;\; 
	d_{i\alpha}^{ab} (t) \equiv  \frac{\hat u_{i\alpha}^a(t)\hat u_{i\alpha}^b(t)-\hat v_{i\alpha}^a(t)\hat v_{i\alpha}^b(t)}{2} \;.
	\label{2arms}
\end{equation}
Analogously, since ${\hat u}_{j\alpha} \left( t \right)$ and  ${\hat v}_{j\alpha} \left( t \right)$ are the unit vectors in the directions of the two arms that start from the $j-$th L-shaped instrument, the signal at the time $t$ at the $j$-th L-shaped detector, located at $\vec{x}_{j}\left( t \right)$, is 
\begin{equation}
	s_{j} (t) = d_{j}^{ab} (t) h_{ab}( t,\vec x_i(t)) \;\;,\;\; 
	d_{j}^{ab} (t) \equiv  \frac{\hat u_{j}^a(t)\hat u_{j}^b(t)-\hat v_{j}^a(t)\hat v_{j}^b(t)}{2}\;,
	\label{2armsLshaped}
\end{equation}
With eqs.~\eqref{2arms}, \eqref{2armsLshaped} and \eqref{eqGWTT} one finds (using a compact notation as in eq.~\eqref{C-def-compact}
\begin{align}
	{\tilde s}_{i(\alpha)} \left( f ,\, t \right) &= \sum_{\lambda=R,L} \int d^2 {\hat n} \int_{-\infty}^{+\infty} d f' 
	{\rm e}^{-2 \pi i \left( f - f' \right)t} \delta_\tau \left( f - f' \right) \nonumber\\
 &\times{\rm e}^{-2 \pi i f' {\hat n} \cdot \vec{x}_{i(\alpha)} \left( t \right)} 
	h_\lambda \left( f' ,\, {\hat n} \right) e_{ab}^\lambda \left( {\hat n} \right) d_{i(\alpha)}^{ab} \left( t \right)\;,
\label{sti}
\end{align} 
where the interferometer location and arm directions can be treated as constant in the time interval $\tau$, provided that its value is sufficiently small. In analogy with \cite{Allen:1996gp} we have introduced the notation
\begin{equation}
	\delta_\tau \left( f \right) \equiv \frac{\sin \left( \pi \tau f \right)}{\pi f} \;\;\;\;,\;\;\;\; 
	\lim_{\tau \to \infty} \delta_\tau \left( f \right) = \delta_D \left( f \right) \;. 
\end{equation} 

We can now compute the correlators 
\begin{align} 
	\left\langle  {\tilde s}_{i(\alpha)}^* \left( f , t \right) {\tilde s}_{j(\beta)} \left( f' , t \right) \right\rangle =&  
	\sum_{\lambda=R,L} \int_{-\infty}^{+\infty} d f' \, \delta_\tau^2 \left( f - f' \right) H_\lambda \left( \left\vert f' \right\vert \right) 
	\int d^2 {\hat n} {\rm e}^{2 \pi i f' {\hat n} \cdot \left( \vec{x}_{i(\alpha)}\left(  t \right) - \vec{x}_{j(\beta)}\left(  t \right) \right)} \nonumber\\ 
	&\times \sum_{\ell m} \delta^{GW}_{\ell m,\lambda} Y_{\ell m} \left( {\hat n} \right) \, 
	e_{ab}^{*\lambda} \left( {\hat n} \right) e_{cd}^\lambda \left( {\hat n} \right) d_{i(\alpha)}^{ab} \left( t \right) d_{j(\beta)}^{cd} \left( t \right)\;.
\end{align}

Since the integration time $\tau$ is set to be much greater than the inverse of the typical measured GW frequencies, one of the two $\delta_\tau$ in this expression can be replaced by a Dirac $\delta-$function, while the other one evaluates to the integration time. We obtain
\begin{align} 
	\left\langle  {\tilde s}_{i(\alpha)}^* \left( f ,\, t \right) {\tilde s}_{j(\beta)} \left( f ,\, t \right) \right\rangle =&  \tau 
	\sum_{\lambda=R,L}  H_\lambda \left( \left\vert f \right\vert \right) 
	\int d^2 {\hat n} \, {\rm e}^{2 \pi i f {\hat n} \cdot \left( \vec{x}_{i(\alpha)}\left(  t \right) - \vec{x}_{j(\beta)}\left(  t \right) \right)} \nonumber\\ 
	&\times \sum_{\ell m} \delta^{GW}_{\ell m,\lambda} Y_{\ell m} \left( {\hat n} \right) \, 
	e_{ab}^{*\lambda} \left( {\hat n} \right) e_{cd}^\lambda \left( {\hat n} \right) d_{i(\alpha)}^{ab} \left( t \right) d_{j(\beta)}^{cd} \left( t \right)\;.
\label{sti-stj} 
\end{align} 
At this point, we can go from the correlation of the signal at the three vertices to the correlator of the signal in the three channels and write the first equation of eqs.~\eqref{sti-stj} as
\begin{align} 
	\left\langle  {\tilde s}_{iO}^* \left( f ,\, t \right) {\tilde s}_{jO'} \left( f ,\, t \right) \right\rangle =&  \tau 
	\sum_{\lambda=R,L}  H_\lambda \left( \left\vert f \right\vert \right) 
	\int d^2 {\hat n} {\rm e}^{2 \pi i f {\hat n} \cdot \left( \vec{x}_{i}\left(  t \right) - \vec{x}_{j}\left(  t \right) \right)} \nonumber\\
	\times&\sum_{\ell m} \delta^{GW}_{\ell m,\lambda} Y_{\ell m} \left( {\hat n} \right) 
	e_{ab}^{*\lambda} \left( {\hat n} \right) e_{cd}^\lambda \left( {\hat n} \right) d_{iO}^{ab} \left( t \right) d_{jO'}^{cd} \left( t \right)\;,
	\label{s-2ET}
\end{align} 
where $d_{iO}^{ab} \equiv c_{O\alpha} d_{i\alpha}^{ab}$. By evaluating these linear combinations, we find 
\begin{equation}
	d_{iA}^{ab} = \frac{2}{\sqrt{3}} \, d_{iX}^{ab} \;\;\;\;\;,\;\;\;\;\; 
	d_{iE}^{ab} =  - \frac{2}{3}  \left( d_{iX}^{ab}+ 2 d_{iY}^{ab} \right)   \;\;\;\;\;,\;\;\;\;\; 
	d_{iT}^{ab} = 0\;,
\end{equation}
namely, only the two channels $A$ and $E$ are non-vanishing in the short-arm approximation.
Inserting all this in eq.~\eqref{<C>}, we obtain (again trivial extension for $\langle{\cal C}_{ij,O,m}\rangle$ and $\langle{\cal C}_{ij,OO',m}\rangle$)
\begin{align} 
	\left\langle C_{ij,m} \right\rangle =&  \frac{\tau}{T} \int_0^T d t \; {\rm e}^{-i m \omega_e t } \sum_{\lambda=R,L} 
	\int_{-\infty}^{+\infty} d f  \, H_\lambda \left( \vert f \vert  \right) 
	\int d^2 {\hat n} \, 
	e_{ab}^{*\lambda} \left( {\hat n} \right) e_{cd}^\lambda \left( {\hat n} \right) \, 
	\sum_{\ell m'} p_{\ell m',\lambda} \, Y_{\ell m'} \left( {\hat n} \right) 
	\nonumber\\ 
	\times&d^{ab}_{i} \left( t \right) \, d^{cd}_{j} \left( t \right) Q_{ij} \left( f \right){\rm e}^{2 \pi i f {\hat n} \cdot \left( \vec{x}_{i}\left(  t \right) - \vec{x}_{j}\left(  t \right) \right)}\;.
	\label{Cm}
\end{align}
The quantities $d^{ab}_{i,A/E} \left( t \right) $ and $d^{ab}_{i} \left( t \right) $ are time-dependent because of the rotation of the Earth in the frame of the fixed stars. Therefore we denote by  $d^{ab}_{i,A/E} $ and $d^{ab}_{i} $ the same quantities in a frame that is fixed with respect to the Earth, 
\begin{align}\label{eq:Rotation-dab}
	&d_{i,A/E}^{ab} \left( t \right) = R_{aa'} \left( t \right) R_{bb'}  \left( t \right) \, d_{i,A/E}^{a'b'}\;,\nonumber\\
	&d_{i}^{ab} \left( t \right) = R_{aa'} \left( t \right) R_{bb'}  \left( t \right) \, d_{i}^{a'b'}\;,
\end{align} 
with $R \left( t \right)$ the rotation matrix of period $T_e$ around the $z-$axis. By changing integration variable ${\hat n} \to R {\hat n}$ in  eq.~\eqref{Cm}, we can reabsorb this rotation and take advantage of the properties of the basis tensors. We see that the rotation matrix does not appear anywhere apart from the argument of the spherical harmonic, where it produces  $Y_{\ell m'} \left( R {\hat n} \right) = {\rm e}^{i m' \omega_e t}  Y_{\ell m'}  \left(  {\hat n} \right) $. At this point, we define the position of the $i$-th interferometer in this reference frame as $\vec x_i$, in such a way that we can set, as in eq.~\eqref{eq:deltaxij},  
\begin{equation}
	\Delta \vec x_{ij}=\vec x_i-\vec x_j\;.
\end{equation}
It then follows immediately that the integration in the time variable forces $m'=m$, and 
\begin{align} 
	\left\langle C_{ij,m} \right\rangle &= \tau \sum_{\lambda=R,L} 
	\int_{-\infty}^{+\infty} d f  \, H_\lambda \left( \vert f \vert  \right) 
	\int d^2 {\hat n}  \,
	e_{ab}^{*\lambda} \left( {\hat n} \right) e_{cd}^\lambda \left( {\hat n} \right) \nonumber\\
 &\times\sum_{\ell =\vert m \vert}^\infty \delta^{GW}_{\ell m,\lambda} \, Y_{\ell m} \left( {\hat n} \right)\,d^{ab}_{i}  d^{cd}_{j} Q_{ij} \left( f \right){\rm e}^{2 \pi i f {\hat n} \cdot \Delta \vec x_{ij}}\;,
\end{align}
which leads to eq.~\eqref{C1-res} of the main text.

\section{Covariance matrix}\label{appVarSignal}

We compute the elements of the covariance matrix $\Sigma_{ij,kl,m'm''}^2$ as defined in eq.~\eqref{rijm}, considering initial contributions coming from both the instrumental noise and the variance of the signal\footnote{As remarked in the main text, for the sake of simplicity, we perform the computation assuming a single non-vanishing (and subdominant) multipole in addition to the monopole. For brevity, we do not report the details of the computation of $\Sigma_{ij(OO'),kl(O''O'''),m'm''}^2$, where triangular-shaped instruments are involved, as it follows analogous steps to the computation presented here.}:
\begin{align}
	\Sigma&_{ij,kl,m'm''}^2=\langle\mathcal{C}^*_{ij,m'}\mathcal{C}_{kl,m''}\rangle+\langle\mathcal{C}^*_{ij,m'}\rangle\langle\mathcal{C}_{kl,m''}\rangle-2\langle\mathcal{C}^*_{ij,m'}\rangle\langle\mathcal{C}_{kl,m''}\rangle=\nonumber\\
	&=\frac{1}{T^{2}} \int_{0}^{T} d t d t^{\prime} \mathrm{e}^{i m' \omega_{e} t} \mathrm{e}^{-i m'' \omega_{e} t^{\prime}}\int_{-\infty}^\infty df\int_{-\infty}^\infty df'\Bigg[\left\langle m_i(f,t)m_k^*(f',t')\right\rangle\left\langle m_j^*(f,t)m_l(f',t') \right\rangle \nonumber\\
	&\quad\quad\quad\quad \quad\quad +\left\langle  m_i(f,t) m_l(f',t')\right\rangle\left\langle m_j^*(f,t) m_k^*(f',t') \right\rangle\Bigg]Q^*_{ij,m'}(f)Q_{kl,m''}(f')\;,
\end{align}
which simplifies to
\begin{align}\label{Likelihood}
	\Sigma&_{ij,kl,m'm''}^2=\frac{1}{T^{2}} \int_{0}^{T} d t d t^{\prime} \mathrm{e}^{i m' \omega_{e} t} \mathrm{e}^{-i m'' \omega_{e} t^{\prime}}\int_{-\infty}^\infty df\int_{-\infty}^\infty df'Q^*_{ij,m'}(f)Q_{kl,m''}(f')\nonumber\\
	&\times\Bigg[\left\langle s_i(f,t)s_k^*(f',t')\right\rangle\left\langle s_j^*(f,t)s_l(f',t') \right\rangle +\left\langle  s_i(f,t) s_l(f',t')\right\rangle\left\langle s_j^*(f,t) s_k^*(f',t') \right\rangle\nonumber\\
	&+\left\langle s_i(f,t)s_k^*(f',t')\right\rangle\left\langle n_j^*(f,t)n_l(f',t') \right\rangle +\left\langle  s_i(f,t) s_l(f',t')\right\rangle\left\langle n_j^*(f,t) n_k^*(f',t') \right\rangle\nonumber\\
	&+\left\langle n_i(f,t)n_k^*(f',t')\right\rangle\left\langle s_j^*(f,t)s_l(f',t') \right\rangle +\left\langle  n_i(f,t) n_l(f',t')\right\rangle\left\langle s_j^*(f,t) s_k^*(f',t') \right\rangle\nonumber\\
	&+\left\langle n_i(f,t)n_k^*(f',t')\right\rangle\left\langle n_j^*(f,t)n_l(f',t') \right\rangle +\left\langle  n_i(f,t) n_l(f',t')\right\rangle\left\langle n_j^*(f,t) n_k^*(f',t') \right\rangle\Bigg]\;.
\end{align}
In terms of the quantities introduced in the main text, this rewrites as
\begin{align}
	&\Sigma_{ij,kl,m'm''}^2=\left(\frac{8\pi}{5}\right)^2\frac{1}{T^{2}} \int_{0}^{T} d t d t^{\prime} \mathrm{e}^{i m' \omega_{e} t} \mathrm{e}^{-i m'' \omega_{e} t^{\prime}} \int_{-\infty}^\infty df df'Q^*_{ij,m'}(f)Q_{kl,m''}(f')\nonumber\\
	&\;\;\;\;\times\int_{-\infty}^\infty df_1df_2df_3df_4\delta_\tau(f-f_1)\delta_\tau(f-f_3)\delta(f_1-f_2)\delta(f_3-f_4)\nonumber\\
	&\;\;\;\;\times\Bigg[e^{2\pi i t(f-f_1)}e^{-2\pi i t'(f'-f_2)}e^{-2\pi i t(f-f_3)}e^{2\pi i t'(f'-f_4)}\delta_\tau(f'-f_2)\delta_\tau(f'-f_4)\, \tilde{\cal T}_{ik}^* \left( f_1, f_2 \right) \, \tilde{\cal T}_{lm} \left( f_3 , f_4 \right) \nonumber\\
	&\;\;\;\;+e^{2\pi i t(f-f_1)}e^{-2\pi i t'(-f'-f_2)}e^{-2\pi i t(f-f_3)}e^{2\pi i t'(-f'-f_4)}\delta_\tau(f'+f_2)\delta_\tau(f'+f_4) \, \tilde{\cal T}_{il}^* \left( f_1, f_2 \right) \, \tilde{\cal T}_{jk} \left( f_3, f_4 \right) \Bigg]\;,
\end{align}
where we defined 
\begin{align}
\mathcal{\tilde T}_{ij} \left( f , g ,t,t'\right) &\equiv\sum_\lambda H_\lambda(f)(\delta^{GW}_{00}\tilde\gamma_{ij,00,\lambda}(f,g,t,t')+ \delta^{GW}_{\ell m}\;\tilde\gamma_{ij,\ell m,\lambda}(f,g,t,t'))+\frac{5}{16\pi}\delta_{ij}P_i(f) \;,\nonumber\\
\tilde\gamma_{\ell m,ij,\lambda}(f,g,t,t')&\equiv\frac{5}{8\pi}\int d^2\hat n e^{2\pi i\hat n(f\textbf{x}_i(t)-g\textbf{x}_j(t'))}Y_{\ell,m}(\hat n)e^{*\lambda}_{ab}(\hat n)e^{\lambda}_{bc}(\hat n)d_i^{ab}(t)d_j^{cd}(t')\;.
\end{align}
After performing the $f_2$, $f_4$ integrals we obtain
\begin{align}
\Sigma&_{ij,kl,m'm''}^2=\left(\frac{8\pi}{5}\right)^2\frac{1}{T^{2}} \int_{0}^{T} d t d t^{\prime} \mathrm{e}^{i m' \omega_{e} t} \mathrm{e}^{-i m'' \omega_{e} t^{\prime}}\nonumber\\
&\times\int_{-\infty}^{+\infty}dfdf'df_1df_3\delta_\tau(f-f_1)\delta_\tau(f-f_3)e^{2\pi i t(f_3-f_1)}e^{-2\pi i t'(f_3-f_1)}Q^*_{ij,m'}(f)Q_{kl,m''}(f')\nonumber\\
&\times\Bigg[\delta_\tau(f'-f_1)\delta_\tau(f'-f_3) \tilde{\cal T}_{ik}^* \left( f_1, f_1 ,t,t'\right) \tilde{\cal T}_{jl} \left( f_3 , f_3 ,t,t'\right)\nonumber\\
&\;\;\;\;+\delta_\tau(f'+f_1)\delta_\tau(f'+f_3) \tilde{\cal T}_{il}^* \left( f_1, f_1,t,t' \right)  \tilde{\cal T}_{jk} \left( f_3 , f_3 ,t,t'\right)	\Bigg]\;,
\end{align}
and with the change of variables $ a=t-t'$ and $b=t+t'$, this simplifies to
\begin{align}
	&\Sigma_{ij,kl,m'm''}^2=\left(\frac{8\pi}{5}\right)^2\frac{1}{2T^{2}} \int_{0}^{2T} d b\, \mathrm{e}^{i (m'-m'') \omega_{e} b}\int_{-T}^{T}d a\, \mathrm{e}^{i (m'+m'') \omega_{e} a}\nonumber\\
	&\times\int_{-\infty}^{+\infty}dfdf'df_1df_3\delta_\tau(f-f_1)\delta_\tau(f-f_3)e^{2\pi i a(f_3-f_1)}Q^*_{ij,m'}(f)Q_{kl,m''}(f')\nonumber\\
	&\times\Bigg[\delta_\tau(f'-f_1)\delta_\tau(f'-f_3)\mathcal{\tilde T}_{ik}^* \left( f_1,a\right)\mathcal{\tilde T}_{jl} \left( f_3,a\right)+\delta_\tau(f'+f_1)\delta_\tau(f'+f_3)\mathcal{\tilde T}_{il}^* \left( f_1,a\right)\mathcal{\tilde T}_{jk} \left( f_3,a\right)	\Bigg]\;,
\end{align}
where we have introduced\footnote{This can be shown to be true after a change of variables as a rotation $R(\theta)$ by an angle $\theta=\omega_{e}t'$.}
\begin{align}
\mathcal{\tilde T}_{ij} \left( f,a\right) &\equiv\sum_\lambda H_\lambda(f)(\delta^{GW}_{00}\tilde\gamma_{ij,00,\lambda}(f,a)+ \delta^{GW}_{\ell m}\;\tilde\gamma_{ij,\ell m,\lambda}(f,a))+\frac{5}{16\pi}\delta_{ij}P_i(f) \;,\nonumber\\
\tilde\gamma_{\ell m,ij,\lambda}(f,a)&\equiv\frac{5}{8\pi}\int d^2\hat n e^{2\pi i f\hat n(\textbf{x}_i(a)-\textbf{x}_j)}Y_{\ell,m}(\hat n)e^{*\lambda}_{ab}(\hat n)e^{\lambda}_{bc}(\hat n)d_i^{ab}(a)d_j^{cd}\;.
\label{def-calT}
\end{align}
The integration in $db$ then becomes trivial. This is because the covariance is built by products of two correlations taken at two different times $t$ and $t'$ between different pairs. As we are assuming stationary statistics of both the signal and the noise, the covariance is therefore insensitive to the ``absolute'' time $\frac{b}{2} = \frac{t+t'}{2}$, so that only the dependence in $a = t-t'$ is nontrivial. This leads to 
\begin{align}
\label{sigmaijkl_general}	&\Sigma_{ij,kl,m'm''}^2=\delta_{m'm''}\left(\frac{8\pi}{5}\right)^2\frac{1}{T}\int_{-T}^{T}d a\, \mathrm{e}^{2i m' \omega_{e} a}\nonumber\\
	&\times\int_{-\infty}^{+\infty}dfdf'df_1df_3\delta_\tau(f-f_1)\delta_\tau(f-f_3)e^{2\pi i a(f_3-f_1)}Q^*_{ij,m'}(f)Q_{kl,m'}(f')\nonumber\\
	&\times\Bigg[\delta_\tau(f'-f_1)\delta_\tau(f'-f_3)\mathcal{\tilde T}_{ik}^* \left( f_1,a\right) \mathcal{\tilde T}_{jl} \left( f_3,a\right)+\delta_\tau(f'+f_1)\delta_\tau(f'+f_3)\mathcal{\tilde T}_{il}^* \left( f_1,a\right)\mathcal{\tilde T}_{jk} \left( f_3,a\right)	\Bigg]\;,
\end{align}
where it is now manifest that the covariance matrix is diagonal in $m',m''$.

The last expression is evaluated in the remainder of this appendix. Specifically, for clarity of exposition, in subsection~\ref{app:subsec_2Lshaped} we evaluate it for the simplest case of just two L-shaped instruments, to outline the procedure adopted to  perform all the integrations. In subsection~\ref{app_variance:subsec_generalnetwork} we then extend this computation to a generic network.
 
\subsection{The simplest network: two L-shaped detectors}
\label{app:subsec_2Lshaped}

In the case of two L-shaped detectors, we can set $i=k=1$ and $j=l=2$ in eq.~\eqref{sigmaijkl_general}. We recall that we are assuming small anisotropies, namely $\delta^{GW}_{\ell m}\ll \delta^{GW}_{00}=\sqrt{4\pi}$. In this case 
\begin{align}\label{2ligos_sigdom_start}
&\Sigma_{m}^2\equiv\left(\frac{8\pi}{5}\right)^2\frac{1}{T}\int_{-T}^{T}d a\, \mathrm{e}^{i m \omega_{e} a}\int_{-\infty}^{+\infty}dfdf'df_1df_3\delta_\tau(f-f_1)\delta_\tau(f-f_3)e^{2\pi i a(f_3-f_1)}Q^*_m(f)Q_m(f')\nonumber\\
	&\times\Bigg[\delta_\tau(f'-f_1)\delta_\tau(f'-f_3)\mathcal{\tilde T}_{11}^* \left( f_1,a\right) \mathcal{\tilde T}_{22} \left( f_3,a\right) +\delta_\tau(f'+f_1)\delta_\tau(f'+f_3)\mathcal{\tilde T}_{12}^* \left( f_1,a\right) \mathcal{\tilde T}_{21} \left( f_3,a\right) 	\Bigg]\;,\nonumber\\
&\mathcal{\tilde T}_{ij} \left( f,a\right) \simeq\sum_\lambda H_\lambda(f)\delta^{GW}_{00}\tilde\gamma_{ij,\lambda}(f,a)+\frac{5}{16\pi}\delta_{ij}P_i(f) \;,
\end{align}
where we used the shorthand notation $\tilde\gamma_{ij,\lambda}=\tilde\gamma_{ij,00,\lambda}$ and $Q_m(f)\equiv Q_{12,m}(f)$. Introducing the dimensionless integration variable $u \equiv \frac{a}{T_e}$
and, making use of the periodicity of the integrand over the time $T_e$, and of the fact that $T=n T_e$ where $n$ is a large integer number, we have
\begin{align}
	\Sigma_{m}^2&=\left(\frac{8\pi}{5}\right)^2\frac{1}{n}\int_{-n}^{n}d u\,\int_{-\infty}^{+\infty}dfdf'df_1df_3\delta_\tau(f-f_1)\delta_\tau(f-f_3)e^{2\pi i u[m+T_e(f_3-f_1)]}Q^*_m(f)Q_m(f')\nonumber\\
	&\times \Bigg[\delta_\tau(f'-f_1)\delta_\tau(f'-f_3)\mathcal{\tilde T}_{11} \left( f_1,a\right) \mathcal{\tilde T}_{22} \left( f_3,a\right) +\delta_\tau(f'+f_1)\delta_\tau(f'+f_3)\mathcal{\tilde T}_{12} \left( f_1,a\right) \mathcal{\tilde T}_{21} \left( f_3,a\right) 	\Bigg]\;, 
\end{align}
where we also used the fact that $\mathcal{\tilde T}_{ij}^* \left( f,a\right)=\mathcal{\tilde T}_{ij} \left( f,a\right)$. We note that the functions $H_\lambda(f)$ and $\tilde\gamma(f)$ vary on a much slower scale than the  $\delta_\tau(f)$ functions, which can therefore be treated as Dirac $\delta$-functions, enforcing $f' \to f$ in the filter function when it multiplies the first term, and $f' \to -f$ when it multiplies the second term. Therefore, we can write
\begin{align}
\Sigma_{m}^2&=\left(\frac{8\pi}{5}\right)^2\frac{1}{n}\int_{-n}^{n}d u\,\int_{-\infty}^{+\infty}dfdf'df_1df_3\delta_\tau(f-f_1)\delta_\tau(f-f_3)e^{2\pi i u[m+T_e(f_3-f_1)]}\nonumber\\
	&\times\Bigg[\delta_\tau(f'-f_1)\delta_\tau(f'-f_3)\mathcal{\tilde T}_{11} \left( f,u\right)\mathcal{\tilde T}_{22} \left( f,u\right)\left\vert Q_m(f)\right\vert^2\nonumber\\&\;\;\;\;\;\;+\delta_\tau(f'+f_1)\delta_\tau(f'+f_3)\mathcal{\tilde T}_{12} \left( f,u\right) 	\mathcal{\tilde T}_{21} \left( f,u\right) 	Q^*_m(f)Q_m(-f)	\Bigg]\;.
 \end{align}
Now let $g=f_3-f$, $h=f_1-f$ and $k=f'-f$, noting that the second term of the sum is symmetric in $f'\to -f'$, giving 
\begin{align}\label{sigmamIku}
\Sigma_{m}^2&=\left(\frac{8\pi}{5}\right)^2\frac{1}{n}\int_{-n}^{n}d u\,\int_{-\infty}^{+\infty}df\Bigg[\mathcal{\tilde T}_{11} \left( f,u\right)\mathcal{\tilde T}_{22} \left( f,u\right)\left\vert Q_m(f)\right\vert^2+\mathcal{\tilde T}_{12} \left( f,u\right)\mathcal{\tilde T}_{21} \left( f,u\right)Q^*_m(f)Q_m(-f)	\Bigg] \nonumber\\
&\quad\quad\quad\quad \times\int_{-\infty}^{+\infty}dk\,dh\,dg\,e^{2\pi i u[m+T_e(g-h)]}\delta_\tau(h)\delta_\tau(g)\delta_\tau(k-h)\delta_\tau(k-g)\nonumber\\
&=\left(\frac{8\pi}{5}\right)\frac{\delta^{2}_{00,GW}}{n}\int_{-n}^{n}d u\,e^{2\pi i um}\int_{-\infty}^{+\infty}df\Bigg[\mathcal{\tilde T}_{11} \left( f,u\right)\mathcal{\tilde T}_{22} \left( f,u\right)\left\vert Q_m(f)\right\vert^2\nonumber\\
&\quad\quad\quad\quad+\mathcal{\tilde T}_{12} \left( f,u\right)\mathcal{\tilde T}_{21} \left( f,u\right)Q^*_m(f)Q_m(-f)	\Bigg]\int_{-\infty}^{+\infty}dk\, I(k,u)\,I(k,-u)\;,
\end{align}
where
\begin{align}\label{Iku}
	I(k,u) \equiv \int_{-\infty}^{+\infty}dg\,e^{2\pi i uT_eg}\delta_\tau(g)\delta_\tau(k-g)\;.
\end{align}
To compute this quantity explicitly we expand the $\delta_\tau(f)$ and $\sin(x)$ functions to obtain
\begin{align}
	I(k,u)&=\frac{2\cos(\pi\tau k)}{(2\pi i)^2}\int_{-\infty}^{+\infty}dg\,\frac{e^{2\pi i uT_eg}}{g(k-g)}-\int_{-\infty}^{+\infty}dg\frac{e^{-i\pi\tau k}}{(2\pi i)^2}\frac{e^{2i\pi\tau g (\frac{uTe}{\tau}+1)}}{g(k-g)}\nonumber\\
	&-\int_{-\infty}^{+\infty}dg\frac{e^{i\pi\tau k}}{(2\pi i)^2}\frac{e^{2i\pi\tau g (\frac{uTe}{\tau}-1)}}{g(k-g)}
	\equiv I_1(k,u)+I_2(k,u)+I_3(k,u)\;.
\end{align}
In the evaluation of the first term, we separate the two factors at the denominator, $\frac{1}{g(k-g)}=\frac{1}{kg}-\frac{1}{k(g-k)}$, we write the two terms in separate integrals, and we change variable $g \to g + k$ in the second one, to obtain
\begin{align}
	I_1(k,u)&=\frac{2\cos(\pi\tau k)}{k(2\pi i)^2}(1-e^{2\pi i uT_ek})\int_{-\infty}^{+\infty}dg\,\frac{e^{2\pi i uT_eg}}{g}\;. 
\end{align}
In the Euler expansion of the exponential, only the sine term provides a non-vanishing contribution to the integral as $\frac{\cos(x)}{x}$ is odd 
\begin{align}
I_1(k,u)&=\frac{2i\cos(\pi\tau k)}{k(2\pi i)^2}(1-e^{2\pi i uT_ek})\int_{-\infty}^{+\infty}dg\,\frac{\sin(2\pi uT_eg)}{g}\;,\nonumber\\
&=\frac{\cos(\pi\tau k)}{2\pi ik}(1-e^{2\pi i uT_ek})\,\text{sgn}(u)\;,
\end{align}
with $\text{sgn}(u)$ being the sign function. In an analogous manner, we find
\begin{align}
	I_2(k,u)&=-\frac 1 2\frac{e^{-i\pi\tau k}}{2\pi ik}(1-e^{2\pi i \tau (\frac{uTe}{\tau}+1)k})\text{sgn}\left(\frac{uTe}{\tau}+1\right)\;,\nonumber\\
	I_3(k,u)&=-\frac 1 2\frac{e^{i\pi\tau k}}{2\pi ik}(1-e^{2\pi i \tau (\frac{uTe}{\tau}-1)k})\text{sgn}\left(\frac{uTe}{\tau}-1\right)\;,
\end{align}
Adding the three terms, one can verify that 
\begin{align}
\int_{-\infty}^{+\infty} d k \, I(k,u)I(k,-u) &= 
\int_{-\infty}^{+\infty} d k \, 
\frac{\sin ^2(\pi  k (\tau-T_e  \left\vert u \right\vert))}{\pi ^2 k^2}
\, \theta \left( \frac{\tau}{T_e} - \left\vert u \right\vert \right) \nonumber\\
&= \left( \tau - T_e \, \left\vert u \right\vert \right) \, \theta \left( \frac{\tau}{T_e} - \left\vert u \right\vert \right) \;, 
\end{align}
where $\theta$ is the Heaviside $\theta$ function.
Therefore, eq.~\eqref{sigmamIku} becomes
\begin{align}
\Sigma_{m}^2&=\left(\frac{8\pi}{5}\right)^2\frac{1}{n}\int_{-\frac{\tau}{T_e}}^{\frac{\tau}{T_e}}d u\,(\tau-T_e  \left\vert u \right\vert)e^{2\pi i um}\nonumber\\
&\times\int_{-\infty}^{+\infty}df\Bigg[\mathcal{\tilde T}_{11} \left( f,u\right)\mathcal{\tilde T}_{22} \left( f,u\right)\left\vert Q_m(f)\right\vert^2+\mathcal{\tilde T}_{12} \left( f,u\right)\mathcal{\tilde T}_{21} \left( f,u\right)Q^*_m(f)Q_m(-f)\Bigg]\nonumber\\
&= \left(\frac{8\pi}{5}\right)^2\frac{\tau^2}{n\,T_e}\int_{-\infty}^{+\infty}df \, V_m(f)\;,
\end{align}
where we used the fact that the functions $\mathcal{\tilde T}$ are even in $u$, and where we introduced the quantity
\begin{align}
&V_m(f) \equiv 2\int_{0}^{1}d u'\,\left(1- u'\right)\cos\left(2\pi u'\frac{\tau}{T_e}m\right)\nonumber\\
&\times\Bigg[\mathcal{\tilde T}_{11} \left( f,u'\frac{\tau}{T_e}\right)\mathcal{\tilde T}_{22} \left( f,u'\frac{\tau}{T_e}\right)\left\vert Q_m(f)\right\vert^2+\mathcal{\tilde T}_{12} \left( f,u'\frac{\tau}{T_e}\right)\mathcal{\tilde T}_{21} \left( f,u'\frac{\tau}{T_e}\right)Q^*_m(f)Q_m(-f)\Bigg]\;. 
\end{align}
Next, we use the fact that the cosine term and the $\mathcal{\tilde T}$ functions vary very slowly in the domain of integration for $u'$. As a consequence, $V_m(f)$ can be well approximated by
\begin{align}
V_m(f)&\simeq \Bigg[\mathcal{\tilde T}_{11} \left( f,0\right)\mathcal{\tilde T}_{22} \left( f,0\right)\left\vert Q(f)\right\vert^2+\mathcal{\tilde T}_{12} \left( f,0\right)\mathcal{\tilde T}_{21} \left( f,0\right)Q^*_m(f)Q_m(-f)\Bigg]2\int_{0}^{1}d u'\,\left(1- u'\right)=\nonumber\\
&=\Bigg[\mathcal{T}_{11} \left( f\right)\mathcal{T}_{22} \left( f\right)\left\vert Q_m(f)\right\vert^2+\mathcal{T}_{12} \left( f\right)\mathcal{T}_{21} \left( f\right)Q^*_m(f)Q_m(-f)\Bigg]\;,\nonumber\\
\mathcal{T}_{ij} \left( f\right) &\equiv\sum_\lambda H_\lambda(f)\delta^{GW}_{00}\gamma_{ij,\lambda}(f)+\frac{5}{16\pi}\delta_{ij}P_i(f) \;,
\end{align}
where we have used that fact that the function ${\tilde \gamma}_{ij,\lambda} \left( f ,\, 0 \right)$, defined in eq.~\eqref{def-calT}, and  evaluated at $a=0$, is the monopole response function $\gamma_{ij,\lambda} \left( f \right)$ (which we evaluate in eq.~\eqref{CalM}).

Finally, we can note that the filters $Q_m(f)$ must satisfy the condition $Q_m^*(-f)=Q_m(f)$ so that the correlators in eq.~\eqref{C-def} are real. Therefore, we can write
\begin{align}
\Sigma_{m}^2&=\left(\frac{8\pi}{5}\right)^2\frac{\tau^2}{T}\int_{-\infty}^{+\infty}df\,\left\vert Q_m(f)\right\vert^2\Bigg[\mathcal{T}_{11} \left( f\right)\mathcal{T}_{22} \left( f\right)+\mathcal{T}_{12} \left( f\right)\mathcal{T}_{21} \left( f\right)\Bigg]\;.
\end{align}
where $T=nT_e$ is the total time of observation.

\subsection{The most general network}\label{app_variance:subsec_generalnetwork}

Extending the above computation to a generic network of $N$ instruments (labelled by $i=1\dots N$) is straightforward. The only nontrivial extension is related to the definition of optimal filters. Proceeding as in \ref{app:subsec_2Lshaped}, we arrive to the expression 
\begin{align}\label{sigmaijkl_computed}
\Sigma_{ij,kl,m}^2 &=\left(\frac{8\pi}{5}\right)^2\frac{\tau^2}{T}\int_{-\infty}^{+\infty}df\,Q^*_{ij,m}(f)Q_{kl,m}(f)\Bigg[\mathcal{T}_{ik} \left( f\right) \mathcal{T}_{jl} \left( f\right)+\mathcal{T}_{il} \left( f\right)\mathcal{T}_{jk} \left( f\right) \Bigg] \nonumber\\
&= \int_{-\infty}^{+\infty} df V'_{ij,kl,m} Q^*_{ij,m}(f)Q_{kl,m}(f)\;,
\end{align}
where we have introduced the quantities 
\begin{align}
V_{ij,kl,m}'(f) &\equiv \left(\frac{8\pi}{5}\right)^2\frac{\tau^2}{T}\int_{-\infty}^{+\infty}df\Bigg[\mathcal{T}_{ik} \left( f\right) \mathcal{T}_{jl} \left( f\right)+\mathcal{T}_{il} \left( f\right)\mathcal{T}_{jk} \left( f\right)	\Bigg]\;,\nonumber\\
\mathcal{T}_{ij} \left( f\right) &\equiv\sum_\lambda H_\lambda(f)\delta^{GW}_{00}\gamma_{ij,\lambda}(f)+\frac{5}{16\pi}\delta_{ij}P_i(f) \;. 
\end{align}
We use the shorthand notation to write all the possible pairs $a \equiv (i,j)$ (where $a$ ranges from $1$ to the combinatorial factor $\binom{N}{2}$). We study the $\chi^2$ (which is diagonal in $m$ and therefore we can study the problem at a fixed value of $m$).
With this notation, the $\chi^2$ is 

\begin{equation}
\chi^2= \sum_{m'=\{0,m\}} \sum_{a,b} 
\int df \, r_{a,m'}^*(f)\, Q_{a,m'}^*(f) \; \Sigma_{ab,m'}^{-2} \; 
\int df \, r_{b,m'} (f)\, Q_{b,m'} (f) \;, 
\label{chi2-app}
\end{equation}
where we recall that $r_{a,m} \equiv 
\mathcal{C}_{a,m} - \left\langle \mathcal{C}_{a,m} \right\rangle$, where the expectation value is obtained with the fiducial parameters, see eq.~\eqref{rijm}. Moreover, 
\begin{align}
\Sigma^2_{ab,m'}\equiv\begin{pmatrix} \int df V_{aa}'(f)\, \left\vert Q_{a,m'}(f)\right\vert^2 & \int df V_{ab}'(f)\,Q_{a,m'}(f)Q_{b,m'}^*(f) \\ \int df V_{ba}'(f)\,Q_{a,m'}^*(f)Q_{b,m'}(f) & \int df V_{bb}'(f)\, \left\vert Q_{b,m'}(f)\right\vert^2
\end{pmatrix}\;.
\label{Sigma-ab}
\end{align}
%$
To find the optimal filters, we set the functional derivatives of the $\chi^2$ with respect to the filters $Q_{a,m'} (f)$ to zero, and we solve the corresponding linear system. We verified through inspection of $V'_{ab}(f)$ and $V'_{ba}(f)$ that the off-diagonal terms of the covariance matrix are much smaller than the diagonal ones. Therefore, we can approximately invert eq.~\eqref{Sigma-ab} to linear order in the off-diagonal $V_{ab}'(f)$ (with $a \neq b$) as 
\begin{align}
\Sigma_{ab,m'}^{-2} \equiv\begin{pmatrix} \frac{1}{\int df V_{aa}'(f)\, \left\vert Q_{a,m'}(f)\right\vert^2} & -\frac{\int df V_{ab}'(f)\,Q_{a,m'}(f)Q_{b,m'}^*(f)}{\int df V_{aa}'(f)\, \left\vert Q_{a,m'}(f)\right\vert^2\int df V_{bb}'(f)\, \left\vert Q_{b,m'}(f)\right\vert^2} \\ -\frac{\int df V_{ba}'(f)\,Q_{a,m'}^*(f)Q_{b,m'}(f)}{\int df V_{aa}'(f)\, \left\vert Q_{a,m'}(f)\right\vert^2\int df V_{bb}'(f)\, \left\vert Q_{b,m'}(f)\right\vert^2} & \frac{1}{\int df V_{bb}'(f)\, \left\vert Q_{b,m'}(f)\right\vert^2}
\end{pmatrix}\;,
\end{align}
where we note that repeated indices such as in $V_{aa}'(f)$ do not imply a summation. We also note that $V_{ba}'(f)=V_{ab}^{*'}(f)$ and therefore the covariance matrix is hermitian. 

Inserting this expression into eq.~\eqref{chi2-app}, and setting to zero  the functional derivatives, $\frac{\delta \chi^2}{\delta Q_{c,m''}}=0$, we obtain
\begin{align}
	Q_{a,m'}^*(f)=\frac{r_{a,m'}(f)}{ V_{aa}'(f)}-\sum_{b \neq a} \left[ \frac{V_{ab}'(f) \, r_{b,m'}(f)}{ V_{aa}'(f)V_{bb}'(f)} + {\cal O }\left( V_{ab}^{'2} \left( f \right) \right) \right] \;,
\end{align}
Inserting this back into the $\chi^2$, we obtain the optimal chi-squared 
\begin{equation}
	\chi^2_{\rm opt}=\sum_{a,m'} \left\{ \int df \frac{\left\vert r_{a,m'}(f)\right\vert^2}{V'_{aa}(f)}-\sum_{b\neq a} \int df \left[ \frac{V'_{ab}(f)}{V'_{aa}(f)V'_{bb}(f)} r^*_{a,m'}(f) r_{b,m'}(f)  + {\cal O }\left( V_{ab}^{'2} \left( f \right) \right) \right] \right\} \;.
\end{equation}
A numerical inspection of this  equation shows that the contributions from the linear terms 
in the off-diagonal $V'_{ab}(f)$ are $O(10^{-4})$ suppressed with respect to the first ($b-$independent) integral, and therefore we can write, with high accuracy
\begin{equation}
\chi^2_{\rm opt}\simeq\sum_{a,m'} \int_{-\infty}^{+\infty} df \frac{\left\vert r_{a,m'} (f)\right\vert^2}{V'_{aa}(f)}\;.
\end{equation}
This shows that the covariance matrix in eq.~\eqref{sigmaijkl_computed} can be well approximated by only its diagonal part in the $(i,j),(k,l)$ pairs, namely
\begin{align}
\Sigma_{ij,ij,m'}^2=\left(\frac{8\pi}{5}\right)^2\frac{\tau^2}{T}\int_{-\infty}^{+\infty}df\,\left|Q_{ij,m'}(f)\right|^2\Bigg[\mathcal{T}_{ii} \left( f\right) \mathcal{T}_{jj} \left( f\right)+\mathcal{T}_{ij} \left( f\right)\mathcal{T}_{ji} \left( f\right)	\Bigg] \;, 
\end{align}
leading to eq.~\eqref{SigmaNoiseOrSignal} of the main text.

\section{Evaluation of the response functions}
\label{app:Gamma-MDQ}
In this appendix, we provide exact analytic expressions for the detector-independent response functions of eq.~\eqref{ga-lm-ab-cd-allf} for the monopole, dipole, and quadrupole. Although the response functions can be evaluated numerically, employing the explicit analytical expressions considerably speed up the computation of the correlators of eq.~\eqref{C1-res}. 

The expressions for the monopole and dipole can be found in \cite{anisotropy-DomckePeloso}, and we report them here for completeness, while the expression for the quadrupole is an original result of this paper. The extension to higher multipoles can be performed proceeding as we do in eqs.~\eqref{gammalmks},  \eqref{Gamma-MD}, and \eqref{Gamma-Q}. From the expression of the spherical harmonics, we can write 
\begin{align}
	&\gamma_{00,abcd,\lambda}(\kappa,\hat s)=\frac{5}{2\sqrt{\pi}}\Gamma_{abcd,\lambda}^M(\kappa,\hat s) \;, \nonumber\\
	&\gamma_{10,abcd,\lambda}(\kappa,\hat s)=-\frac{5i}{2}\sqrt{\frac{3}{\pi}}\Gamma_{abcd,\lambda}^D(\kappa,\hat s, \hat z) \;, \nonumber\\
	&\gamma_{1\pm 1,abcd,\lambda}(\kappa,\hat s)=\frac{5i}{2}\sqrt{\frac{3}{2\pi}}\left[\pm\Gamma_{abcd,\lambda}^D(\kappa,\hat s, \hat x)+i\Gamma_{abcd,\lambda}^D(\kappa,\hat s, \hat y)\right] \;, \nonumber\\
	&\gamma_{20,abcd,\lambda}(\kappa,\hat s)=\frac{5}{4}\sqrt{\frac{5}{\pi}}\left[-3\Gamma_{abcd,\lambda}^Q(\kappa,\hat s, \hat z,\hat z)-\Gamma_{abcd,\lambda}^M(\kappa,\hat s)\right] \;, \nonumber\\
	&\gamma_{2\pm 1,abcd,\lambda}(\kappa,\hat s)=\frac{5}{2}\sqrt{\frac{15}{2\pi}}\left[\pm\Gamma_{abcd,\lambda}^Q(\kappa,\hat s, \hat x,\hat z)+i\Gamma_{abcd,\lambda}^Q(\kappa,\hat s, \hat y,\hat z)\right] \;, \nonumber\\
	&\gamma_{2 \pm2,abcd,\lambda}(\kappa,\hat s)=\frac{5}{4}\sqrt{\frac{15}{2\pi}}\left[-\Gamma_{abcd,\lambda}^Q(\kappa,\hat s, \hat x,\hat x)\mp 2i\Gamma_{abcd,\lambda}^Q(\kappa,\hat s,\hat x, \hat y)+\Gamma_{abcd,\lambda}^Q(\kappa,\hat s,\hat y, \hat y)\right] \;, 
	\label{gammalmks}
\end{align}
where we introduced the following functions:
\begin{align}
	&\Gamma_{abcd,\lambda}^M(\kappa, \hat s)=\frac{1}{8 \pi} \int d^2 {\hat n}\,{\rm e}^{i\kappa {\hat n} \cdot \hat s} \, 
	e_{ab}^{*\lambda} \left(  {\hat n} \right) e_{cd}^\lambda \left(  {\hat n} \right) \;, \nonumber\\
	&\Gamma_{abcd,\lambda}^D(\kappa, \hat s,\hat v)=\frac{1}{8 \pi} \int d^2 {\hat n}\,{\rm e}^{i\kappa {\hat n} \cdot \hat s} \,	e_{ab}^{*\lambda} \left(  {\hat n} \right) e_{cd}^\lambda \left(  {\hat n} \right)\,(i\hat v\cdot\hat n) \;, \nonumber\\
	&\Gamma_{ab,cd,\lambda}^Q(\kappa, \hat s,\hat v,\hat w)=\frac{1}{8 \pi} \int d^2 {\hat n}\,{\rm e}^{i\kappa {\hat n} \cdot \hat s} \,	e_{ab}^{*\lambda} \left(  {\hat n} \right) e_{cd}^\lambda \left(  {\hat n} \right)\,(i\hat v\cdot\hat n)\,(i\hat w\cdot\hat n) \;. 
\end{align}
In \cite{anisotropy-DomckePeloso} it has been proven that the first two expressions can be rewritten in the form 
\begin{align}
	&\Gamma_{abcd,\lambda}^M(\kappa, \hat s)=\Gamma_{abcd,\lambda}(\kappa, \vec s)_{\rvert s=1} \;, \nonumber\\
	&\Gamma_{abcd,\lambda}^D(\kappa, \hat s,\hat v)=\left[\frac{1}{\kappa}\hat v_i \frac{\partial}{\partial s_i}\Gamma_{abcd,\lambda}(\kappa, \vec s)\right]_{\rvert s=1} \;, 
	\label{Gamma-MD}
\end{align}
where it was found that 
\begin{align}
	\Gamma_{abcd,\lambda}(\kappa, \vec s)&=f_A(\kappa s)\frac{\delta_{ac}\delta_{bd}+\delta_{ad}\delta_{bc}}{2}+\frac{f_B(\kappa s)}{s^2}\frac{\delta_{ac}s_bs_d+\delta_{bd}s_as_c+\delta_{ad}s_bs_c+\delta_{bc}s_as_d}{4}+\nonumber\\
	&+\frac{f_C(\kappa s)}{s^4}s_as_bs_cs_d+\lambda\frac{f_D(\kappa s)}{s}\frac{\delta_{ac}s_e\epsilon_{bde}+\delta_{bd}s_e\epsilon_{ace}+\delta_{ad}s_e\epsilon_{bce}+\delta_{bc}s_e\epsilon_{ade}}{4}+\nonumber\\
	&+\lambda\frac{f_E(\kappa s)}{s^3}\frac{s_as_cs_e\epsilon_{bde}+s_bs_ds_e\epsilon_{ace}+s_as_ds_e\epsilon_{bce}+s_bs_cs_e\epsilon_{ade}}{4} \;, 
\end{align}
and
\begin{align}\label{eq-fABCDE}
	&f_A(y)=\frac{j_1(y)}{2y}+\frac{1-y^2}{2y^2}j_2(y) \;\;,\;\; 
	f_B(y)=\frac{j_1(y)}{y}+\frac{5-y^2}{y^2}j_2(y) \;\;,\;\; 
	f_C(y)=-7\frac{j_1(y)}{4y}+\frac{35-y^2}{y^2}j_2(y)\nonumber\\
	&f_D(y)=\frac{j_1(y)}{2}-\frac{j_2(y)}{2y} \;\;,\;\; 
	f_E(y)=-\frac{j_1(y)}{2}+\frac{5j_2(y)}{2y} \;, 
\end{align}
where $j_1(y)$ and $j_2(y)$ are the spherical Bessel Functions, of, respectively, the first and the second type. On being contracted with the detector geometry, the response function acquires the form~\footnote{We denote the two response functions as $\mathcal{M}_{ij}^{\lambda}$ and $\mathcal{D}_{ij}^{\lambda}$ in the notation of \cite{anisotropy-DomckePeloso}. We note that the former quantity coincides with the monopole response function that we have introduced in eq.~\eqref{gamma} up to a constant factor, $\gamma_{00,ij,\lambda} = \frac{5}{2 \sqrt{\pi}} \, {\cal M}_{ij}^\lambda$, while $\mathcal{D}_{ij}^{\lambda}$ is a linear combination of $\gamma_{1m,ij,\lambda}$.}
\begin{align}
	\mathcal{M}_{ij}^{\lambda}(k,\hat s)=\Gamma_{abcd,\lambda}^{M}(\kappa, \hat s)d_i^{ab}d_j^{cd} \;\;\;,\;\;\; 
	\mathcal{D}_{ij}^{\lambda}(k,\hat s,\hat v)=\Gamma_{abcd,\lambda}^{D}(\kappa, \hat s,\hat v)d_i^{ab}d_j^{cd} \;, 
	\label{MD-projected}
\end{align}
where \cite{anisotropy-DomckePeloso} 
\begin{align}\label{CalM}
	\mathcal{M}_{i j}^{\lambda}(k) &=f_{A}(\kappa) \operatorname{tr}\left[d_{i} d_{j}\right]+f_{B}(\kappa)\left(d_{i} \hat{s}_{i j}\right)^{a}\left(d_{j} \hat{s}_{i j}\right)^{a}+f_{C}(\kappa)\left(d_{i} \hat{s}_{i j} \hat{s}_{i j}\right)\left(d_{j} \hat{s}_{i j} \hat{s}_{i j}\right) \nonumber\\
	&+\lambda f_{D}(\kappa)\left[d_{i} d_{j}\right]^{a b} \epsilon_{a b c} \hat{s}_{i j}^{c}+\lambda f_{E}(\kappa)\left(d_{i} \hat{s}_{i j}\right)^{a}\left(d_{j} \hat{s}_{i j}\right)^{b} \epsilon_{a b c} \hat{s}_{i j}^{c} \;,  
\end{align}
and
\begin{align}\label{CalD}
	\mathcal{D}_{ij, \lambda}&(\kappa, \hat{s}, \hat{v})=f_{A}^{\prime}(\kappa)(\hat{v} \hat{s})\left(d_id_j\right)^{a a} \nonumber\\
	&+\left[f_{B}^{\prime}(\kappa)-\frac{2f_{B}(\kappa)}{\kappa}\right](\hat{v} \hat{s})\left(d_i \hat{s}\right)^{a}\left(d_j \hat{s}\right)^{a}+\frac{f_{B}(\kappa)}{\kappa}\left[\left(d_i \hat{v}\right)^{a}\left(d_j \hat{s}\right)^{a}+\left(d_i \hat{s}\right)^{a}\left(d_j \hat{v}\right)^{a}\right] \nonumber\\
	&+\left[f_{C}^{\prime}(\kappa)-\frac{4f_{C}(\kappa)}{\kappa}\right](\hat{v} \hat{s})\left(d_i \hat{s} \hat{s}\right)\left(d_j \hat{s} \hat{s}\right)+\frac{2 f_{C}(\kappa)}{\kappa}\left[\left(d_i \hat{v} \hat{s}\right)\left(d_j \hat{s} \hat{s}\right)+\left(d_i \hat{s} \hat{s}\right)\left(d_j \hat{v} \hat{s}\right)\right] \nonumber\\
	&+\lambda(\hat{v} \hat{s})\left[f_{D}^{\prime}(\kappa)-\frac{f_{D}(\kappa)}{\kappa}\right]\left(d_i d_j \times \hat{s}\right)+\lambda \frac{f_{D}(\kappa)}{\kappa}\left(d_i d_j \times \hat{v}\right) \nonumber\\
	&+\lambda\left[f_{E}^{\prime}(\kappa)-\frac{3 f_{E}(\kappa)}{\kappa}\right](\hat{v} \hat{s})\left[\left(d_i \hat{s}\right)\left(d_j \hat{s}\right) \times \hat{s}\right]\nonumber\\
	&+\lambda \frac{f_{E}(\kappa)}{\kappa}\left\{\left[\left(d_i \hat{v}\right)\left(d_j \hat{s}\right) \times \hat{s}\right]+\left[\left(d_i \hat{s}\right)\left(d_j \hat{v}\right) \times \hat{s}\right]+\left[\left(d_i \hat{s}\right)\left(d_j \hat{s}\right) \times \hat{v}\right]\right\} \;. 
\end{align}
In these expressions, we defined  
\begin{align}
	(d_id_j)^{a b} & \equiv d_i^{a c} d_j^{c b} \;\;,\;\; 
	(d_i \hat{s})^{a} \equiv d_i^{a b} \hat{s}^{b} \;\;,\;\; 
	(d_i \hat{s} \hat{s}) \equiv d_i^{a b} \hat{s}^{a} \hat{s}^{b} \;, \nonumber\\
	\left(d_i d_j \times \hat{s}\right) & \equiv d_i^{a e} d_j^{e b} \epsilon_{a b c} \hat{s}_{c} \;\;,\;\; 
	\left[\left(d_i \hat{s}\right)\left(d_j \hat{s}\right) \times \hat{s}\right] \equiv\left(d_i \hat{s}\right)^{a}\left(d_j \hat{s}\right)^{b} \epsilon_{a b c} \hat{s}_{c} \;, \nonumber\\
	(\hat{v} \hat{s}) & \equiv \hat{v}^{i} \hat{s}^{i} \;\;,\;\; 
	(d_i \hat{v})^{a}  \equiv d_i^{a b} \hat{v}^{b} \;\;,\;\; 
	(d_i \hat{v} \hat{s})  \equiv d_i^{a b} \hat{v}^{a} \hat{s}^{b} \;, \nonumber\\
	\left(d_i d_j \times \hat{v}\right) & \equiv d_i^{a e} d_j^{e b} \epsilon_{a b c} \hat{v}_{c} \;\;,\;\; \left[\left(d_i \hat{s}\right)\left(d_j \hat{s}\right) \times \hat{v}\right]  \equiv \left(d_i \hat{s}\right)^{a}\left(d_j \hat{s}\right)^{b} \epsilon_{a b c} \hat{v}_{c} \;, \nonumber\\
	\left[\left(d_i \hat{v}\right)\left(d_j \hat{s}\right) \times \hat{s}\right] & \equiv\left(d_i \hat{v}\right)^{a}\left(d_j \hat{s}\right)^{b} \epsilon_{a b c} \hat{s}_{c} \;\;,\;\; 
	\left[\left(d_i \hat{s}\right)\left(d_j \hat{v}\right) \times \hat{s}\right]  \equiv \left(d_i \hat{s}\right)^{a}\left(d_j \hat{v}\right)^{b} \epsilon_{a b c} \hat{s}_{c} \;. 
 \label{d-v-s}
\end{align}
We extend the procedure of eq.~\eqref{Gamma-MD} to the quadrupole contribution~\footnote{Analytic response functions to higher multipoles can be obtained from higher derivatives of $\Gamma_{abcd,\lambda}$ with respect to $\vec{s}$.}
\begin{align}
	&\Gamma_{ab,cd,\lambda}^Q(\kappa, \hat s,\hat v,\hat w)=\left[\frac{1}{\kappa^2}\hat v_i\hat w_j \frac{\partial^2}{\partial s_i\partial s_j}\Gamma_{abcd,\lambda}(\kappa, \vec s)\right]_{\rvert s=1} \;, 
	\label{Gamma-Q}
\end{align}
obtaining, after projecting on the geometry of the pair of detectors 
\begin{align}
\mathcal{Q}_{ij,\lambda}(\kappa, \hat s,\hat v,\hat w) & =\Gamma_{abcd,\lambda}^{Q}(\kappa, \hat s,\hat v,\hat w)d_i^{ab}d_j^{cd} \nonumber\\ 
&=\left.\mathcal{Q}_{ij, \lambda}(\kappa, \hat{s}, \hat{v}, \hat{w})\right|_{\text {Parity invariant }}+\left.\mathcal{Q}_{ij, \lambda}(\kappa, \hat{s}, \hat{v}, \hat{w})\right|_{\text {Parity violating}} \;, 
\label{Q-projected}
\end{align}
where the parity-invariant and parity-violating terms are given, respectively, by 
\begin{align}
	&\left.\mathcal{Q}_{ij, \lambda}(\kappa, \hat{s}, \hat{v}, \hat{w})\right|_{\text {Parity invariant }}=\nonumber\\
	&\left\{\frac{f_{A}^{\prime}(\kappa)}{\kappa}(\hat{v} \hat{w})+\left[f_{A}^{\prime \prime}(\kappa)-\frac{f_{A}^{\prime}(\kappa)}{\kappa}\right](\hat{v} \hat{s})(\hat{w} \hat{s})\right\}\left(d_id_j\right)^{a a} \nonumber\\
	+&\left\{\left[\frac{f_{B}^{\prime}(\kappa)}{\kappa}-\frac{2 f_{B}(\kappa)}{\kappa^{2}}\right](\hat{v} \hat{w})+\left[f_{B}^{\prime \prime}(\kappa)-\frac{5 f_{B}^{\prime}(\kappa)}{\kappa}+\frac{8 f_{B}(\kappa)}{\kappa^{2}}\right](\hat{v} \hat{s})(\hat{w} \hat{s})\right\}\left(d_i \hat{s}\right)^{a}\left(d_j \hat{s}\right)^{a} \nonumber\\
	+&\left[\frac{f_{B}^{\prime}(\kappa)}{\kappa}-\frac{2 f_{B}(\kappa)}{\kappa^{2}}\right](\hat{v} \hat{s})\left[\left(d_i \hat{w}\right)^{a}\left(d_j \hat{s}\right)^{a}+\left(d_i \hat{s}\right)^{a}\left(d_j \hat{w}\right)^{a}\right] \nonumber\\
	+&\left[\frac{f_{B}^{\prime}(\kappa)}{\kappa}-\frac{2 f_{B}(\kappa)}{\kappa^{2}}\right](\hat{w} \hat{s})\left[\left(d_i \hat{v}\right)^{a}\left(d_j \hat{s}\right)^{a}+\left(d_i \hat{s}\right)^{a}\left(d_j \hat{v}\right)^{a}\right] \nonumber\\
	+& \frac{f_{B}(\kappa)}{\kappa^{2}}\left[\left(d_i \hat{v}\right)^{a}\left(d_j \hat{w}\right)^{a}+\left(d_i \hat{w}\right)^{a}\left(d_j \hat{v}\right)^{a}\right] \nonumber\\
	+&\left\{\left[\frac{f_{C}^{\prime}(\kappa)}{\kappa}-\frac{4 f_{C}(\kappa)}{\kappa^{2}}\right](\hat{v} \hat{w})+\left[f_{C}^{\prime \prime}(\kappa)-\frac{9 f_{C}^{\prime}(\kappa)}{\kappa}+\frac{24 f_{C}(\kappa)}{\kappa^{2}}\right](\hat{v} \hat{s})(\hat{w} \hat{s})\right\}\left(d_i \hat{s} \hat{s}\right)\left(d_j \hat{s} \hat{s}\right) \nonumber\\
	&+2\left[\frac{f_{C}^{\prime}(\kappa)}{\kappa}-\frac{4 f_{C}(\kappa)}{\kappa^{2}}\right](\hat{v} \hat{s})\left[\left(d_i \hat{w} \hat{s}\right)\left(d_j \hat{s} \hat{s}\right)+\left(d_i \hat{s} \hat{s}\right)\left(d_j \hat{w} \hat{s}\right)\right] \nonumber\\
	&+2\left[\frac{f_{C}^{\prime}(\kappa)}{\kappa}-\frac{4 f_{C}(\kappa)}{\kappa^{2}}\right](\hat{w} \hat{s})\left[\left(d_i \hat{v} \hat{s}\right)\left(d_j \hat{s} \hat{s}\right)+\left(d_i \hat{s} \hat{s}\right)\left(d_j \hat{v} \hat{s}\right)\right] \nonumber\\
	& \quad+\frac{2 f_{C}(\kappa)}{\kappa^{2}}\left[\left(d_i \hat{v} \hat{w}\right)\left(d_j \hat{s} \hat{s}\right)+\left(d_i \hat{s} \hat{s}\right)\left(d_j \hat{v} \hat{w}\right)+2\left(d_i \hat{v} \hat{s}\right)\left(d_j \hat{w} \hat{s}\right)+2\left(d_i \hat{w} \hat{s}\right)\left(d_j \hat{v} \hat{s}\right)\right] \;, 
\end{align}
and by 
\begingroup
\allowdisplaybreaks
\begin{align}
	&\left.\mathcal{Q}_{ij, \lambda}(\kappa, \hat{s}, \hat{v}, \hat{w})\right|_{\text {Parity violating }}= \nonumber\\
	&\lambda\left\{\left[\frac{f_{D}^{\prime}(\kappa)}{\kappa}-\frac{f_{D}(\kappa)}{\kappa^{2}}\right](\hat{v} \hat{w})+\left[f_{D}^{\prime \prime}(\kappa)-\frac{3 f_{D}^{\prime}(\kappa)}{\kappa}+\frac{3 f_{D}(\kappa)}{\kappa^{2}}\right](\hat{v} \hat{s})(\hat{w} \hat{s})\right\}\left(d_id_j \times \hat{s}\right) \nonumber\\
	&+\lambda\left[\frac{f_{D}^{\prime}(\kappa)}{\kappa}-\frac{f_{D}(\kappa)}{\kappa^{2}}\right](\hat{v} \hat{s})\left(d_id_j \times \hat{w}\right)+\lambda\left[\frac{f_{D}^{\prime}(\kappa)}{\kappa}-\frac{f_{D}(\kappa)}{\kappa^{2}}\right](\hat{w} \hat{s})\left(d_id_j \times \hat{v}\right) \nonumber\\
	&+\lambda\left\{\left[\frac{f_{E}^{\prime}(\kappa)}{\kappa}-\frac{3 f_{E}(\kappa)}{\kappa^{2}}\right](\hat{v} \hat{w})+\left[f_{E}^{\prime \prime}(\kappa)-\frac{7 f_{E}^{\prime}(\kappa)}{\kappa}+\frac{15 f_{E}(\kappa)}{\kappa^{2}}\right](\hat{v} \hat{s})(\hat{w} \hat{s})\right\}\left[\left(d_i \hat{s}\right)\left(d_j \hat{s}\right) \times \hat{s}\right] \nonumber\\
	&+\lambda\left[\frac{f_{E}^{\prime}(\kappa)}{\kappa}-\frac{3 f_{E}(\kappa)}{\kappa^{2}}\right](\hat{v} \hat{s})\left\{\left[\left(d_i \hat{w}\right)\left(d_j \hat{s}\right) \times \hat{s}\right]+\left[\left(d_i \hat{s}\right)\left(d_j \hat{w}\right) \times \hat{s}\right]+\left[\left(d_i \hat{s}\right)\left(d_j \hat{s}\right) \times \hat{w}\right]\right\} \nonumber\\
	&+\lambda\left[\frac{f_{E}^{\prime}(\kappa)}{\kappa}-\frac{3 f_{E}(\kappa)}{\kappa^{2}}\right](\hat{w} \hat{s})\left\{\left[\left(d_i \hat{v}\right)\left(d_j \hat{s}\right) \times \hat{s}\right]+\left[\left(d_i \hat{s}\right)\left(d_j \hat{v}\right) \times \hat{s}\right]+\left[\left(d_i \hat{s}\right)\left(d_j \hat{s}\right) \times \hat{v}\right]\right\}\nonumber\\
	&+\lambda\frac{f_{E}^{\prime}(\kappa)}{\kappa^2}\Bigg\{\left[\left(d_i \hat{v}\right)\left(d_j \hat{s}\right) \times \hat{w}\right]+\left[\left(d_i \hat{s}\right)\left(d_j \hat{v}\right) \times \hat{w}\right]+\left[\left(d_i \hat{w}\right)\left(d_j \hat{s}\right) \times \hat{v}\right]+\left[\left(d_i \hat{s}\right)\left(d_j \hat{w}\right) \times \hat{v}\right]\nonumber\\
	&+\left[\left(d_i \hat{v}\right)\left(d_j \hat{w}\right) \times \hat{s}\right]+\left[\left(d_i \hat{w}\right)\left(d_j \hat{v}\right) \times \hat{s}\right]\Bigg\} \;. 
\end{align}
\endgroup
Beside the quantities defined in eq.~\eqref{d-v-s}, we have also introduced 
\begin{align}
	(\hat{v} \hat{w}) & \equiv \hat{v}^{i} \hat{w}^{i} \;\;,\;\; 
	(d_i \hat{v} \hat{w})  \equiv d_i^{a b} \hat{v}^{a} \hat{w}^{b} \;\;,\;\; 
	\left[\left(d_i \hat{v}\right)\left(d_j \hat{s}\right) \times \hat{w}\right]  \equiv\left(d_i \hat{v}\right)^{a}\left(d_j \hat{s}\right)^{b} \epsilon_{a b c} \hat{w}_{c} \;, \nonumber\\
	\left[\left(d_i \hat{s}\right)\left(d_j \hat{v}\right) \times \hat{w}\right] & \equiv\left(d_i \hat{s}\right)^{a}\left(d_j \hat{v}\right)^{b} \epsilon_{a b c} \hat{w}_{c} \;\;,\;\; 
	\left[\left(d_i \hat{v}\right)\left(d_j \hat{w}\right) \times \hat{s}\right]  \equiv\left(d_i \hat{v}\right)^{a}\left(d_j \hat{w}\right)^{b} \epsilon_{a b c} \hat{s}_{c} \;. 
\end{align}
For a network that also contains triangular-shaped interferometers, similar to eqs.~\eqref{MD-projected} and \eqref{Q-projected} we introduce 
\begin{align}
	\mathcal{M}_{ij,O}^{\lambda}(k,\hat s)=\Gamma_{abcd,\lambda}^{M}(\kappa, \hat s)d_{i,O}^{ab}d_j^{cd} \;\;,\;\; 
	\mathcal{M}_{ij,OO'}^{\lambda}(k,\hat s)=\Gamma_{abcd,\lambda}^{M}(\kappa, \hat s)d_{i,O}^{ab}d_{j,O'}^{cd} \;, 
	\label{MDQ-projectedLtri}
\end{align}
and analogously for the dipole and quadrupole terms. 

\subsection{Monopole response function for a chiral SGWB}\label{subapp:parity_gammas}

We conclude this appendix with the explicit expression for the response function to a chiral isotropic SGWB, which we separate in a parity even and a parity odd contribution 
\begin{align}
\gamma_{ij,00,\lambda}(f)&\equiv \gamma_{ij,I}(f)+\lambda\gamma_{ij,V}(f) \;, 
\end{align}
From the first of eq.~\eqref{gammalmks}, the first of eq.~\eqref{MD-projected}, and eq.~\eqref{CalM}, the two terms acquire the form 
\begin{align}
\gamma_{ij,I}(f)&\equiv\frac{5}{2\sqrt{\pi}}\Bigg[f_{A}(\kappa) \operatorname{tr}\left[d_{i} d_{j}\right]+f_{B}(\kappa)\left(d_{i} \hat{s}_{i j}\right)^{a}\left(d_{j} \hat{s}_{i j}\right)^{a}+f_{C}(\kappa)\left(d_{i} \hat{s}_{i j} \hat{s}_{i j}\right)\left(d_{j} \hat{s}_{i j} \hat{s}_{i j}\right) \Bigg] \;,\nonumber\\
\gamma_{ij,I}(f)&\equiv\frac{5}{2\sqrt{\pi}}\Bigg[ f_{D}(\kappa)\left[d_{i} d_{j}\right]^{a b} \epsilon_{a b c} \hat{s}_{i j}^{c}+ f_{E}(\kappa)\left(d_{i} \hat{s}_{i j}\right)^{a}\left(d_{j} \hat{s}_{i j}\right)^{b} \epsilon_{a b c} \hat{s}_{i j}^{c}\Bigg] \;, 
\end{align}
where we recall that $\hat s_{ij}$ is the unit vector in the direction of the baseline between the two interferometers, see eq.~\eqref{eq:deltaxij}.

\section{Geometry of the detectors}
\label{app:geom}

\begin{figure}[ht!]
	\centerline{
		\includegraphics[width=0.6\textwidth,angle=0]{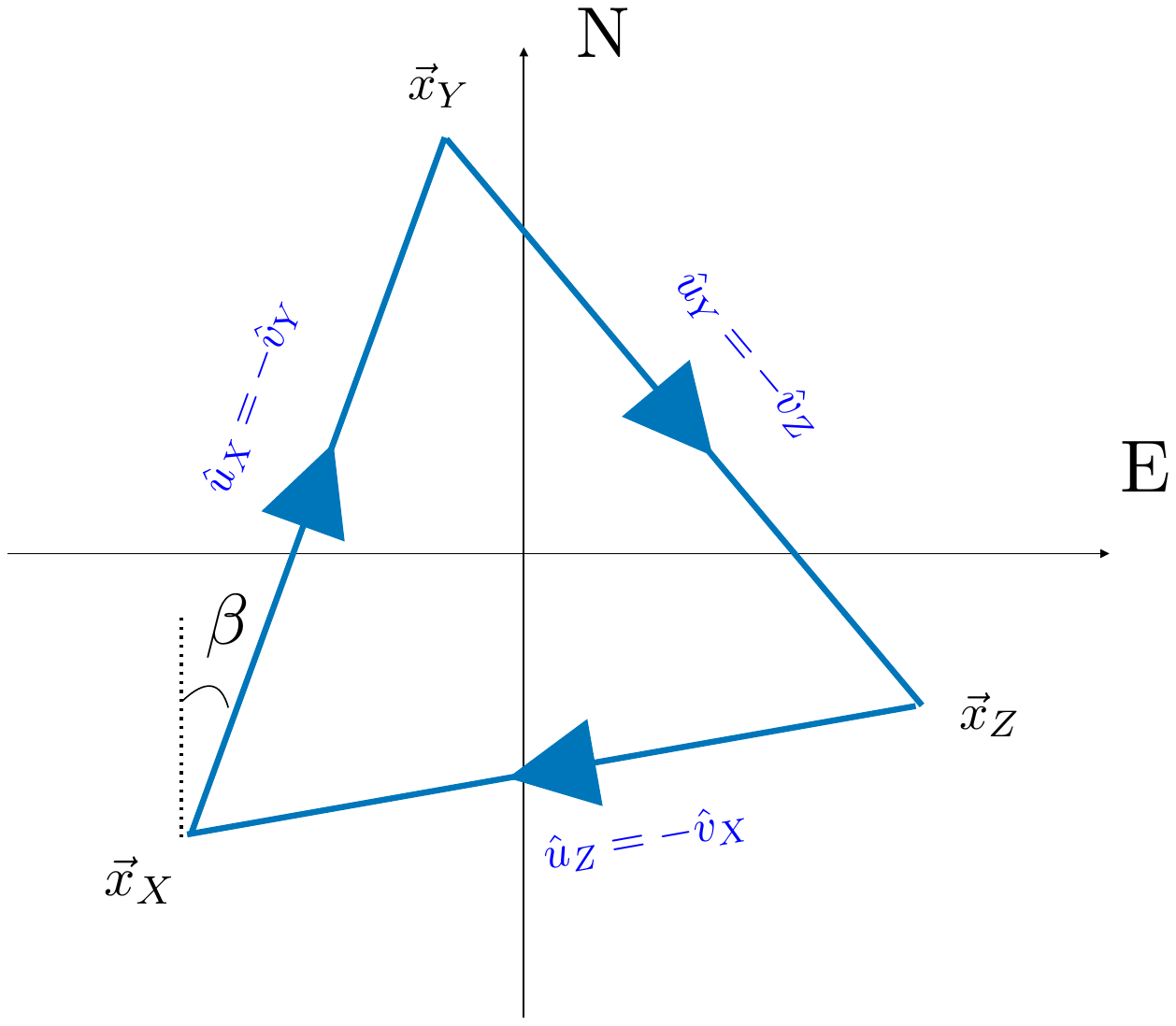}
	}
	\caption{Our convention for the orientation of the three ET arms, with the labels ``N'' and ``E'' indicating, respectively, the north and east direction at the location of the interferometer, and with $\beta$ being the angle formed by the direction ${\hat u}_X$ of the ``first'' arm and the north direction.  
	}
	\label{fig:ET2-orientation}
\end{figure}

In this appendix, we specify the position and orientation of the detectors considered in this work. Let $\theta_i$, and $\phi_i$ be the latitude and longitude of the $i$-th detector, respectively. We define latitude so that  the north pole is at $\theta =0$, while the equator is at $\theta = \pi/2$, while we use the standard notation for longitude, with $\phi =0$ corresponding to Greenwich meridian. We then consider a Cartesian system centred at the centre of the Earth (assumed to be a sphere), with the $x-$ axis (resp., the $y-$axis) points toward the location of $0$ longitude (resp., $\pi/2$ longitude) on the equator, and with the $z-$axis pointing toward the north pole. In these Cartesian coordinates, the north and east directions for a location specified by latitude and longitude $\left\{ \theta ,\, \phi \right\}$ on the surface of the Earth are given by 
\begin{align}
{\hat v}_{{\rm north}}  = 
\left\{ - \sin \theta \, \cos \phi ,\, - \sin \theta \, \sin \phi ,\, \cos \theta \right\} \;\;,\;\; 
{\hat v}_{{\rm east}} = 
\left\{ - \sin \phi ,\,  \cos \phi ,\, 0 \right\} \;. 
\end{align}
We can then specify the orientation of the detector arm ${\hat u}_{X}$ through an angle $\beta$ that it forms with the north direction, 
\begin{equation}
	{\hat u}_{X} = \cos \left( \beta \right) \, {\hat v}_{\rm north} +  \sin \left( \beta \right) \, {\hat v}_{\rm east} \;, 
	\label{orientation-plane} 
\end{equation} 
and for what concerns the L-shaped detectors, the other arm direction will be
\begin{equation}
	{\hat u}_{Y}=-\sin \left( \beta \right) \, {\hat v}_{\rm north} +  \cos \left( \beta \right) \, {\hat v}_{\rm east} \;,
\end{equation}
while for triangular-shaped interferometers, we follow the conventions indicated in figure~\ref{fig:ET2-orientation} to define the other two arm directions ${\hat u}_{Y}$ and ${\hat u}_{Z}$.

Using this convention, in table~\ref{tab:direct}, we display the actual locations of the existing interferometers LIGO, Virgo and KAGRA, with the ET in the proposed site in Sardinia, Italy (\cite{Punturo_2010}). In the table, we set $\beta=0$, with $\beta$ defined in eq.~\eqref{orientation-plane}.

\begin{figure}[ht!]
	\begin{center}
		\begin{tabular}{|c|c|c|}
			\hline  & Central location & $\{-0.338,-0.600,0.725\}$ \\
			\cline { 2 - 3 } { LIGO Hanford (or CE) } & First Arm & $\{-0.224,0.799,0.557\}$ \\
			\cline { 2 - 3 } & Second Arm & $\{-0.914,0.0261,-0.405\}$ \\
			\hline & Central location & $\{-0.0116,-0.861,0.508\}$ \\
			\cline { 2 - 3 } { LIGO Livingston } & First Arm & $\{-0.953,-0.144,-0.266\}$ \\
			\cline { 2 - 3 } & Second Arm & $\{0.302,-0.488,-0.819\}$ \\
			\hline & Central location & $\{0.712,0.132,0.690\}$ \\
			\cline { 2 - 3 } { Virgo } & First Arm & $\{-0.701,0.201,0.684\}$ \\
			\cline { 2 - 3 } & Second Arm & $\{-0.0485,-0.971,0.236\}$ \\
			\hline & Central location & $\{-0.591,0.546,0.594\}$ \\
			\cline { 2 - 3 } { KAGRA } & First Arm & $\{-0.390,-0.838,0.382\}$ \\
			\cline { 2 - 3 } & Second Arm & $\{0.706,-0.00580,0.709\}$ \\
			\hline & Central location & $\{0.751, 0.125, 0.649\}$ \\
			\cline { 2 - 3 } { ET - Sardinia } & First Arm & $\{-0.640, -0.106, 0.761\}$ \\
			\cline { 2 - 3 } & Second Arm & $\{0.178, 0.908, -0.381\}$ \\
			\cline { 2 - 3 } & Third Arm & $\{0.462, -0.801, -0.381\}$ \\
			\hline
		\end{tabular}
	\end{center}
	\caption{Cartesian coordinates of the unit-vectors specifying the positions of the interferometers	and the direction of their arms in the coordinate system described in this appendix. For ET, we consider the configuration with $\beta=0$, with $\beta$ defined in eq.~\eqref{orientation-plane}, located in the proposed site in Sardinia, Italy (\cite{Punturo_2010}).}
	\label{tab:direct}
\end{figure}

\newpage
\bibliographystyle{apsrev}
\bibliography{paper-biblio}

\end{document}